\documentclass{article}

\PassOptionsToPackage{numbers, compress}{natbib}

\usepackage[preprint]{neurips_data_2025_arxiv}
\usepackage[utf8]{inputenc} % allow utf-8 input
\usepackage[T1]{fontenc}    % use 8-bit T1 fonts
\usepackage{hyperref}       % hyperlinks
\usepackage{url}            % simple URL typesetting
\usepackage{booktabs}       % professional-quality tables
\usepackage{amsfonts}       % blackboard math symbols
\usepackage{nicefrac}       % compact symbols for 1/2, etc.
\usepackage{microtype}      % microtypography
\usepackage[dvipsnames,table]{xcolor}
\usepackage{amsmath}
\usepackage{multirow}
\usepackage[breakable]{tcolorbox}
\usepackage{graphicx}
\usepackage{multicol}
\usepackage{cleveref} 
\usepackage{listings}
\usepackage{subfigure}
\usepackage{enumitem}

\usepackage{twemojis}

\tcbset{
  mypromptbox/.style={
    colback=gray!5,
    colframe=gray!80,
    boxrule=0.5pt,
    arc=2pt,
    outer arc=1pt,
    fontupper=\ttfamily,    % <<< monospaced font!
    listing only,
    listing options={
      basicstyle=\ttfamily\small,
      breaklines=true,
      breakatwhitespace=false,
    },
  },
}

\usepackage{mdframed}

\definecolor{cornflowerblue}{rgb}{0.39, 0.58, 0.93}
\hypersetup{
    colorlinks=true,
    linkcolor=cornflowerblue,
    filecolor=magenta,      
    urlcolor=teal,
    citecolor=cornflowerblue,%teal,
    pdftitle={Community Sim},
    pdfpagemode=FullScreen,
    }

\title{PHORECAST: Enabling AI Understanding of Public Health Outreach Across Populations}

\author{
    \textbf{Rifaa Qadri} \hspace{0.5em} % Reduced space
    \textbf{Anh Nhat Nhu} \hspace{0.5em} % Reduced space
    \textbf{Swati Ramnath} \hspace{0.5em} % Reduced space
    \textbf{Laura Yu Zheng} \\ % Fourth name now on the first line
    \textbf{Raj Bhansali} \hspace{1.5em}
    \textbf{Sylvette La Touche-Howard} \hspace{1.5em}
    \textbf{Tracy Marie Zeeger} \\
    \textbf{Tom Goldstein} \hspace{1.5em}
    \textbf{Ming Lin}  \\
    University of Maryland \thanks{Correspondence to \texttt{rqadri@umd.edu}. \\ Dataset available at \url{https://huggingface.co/datasets/tomg-group-umd/PHORECAST}.}
}

\begin{document}

\maketitle
\begin{abstract}
  Understanding how diverse individuals and communities respond to persuasive messaging holds significant potential for advancing personalized and socially aware machine learning. While Large Vision and Language Models (VLMs) offer promise, their ability to emulate nuanced, heterogeneous human responses, particularly in high stakes domains like public health, remains underexplored due in part to the lack of comprehensive, multimodal dataset. We introduce {\bf PHORECAST} ({\bf P}ublic {\bf H}ealth {\bf O}utreach {\bf RE}ceptivity and {\bf CA}mpaign {\bf S}ignal {\bf T}racking), a multimodal dataset curated to enable fine-grained prediction of both individual-level behavioral responses and community-wide engagement patterns to health messaging.
  This dataset supports tasks in multimodal understanding, response prediction, personalization, and social forecasting, allowing rigorous evaluation of how well modern AI systems can emulate, interpret, and anticipate heterogeneous public sentiment and behavior. By providing a new dataset to enable AI advances for public health, PHORECAST aims to catalyze the development of models that are not only more socially aware but also aligned with the goals of adaptive and inclusive health communication.\\
  \textbf{Data and Code}: \hyperlink{Github}{github.com/rifaaQ/PHORECAST} \\
  \textbf{Dataset}: \hyperlink{Hugginface}{https://huggingface.co/datasets/tomg-group-umd/PHORECAST}
\end{abstract}

\section{Introduction}
Predictive models of human responses to persuasive messaging are a foundational challenge in behavioral modeling, with applications spanning social science, policy, and AI alignment. A key obstacle is simulating how individuals with diverse demographics, personalities, and cultural backgrounds, react to the same stimulus (e.g., an image, text, or video). While vision-language models (VLMs) offer a potential solution, it is not clear how well calibrated these models are, or how well they simulate differences between demographic groups. This gap stems from a misalignment between standard VLM training objectives (e.g., benchmark accuracy \cite{xiao2025miebmassiveimageembedding}) and the nuanced demands of behavioral simulation, a task requiring fine-grained preference elicitation and demographic-aware calibration \cite{https://doi.org/10.1002/hec.4754,LANGELLIER2016757}. To address this, we argue for domain-specific tuning of VLMs using human response data, which we demonstrate through the lens of public health messaging, a high-impact domain where tailored messaging can effectively promote awareness, shift attitudes, and inspire healthier behaviors at scale \citep{Adegoke2024,Conway_2025,Ghioe048750}.

Despite advances in behavioral science, there remains no comprehensive dataset capturing how individuals across diverse demographic backgrounds and personality profiles respond to real-world health messages. To address this gap and catalyze research on understanding individual public health preferences, we introduce a novel dataset derived from a large-scale study of over 1,000 participants. 
This dataset comprises of \textbf{30,000+} rich, granular responses to 37 public health posters spanning 7 urgent health topics (e.g., COPD, Mental Health, Nutrition, and more). Each response reflects sentiment, emotional reactions, and behavioral intent, offering unprecedented insight into the interplay between messaging design, individual differences, and community-level engagement. By pairing these responses with detailed demographic and psychometric data, we empower researchers to build predictive models of how public health campaigns are perceived by different groups.

We provide insights to study how unique individuals interact with and react to various multi-media marketing content.
We analyze and present the correlation between different demographic factors and personality traits, as well as with their individual responses to varying public health messaging. We demonstrate the utility of this dataset through two important use cases: (1) training predictive models to simulate response to public health messaging based on demographic and psychographic factors, and (2) establishing the first benchmark for evaluating how personality traits modulate emotional responsiveness to visual persuasion.  
PHORECAST enables LLM models to better align with individual preferences and values, given their demographics and/or personality traits, showcasing its strong generalization capabilities.

\begin{figure}
\vspace*{-0.75em}
    \centering
    \includegraphics[trim={5.1cm 0.6cm 0.8cm 3.5cm},clip,width=0.98\linewidth]{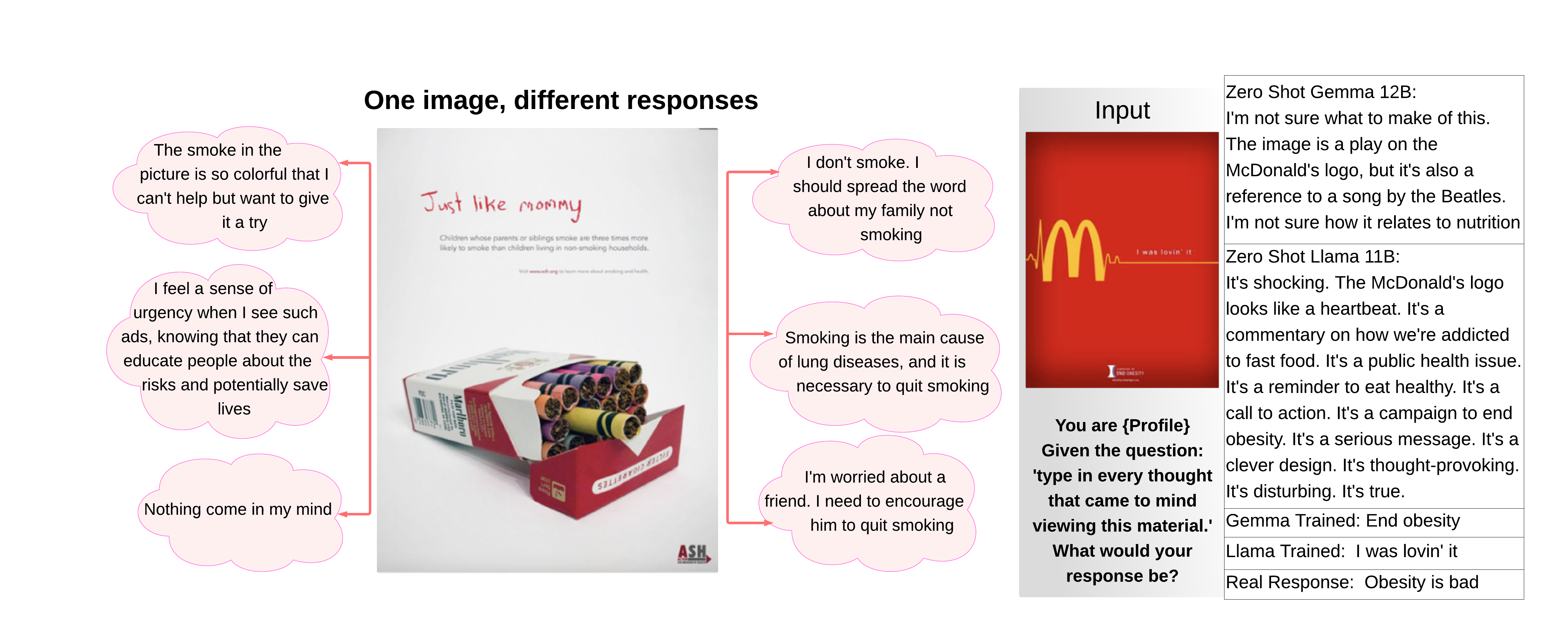}
    \vspace*{-1em}
    \caption{\textbf{Human Nuance vs Model Limitation}: Qualitative analysis demonstrates the diverse human reactions and interpretations evoked by a single image, spanning personal concern to broader advocacy. This underscore the significant influence an individual's background and context has on their perception. The right panel shows that current popular models struggle to capture this rich spectrum of human responses and language, often defaulting to repetitive language (e.g., Llama's "I'm not sure..." in 80\% of cases). By training with PHORECAST, our models learn to emulate real human language, effectively capturing these subtle distinctions in the data.}
    \label{fig:enter-label}
    \vspace*{-1.25em}
\end{figure}

\vspace*{-0.5em}
\section{Related Work}
\vspace*{-0.5em}

\textbf{Simulating Human Behavior with Language Models.}
Recently, a lot of work explores the idea of using large language models (LLMs) as simulators for human behavior. 
Park et al.~\cite{Park2023GenerativeAgents} is one of the first works to investigate emergent human interaction behavior by simulating a sandbox human community with multiple LLM instances. 
This inspired many branch-off topics involving LLM agents, especially for human behavior simulation, a popular and well-motivated area, particularly for healthcare studies or for commerce platform optimization~\cite{lu2025llmagentsactlike}.
One large motivator for human behavior simulation research is the prospect of being able to simulate large-scale social media populations to major political events or digital campaigns. 
Qiu et al.~\cite{qiu2025can} investigates the scope of simulating social media behavior through action-conditioned free text responses, where the actions can be either ``like", ``reply", or ``quote". The human data scraped from X revolved around major political events. 
They find that baseline GPT and Deepseek models are biased heavily towards selecting ``quote" over other actions, which may suggest that complementary text is preferred over direct replies or text-free likes. 
Another study on social media simulation~\cite{li2024fine} found that historical context was by-far the most important information for accurate simulation of human responses, compared to user interests and user info (such as demographics). 
Unlike previous work, Xie et al.~\cite{xie2025human} built a cognitive science-inspired framework for the simulation of detailed human backgrounds, offering a much more explicit and robust way to construct simulated human personas.
Instead of other works using personality tests like the Meyers-Briggs test~\cite{song2024identifyingmultiplepersonalitieslarge}, Xie et al. construct the first framework that uses Jung's psychology theory. In contrast to previous work, our dataset introduces a domain-specific behavior-dependent prediction in the public health domain. The responses include not only free responses, but also self-reported personality evaluations according to the Big Five Inventory~\cite{john1991big_bigfive}, as well as detailed demographic information.

\begin{figure}
    \centering
    \includegraphics[trim={0.25cm 0.8cm 0.3cm 0.3cm},clip,width=1\linewidth]{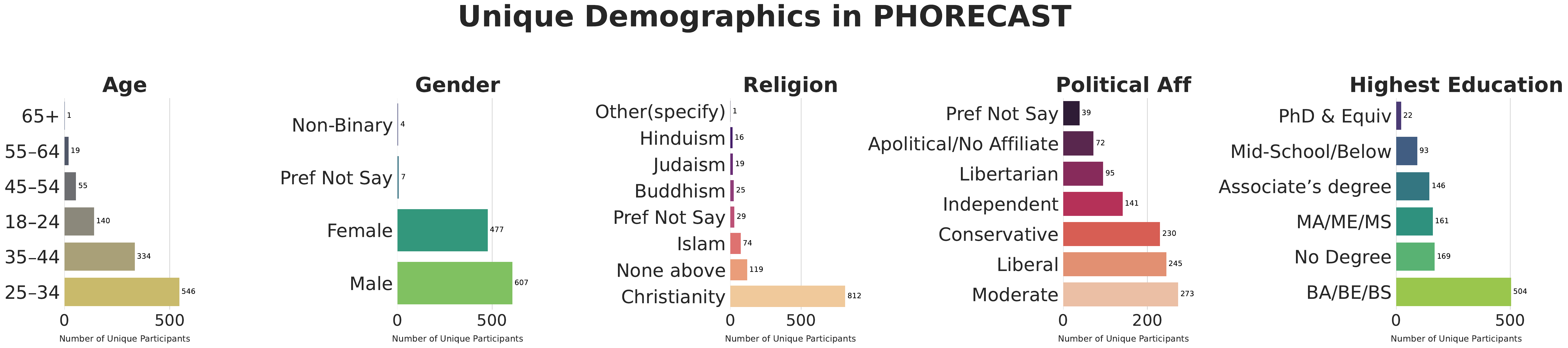}
    \vspace*{-0.5em}
    \caption{Demographic distribution of participants (N=1095) in our study, showing age groups, gender identity, religion, political affiliation, and highest education attainment. The dataset reflects a broad representation across ages (predominantly 18-44), gender (balanced male/female, inclusive non-binary options), and political views (moderate, liberal, and conservative as most frequent). }
    \label{fig:enter-label}
    \vspace*{-1em}
\end{figure}

\section{The PHORECAST Dataset}
\begin{figure}[hb]
    \centering
    \includegraphics[trim={2.5cm 2cm 3cm 3cm},clip,width=0.7\linewidth]{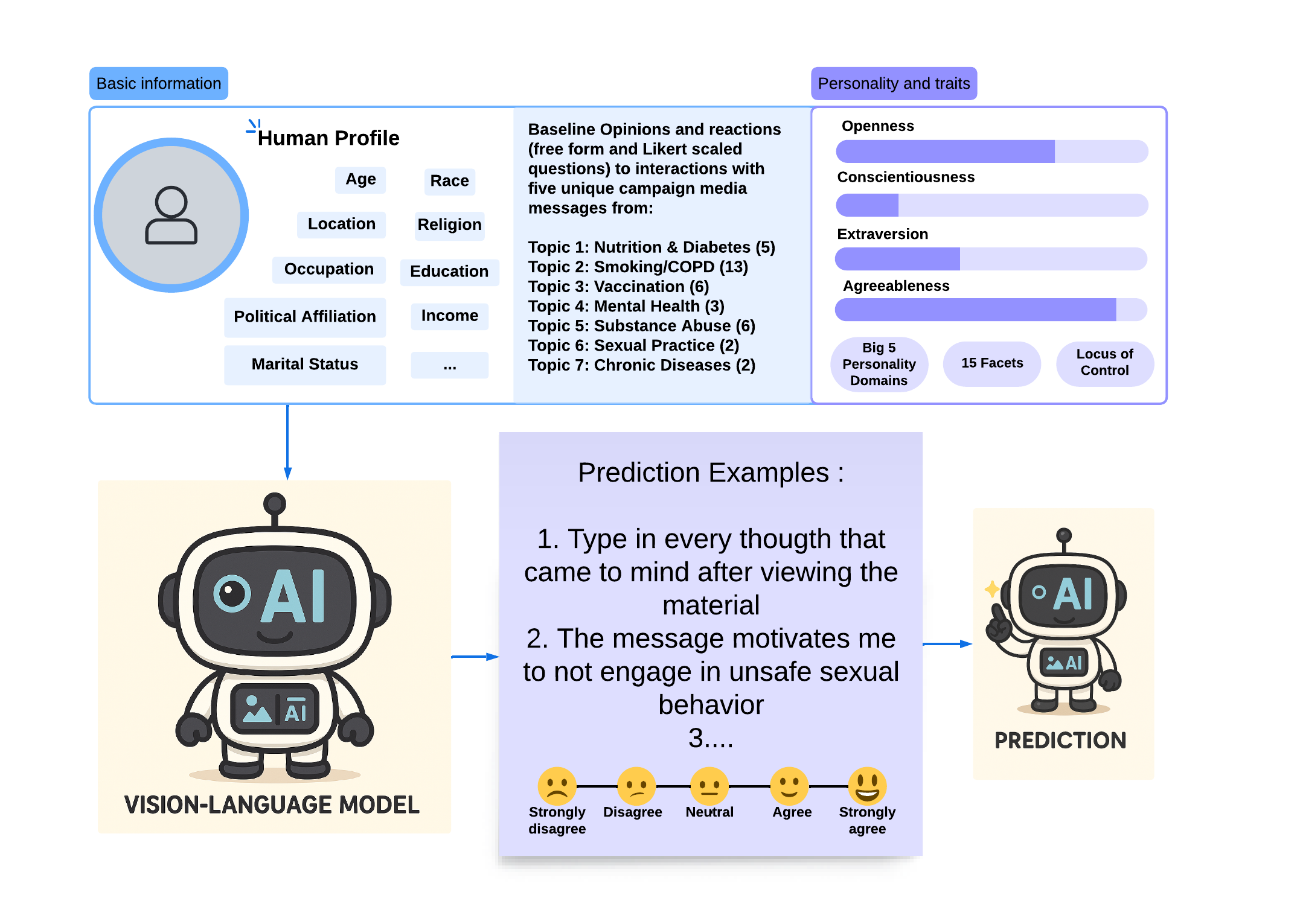}
    \vspace*{-0.5em}
    \caption{{\bf Overview of PHORECAST Pipeline}: Via our Survey, we collect human profiles including demographics, personality, locus of control, and opinions on public health topics before and after their interaction with the campaign message. We then train LLM/VLM models to predict different reactions of an individual given a stimuli.
    }
    \label{fig:overview}
    \vspace*{-1em}
\end{figure}

\vspace*{-0.5em}
\label{dataset}
PHORECAST ({\bf P}ublic {\bf H}ealth {\bf O}utreach {\bf RE}ceptivity and {\bf CA}mpaign {\bf S}ignal {\bf T}racking) is a multimodal dataset designed to evaluate how vision-language models (VLMs) predict human reactions to public health campaigns, conditioned on demographic and psychological factors. It comprises survey responses from diverse U.S. participants, linking structured annotations of health media to rich individual profiles. We recruit participants (age $\geq$ 18, U.S. residents) who provide informed consent and complete a 30-minute anonymized survey. Duplicate IP addresses are filtered to ensure uniqueness. Each participant (1) Profiles their Background: Reports demographics, personality traits, the locus of control and baseline health opinions on five randomly selected topics (Section~\ref{survey}), (2) Reviews Campaigns: Reacts to five randomly assigned public health campaigns (based on the randomly selected topics) via free-form text and Likert-scale ratings (Section~\ref{opinions}), as shown in Fig.~\ref{fig:overview}.

\subsection{Survey details}
\label{survey}
\vspace*{-0.5em}
\paragraph{Health Topics:}\label{health_topics}
% (from our team)
Public health experts from our team curate campaigns from the web and annotate with target behavior (e.g., smoking cessation), target population, and message type (Informative, Persuasive-Efficacy, or Persuasive-Threat).
Each participant is assigned five random topics at the start of the survey from seven categories: 
Nutrition \& Diabetes, Vaccination/HIV/AIDs, Mental Health, Substance Abuse, Sexual Practices, COPD/Smoking, and Chronic Diseases (which includes Heart Disease, Cystic Fibrosis, and Arthritis).

\begin{figure}
    \centering
    \includegraphics[trim={0.3cm 0.2cm 0.2cm 0.0cm},clip,width=0.49\linewidth]{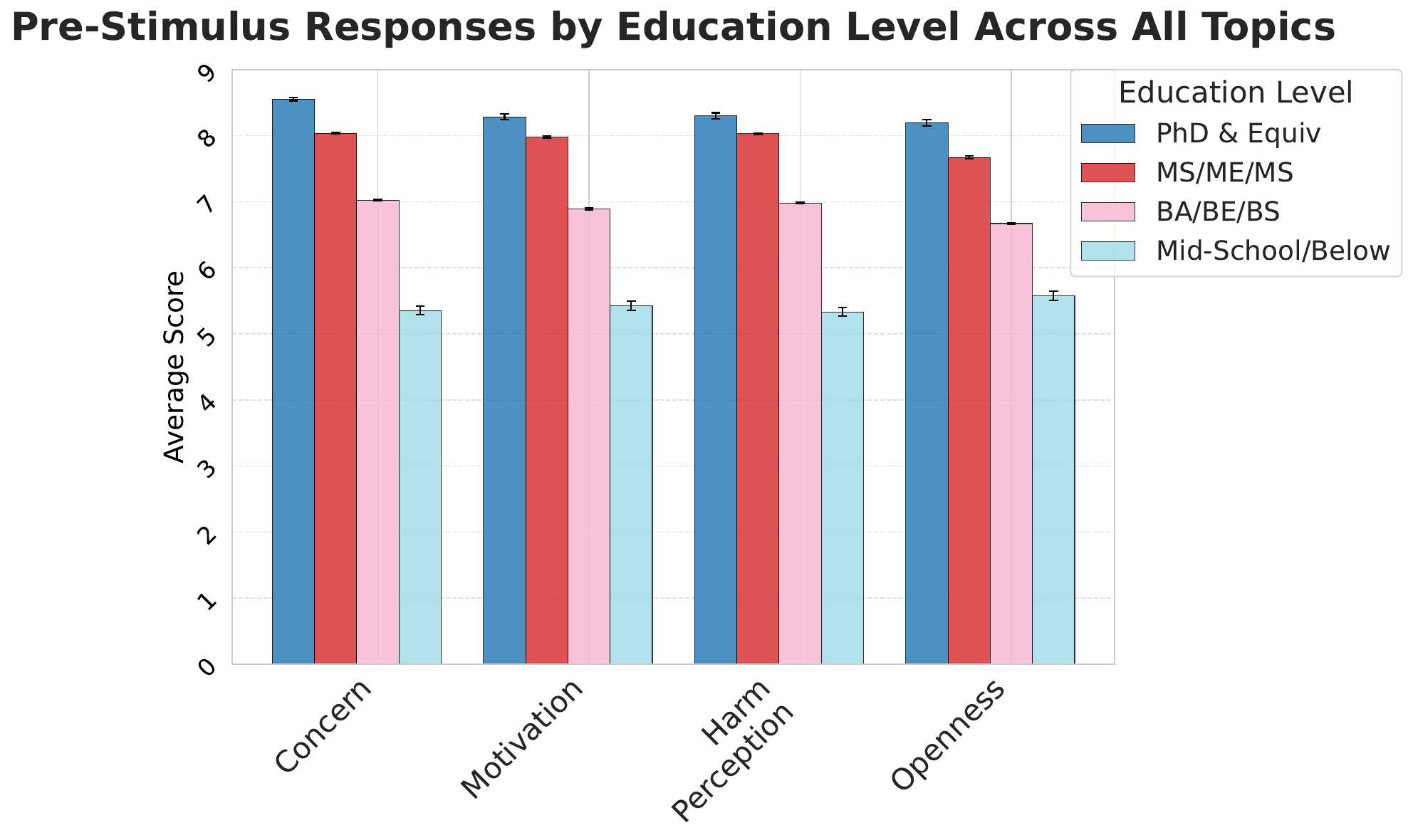}
    \includegraphics[trim={0.3cm 0.55cm 0.2cm 0.0cm},clip,width=0.50\linewidth]{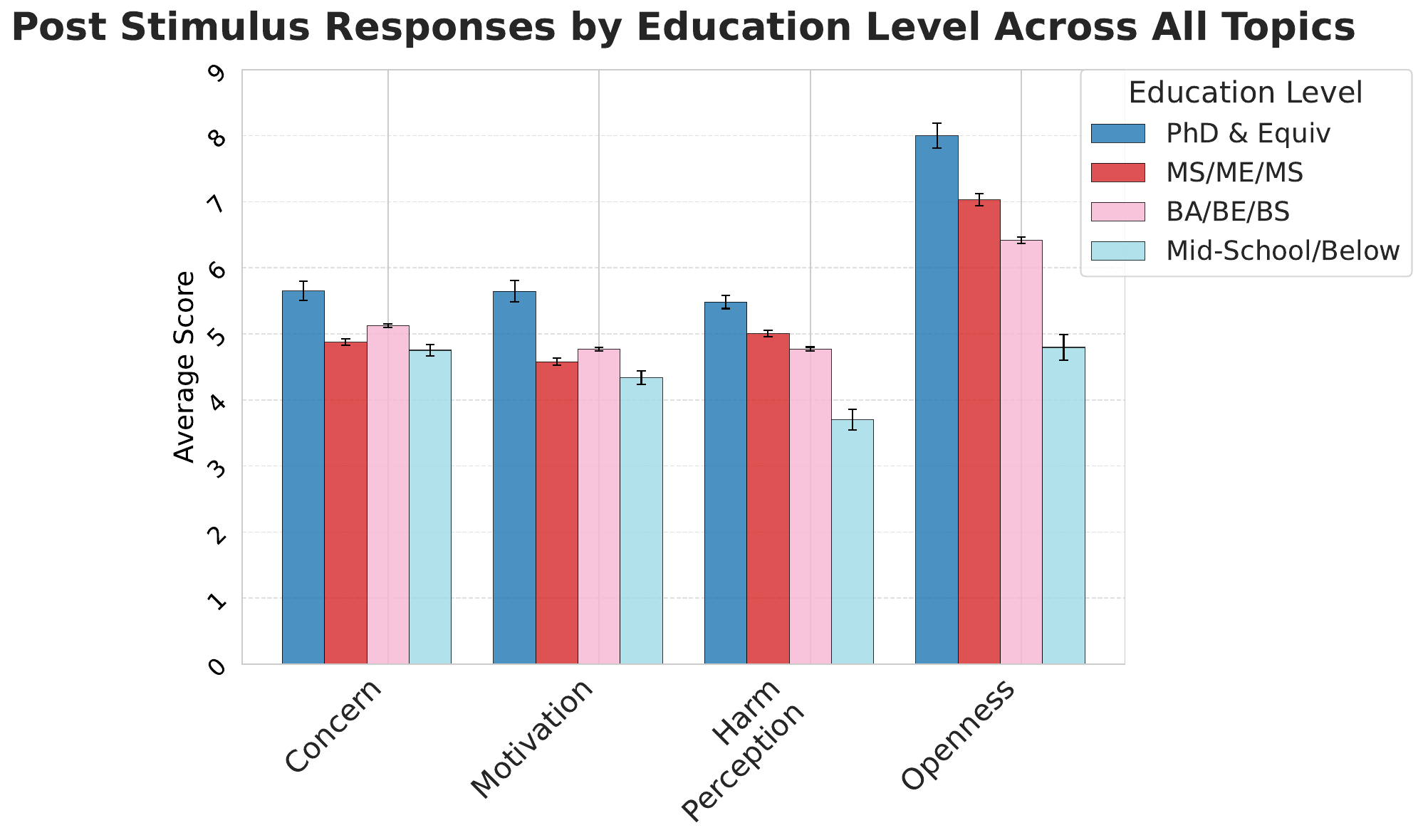}
    \vspace*{-2em}
    \caption{Differential opinion patterns by education level before and after interacting with stimuli across all topics. Generally, individuals with higher education attainment (Doctoral and Masters; N=183) demonstrate (1) significantly greater concern about different aspects of their health, (2) stronger harm perception of health harms, (3), paradoxically, greater self-reported willingness to engage in harmful behavior such as substance use or smoking. This attitude behavior gap suggests that while higher education enhances risk awareness, it may simultaneously increase behavioral intentions, possibly through increase perceived self behavioral control \cite{Assari2019}. A detailed demographic and psychographic analysis for each topic is provided in the appendix. 
    }
    \label{fig:opinions-by-education}
    \vspace*{-1em}
\end{figure}

\vspace*{-0.5em}
\paragraph{Basic demographics:}
\label{demographics}
We request the following demographics features from each participant: age, gender, assigned sex at birth, religious or cultural affiliation, political affiliation, race, ethnicity, primary \& first language, educational attainment, employment status, current profession, annual household income, marital status, family status, physical or health conditions, and zip code. A “Prefer not to say” option is offered for each demographic question. For gender, participants can select \textit{Male}, \textit{Female}, \textit{Non-Binary/Third Gender}, or \textit{self-identify}. We collect self-described ethnicity, religion and political affiliations, employment, marital and family status, as well as first language. This set of demographic information captures the subtle differences in background and current factors that may influence an individual's perception and/or opinions on diverse issues. 

\vspace*{-0.5em}
\paragraph{Personality:} 
\label{personality}
Each participant completes The Big Five Inventory-2 (BFI-2) \cite{Soto2017BFI2}, a 60-item questionnaire designed to measure the Big Five personality domains (Extraversion, Agreeableness, Conscientiousness, Negative Emotionality, and Open-Mindedness) and their 15 facets: sociability, assertiveness, energy, compassion, respectful, trust, organization, productive, responsibility, anxiety, depression, emotional volatility, intellectual curiosity, aesthetic sensibility and creative imagination.
% \rifaa{} facets and questions can be put in appendix

\vspace*{-0.5em}
\paragraph{Locus of Control:} The locus of control measures one's tendency to perceive events as internally or externally controlled. Prior research links an internal locus of control to greater psychological well-being \cite{shojaee2014mentalhealth}. Participants rate four statements ("\textit{I’m my own boss}", "If \textit{I work hard, I succeed}", "W\textit{hat I do is determined by others}", "\textit{Fate often gets in the way of my plans}") on a five-point scale ("\textit{Does not apply at all}" to "\textit{Applies completely}"). Total scores (4-20) categorize respondents into: Low internal (high external) locus (4–11), moderate balance (12-15), and high internal locus (16-20). 

\vspace*{-0.5em}
\paragraph{Baseline Opinions:} 
\label{baseline_q}
% Each participants is randomly assigned five topics at the start of the survey (Sec.~\ref{health_topics}), 
Each participant rate pre-existing concern, motivation, harm perception, and openness, on a 9-point Likert scale (1 = Not at all, 9 = Extremely) on the five topics randomly assigned at the start of the survey. Concern reflects worry about risks from unhealthy behaviors (e.g. poor diet), motivation captures intent to adopt healthy behaviors (e.g., vaccinations), harm perception assesses beliefs about the consequences of neglecting safe practices, and openness gauges willingness to engage in health-supportive behaviors. For inherently harmful topics (e.g., smoking), openness instead reflects receptivity to those behaviors.

\subsection{Opinion Indicators}
\label{opinions}
\vspace{-0.1cm}
We assess participants' reactions to health campaigns through structured and open-ended measures. Emotional responses are quantified via eight discrete emotions (sadness, anger, fear, guilt, disgust, worry, shame, and hope), rated on a 9-point Likert scale (1 = Not at all, 9 = Extremely). We reevaluate the four baseline constructs --concern, motivation, harm perception, and openness-- but now specifically framed by the campaign content: (1) concern reflects message-induced worry about health risks, (2) motivation captures behavior change intent, (3) harm perception assesses consequences of non-compliance, and (4) openness measures receptivity to recommendation (reverse coded for harmful behaviors like smoking). Finally, participants provide free-form responses to '{\em type every thought that came to mind when viewing this material}', yielding qualitative data that complements quantitative ratings.

\section{Training and Benchmarking LLM/VLM Models Using PHORECAST}
\label{training-section}
\vspace*{-0.2em}

We evaluate standard VLMs on their ability to predict human responses both with and without PHORECAST.
We employ a stratified hold-out strategy across gender, religion, and race/ethnicity demographics, as described in detail in the supplementary. This approach allows for the assessment of model performance on underrepresented groups and the quantification of any systematic biases in response predictions, shown and discussed in \ref{tab:bias_summary}. Additionally, a representative image for each topic is excluded from the training and included in the validation set to evaluate models' generalization capabilities to novel campaign visuals across all public health campaigns. We employ Low-Rank Adaptation (LoRA) for efficient fine-tuning of models, optimizing two parallel objectives: a) Free-form natural language response generation, b) Likert-scale opinion prediction. To enhance robustness for real-world deployment scenarios with partial user data, we implement feature randomization protocol during training: (1) Demographic/Psychographic Features: 90\% random sampling per participant (2) Locus of Control: 75\% random sampling per participant, (3) Contextual Q/A pairs: 30\% sampling of question-answer pairs with variable feature subsets (randomly selected combinations of available features). This stochastic regime forces models to operate under partial information, simulating real world constraints where complete user profiles are unavailable.

\begin{table}[ht]
% \vspace*{-0.5em}
\centering
\caption{\textbf{Demographic Analysis Before and After Training.} We prompt the model with personality and demographics information and calculate the accuracy across subgroups, revealing demographic disparities and changes after training using $\pm2$ tolerance threshold. We compare Gemma and Llama pre- and post-training. Notable trends: (1) accuracy improves overall, with Llama benefiting much more (over $17\%$) increase on average, (2) disparities persist for some groups, especially underrepresented samples, such as agnostics (with only 24 samples) and non-binary ($36$ samples).}
% \rifaa{actually not sure if this should be called bias}
\label{tab:bias_summary}
\resizebox{\textwidth}{!}{%
\small
\setlength{\tabcolsep}{6pt}
\renewcommand{\arraystretch}{1.15}
\begin{tabular}{lccccc}
\toprule
\textbf{Demographic Group} & \textbf{Subgroup (Samples)} & \textbf{Gemma} (Pre $\rightarrow$ Post) & \textbf{Gemma $\Delta$} & \textbf{Llama} (Pre $\rightarrow$ Post) & \textbf{Llama $\Delta$} \\
\midrule

\rowcolor{gray!8}

\rowcolor{gray!8}
\textbf{Age} & 18--24 (994) & 65.39\% $\rightarrow$ 63.48\% & -1.91 & 48.09\% $\rightarrow$ 65.59\% & +17.50 \\
& 25--34 (3867) & 68.37\% $\rightarrow$ 64.70\% & -3.67 & 55.29\% $\rightarrow$ 65.62\% & +10.33 \\
& 35--44 (2419) & 66.76\% $\rightarrow$ 71.64\% & +4.88 & 49.81\% $\rightarrow$ 73.97\% & +24.16 \\
& 45--54 (479) & 66.81\% $\rightarrow$ 82.05\% & \underline{+15.24} & 51.98\% $\rightarrow$ 85.62\% & \underline{+33.64} \\
& 55--64 (108) & \textbf{63.89\%} $\rightarrow$ 86.11\% & \underline{+22.22} & \textbf{50.00\%} $\rightarrow$ 86.11\% & \underline{+36.11} \\
& 65+ (12) & \textbf{50.00\%} $\rightarrow$ 66.67\% & +16.67 & 66.67\% $\rightarrow$ \textbf{58.33\%} & -8.34 \\

\rowcolor{gray!3}
\textbf{Gender} & Male (4081) & 67.36\% $\rightarrow$ 67.51\% & +0.15 & 66.80\% $\rightarrow$ 69.92\% & +3.12 \\
& Female (3738) & 67.07\% $\rightarrow$ 68.65\% & +1.58 & \textbf{52.17\%} $\rightarrow$ 69.54\% & +17.37 \\
& Non-Binary (36) & 88.89\% $\rightarrow$ \textbf{58.33\%} & \textbf{-30.56} & 72.22\% $\rightarrow$ \textbf{52.78\%} & -19.44 \\
& Prefer Not to Say (24) & 66.67\% $\rightarrow$ \textbf{52.60\%} & -14.07 & \textbf{50.00\%} $\rightarrow$ 53.65\% & +3.65 \\

\rowcolor{gray!8}
\textbf{Religion} & Christianity (5936) & 64.60\% $\rightarrow$ 67.05\% & +2.45 & 53.82\% $\rightarrow$ 69.34\% & +15.52 \\
& Judaism (131) & \textbf{52.67\%} $\rightarrow$ 58.02\% & +5.35 & \textbf{41.98\%} $\rightarrow$ 62.60\% & +20.62 \\
& Islam (612) & 62.91\% $\rightarrow$ 67.65\% & +4.74 & 50.98\% $\rightarrow$ 68.84\% & +17.86 \\
& Buddhism (192) & 73.96\% $\rightarrow$ 75.52\% & +1.56 & 49.48\% $\rightarrow$ 69.79\% & +20.31 \\
& Hinduism (48) & 60.42\% $\rightarrow$ 62.50\% & +2.08 & 58.33\% $\rightarrow$ \textbf{47.92\%} & -10.41 \\
& Agnostic (24) & 91.67\% $\rightarrow$ 62.50\% & \textbf{-29.17} & 70.83\% $\rightarrow$ 54.17\% & -16.66 \\
& None (744) & 65.32\% $\rightarrow$ 80.51\% & \underline{+15.19} & \textbf{45.83\%} $\rightarrow$ 80.13\% & \underline{+34.30} \\
& Prefer Not to Say (192) & 61.46\% $\rightarrow$ \textbf{52.60\%} & -8.86 & 46.35\% $\rightarrow$ 53.65\% & +7.30 \\
\rowcolor{gray!8}
\textbf{Overall} & All (7879) & 66.92\% $\rightarrow$ 68.02\% & +1.10 & 52.44\% $\rightarrow$ 69.66\% & \underline{+17.22} \\
\bottomrule
\end{tabular}%
}
% \vspace{2mm}
\footnotesize
\textit{Sample counts in parentheses. \textbf{Bold} indicates lowest accuracy in subgroup. \underline{Underlined} $\Delta$ values show top 3 improvements per model. Overall shows weighted average accuracy across all samples in the validation set.}
\vspace*{-1.5em}
\end{table}

\subsection{Predicting Opinion Indicators}
\vspace*{-0.5em}

We assess performance through both exact and approximate matching ($\pm2$) of predicted numerical responses across five psychological dimensions --emotion, concern, openness, motivation, and harm perception-- providing comprehensive and category-specific assessments.  Using PHORECAST, we train and evaluate two state-of-the-art VLMs: Llama 3.2-Vision 11B and Gemma 3 12B. Our baseline results (Table~\ref{tab:model_accuracy_all_input}) reveal Llama-11B achieving 16.82\% exact-match accuracy (60.66\% at $\pm2$ off), while Gemma-12B exhibits marginally better baseline performance at $23.89\%$ exact match (78.26\% at $\pm2$ off). Both models struggle particularly with emotion and openness, suggesting these dimensions present greater challenges for current architectures. The consistent performance gap between exact and approximate matching underscores the inherent difficulty in precisely modeling human psychological responses.

\begin{table}[ht]
\centering
\resizebox{\textwidth}{!}{%
\begin{tabular}{lccccc|c}
\toprule
\textbf{Model} & \textbf{Concern} & \textbf{Emotion} & \textbf{Harm Perception} & \textbf{Motivation} & \textbf{Openness} & \textbf{Total Average} \\
\midrule
Random Baseline & 14.29 / 54.86 / 68.54 & 17.69 / 30.42 / 41.66 & 21.73 / \textbf{77.96} / \textbf{88.75} & 14.13 / 50.00 / 62.92 & 14.59 / 47.57 / 66.11 & 16.49 / 52.16 / 65.60 \\
\addlinespace 
\multicolumn{7}{l}{\textbf{Llama 11B}} \\ % Model Group Header
Base & 16.57 / 48.18 / 62.46 & 14.03 / 32.55 / 45.53 & 21.88 / 66.57 / 78.88 & 15.81 / 46.35 / 60.03 & 15.81 / 41.03 / 56.38 & 16.82 / 46.94 / 60.66 \\
Trained with PHORECAST & \textbf{42.71} / \textbf{72.49} / \textbf{87.69} & 37.39 / \textbf{56.98} / \textbf{ 70.94} & \textbf{43.16} / 64.89 / 85.41 & 40.88 / \textbf{67.63} / \textbf{83.89} & \textbf{36.47} / \textbf{66.26} / \textbf{83.28} & \textbf{40.12} / \textbf{65.65} / \textbf{82.24} \\
\addlinespace 
\multicolumn{7}{l}{\textbf{Gemma 12B}} \\
Base & 26.44 / 63.22 / 86.78 & 13.97 / 40.88 / 60.30 & 29.33 / 71.43 / 88.45 & 24.77 / 60.18 / 79.94 & 24.92 / 55.62 / 75.84 & 23.89 / 58.27 / 78.26 \\
Trained with PHORECAST & 42.40 / 67.93 / 84.80 & \textbf{37.55} / 54.35 / 69.45 & 42.25 / 65.20 / 81.76 & \textbf{41.64} / 64.13 / 81.91 & 36.32 / 61.25 / 80.24 & 40.03 / 62.57 / 79.63 \\
\bottomrule
\end{tabular}%
}
\vspace{0.1cm}
\caption{\textbf{Evaluation Accuracy Using Partial Profiles:} Model accuracy across five response dimensions using partial profiles before and after training: this case tests the most applicable real use case, where we may only have a partial profile of an individual we wish to predict an opinion for. We use the procedure described above to create a split consisting of partial features (personality, demographics, and in-context Q/As), with the stratified strategy, creating over 8k evaluation samples of 521 unique individuals. Each cell reports accuracy at exact / $\pm$1 / $\pm$2 tolerance levels. With $\pm$2 tolerance threshold, Llama Base achieves 45\% on emotion prediction, which are measured based on 8 distinct emotional predictions. After training with our dataset, Llama 11B reaches 70.94\% accuracy in the emotion category, contributing to its superior overall performance of 82.24\% across all question categories, outperforming Gemma 12B.  These results showcase a notable {\bf 2.4×} {\em increase in exact-match accuracy improvement using PHORECAST for training}.}
\label{tab:model_accuracy_all_input}
\vspace*{-1.5em}
\end{table}
% \vspace*{-0.5em}
\subsection{Benchmarking Performance on Freeform Responses}
\vspace*{-0.5em}
Above, we measure model performance on discrete opinion scales, making accuracy easy to quantify.  Measuring the accuracy of freeform text responses is more complex. We don't expect a model to perfectly match a participant's freeform responses, but rather we hope for model outputs to be distributionally similar to humans, conditioned on their demographics/personality.

To this end, we use multiple evaluations approaches: \textbf{Semantic Text Similarity (STS)} to measure how closely model responses align with human answers in meaning, \textbf{Statistical Distribution of Embeddings (SDE)} --stratified by personality traits to assess distributional alignment across different subgroups, and a \textbf{discriminator-based} accuracy metric to determine how well a model can distinguish between human and machine-generated responses. We focus on semantic similarly and refer readers to the appendix for additional measures of distributional fidelity in the Supplementary Materials.

% \paragraph{Semantic Textual Similarity} We measure the semantic similarity between model outputs and human-written reference responses using \textit{all-mpnet-base-v2}. For each participant, we encode the expected human response and corresponding model-generated responses into dense vector representations, then calculate their pairwise cosine similarity scores. A higher similarity score indicates a higher similarity between the two pairs of responses. 

We compute the similarity (0-1 score) between each machine generated response and its expected (ground truth) response using \textit{all-mpnet-base-v2}. We evaluate the semantic similarity scores for different subgroups to analyze which individuals or groups are \textit{easier} to emulate. Table~\ref{tab:combined_analysis} shows the results from benchmarking Gemma and Llama across different demographic groups and the Big 5. As seen, both models benefit from our dataset. In particular, our ability to emulate females, individuals aged 45-54, and Muslims, greatly enhances after training. Interestingly, Asians, a group underrepresented in our study, remain difficult to emulate. 

\begin{table}[h]
\centering
\caption{\textbf{Analysis of Model Response Similarity Across Demographic and Personality Groups}: We present the response similarity scores (range: 0-1) for Gemma and Llama models, segmented by key traits. \textbf{Bold} indicates the highest score per group; \underline{underlined} indicates the lowest. Post-training, both models show improved alignment, with Gemma's similarity improving from 0.32 to 0.37 and Llama from 0.28 to 0.34.}
\label{tab:combined_analysis}
\scriptsize
\begin{tabular}{@{\extracolsep{2pt}} l l cccccl @{}}
\toprule
\multirow{2}{*}{\textbf{Group}} & \multirow{2}{*}{\textbf{Subcategory}} & \multicolumn{2}{c}{\textbf{Gemma 12B}} & \multicolumn{2}{c}{\textbf{Llama 11B}} & \multirow{2}{*}{\textbf{Max $\Delta$}} \\
& & Before & After & Before & After & \\
\midrule

\multirow{5}{*}{\textbf{Age}}
& 18--24 (994) & 0.30 & 0.31 & 0.27 & 0.34 & +0.07 \\
& 25--34 (3867) & 0.31 & 0.33 & 0.27 & 0.33 & +0.06 \\
& 35--44 (2419) & 0.30 & 0.33 & 0.29 & 0.32 & +0.03 \\
& 45--54 (479) & \textbf{0.41} & \textbf{0.47} & \textbf{0.37} & \textbf{0.42} & +0.06 \\
& 55--64 (108) & 0.31 & 0.37 & \underline{0.29} & \underline{0.29} & +0.06 \\
\rowcolor{gray!10}
& \textbf{Avg.} & 0.31 & 0.34 & 0.29 & 0.33 & +0.04 \\
\midrule
\multirow{2}{*}{\textbf{Gender}}
& Male (4081) & 0.33 & 0.34 & 0.29 & 0.34 & +0.05 \\
& Female (3738) & \underline{0.30} & \underline{0.33} & \underline{0.28} & 0.33 & +0.05 \\
\rowcolor{gray!10}
& \textbf{Avg.} & 0.31 & 0.33 & 0.28 & 0.33 & +0.05 \\
\midrule
\multirow{5}{*}{\textbf{Religion}}
& Christianity (5936) & 0.31 & 0.34 & 0.29 & 0.33 & +0.04 \\
& Islam (612)& \textbf{0.36} & \textbf{0.42} & 0.29 & \textbf{0.41} & \textbf{+0.12} \\
& Judaism (131) & 0.36 & 0.36 & 0.32 & \underline{0.29} & +0.00 \\
& Buddhism (192) & 0.38 & \underline{0.34} & \underline{0.28} & 0.31 & +0.03 \\
& Other (936) & \underline{0.32} & \underline{0.30} & \underline{0.25} & 0.37 & \textbf{+0.12} \\
\rowcolor{gray!10}
& \textbf{Avg.} & 0.34 & 0.35 & 0.28 & 0.34 & +0.06 \\
\midrule
\multirow{5}{*}{\textbf{Race}}
& White (434) & 0.31 & 0.33 & 0.28 & 0.33 & +0.05 \\
& Black (177) & 0.33 & 0.35 & 0.30 & \textbf{0.37} & +0.07 \\
& Hispanic (24) & 0.33 & \underline{0.30} & 0.32 & 0.36 & +0.04 \\
& Asian (14) & \underline{0.14} & \underline{0.19} & \underline{0.18} & \underline{0.22} & +0.05 \\
& Other (481) & \textbf{0.34} & \textbf{0.37} & 0.29 & 0.36 & \textbf{+0.11} \\
\rowcolor{gray!10}
& \textbf{Avg.} & 0.29 & 0.31 & 0.27 & 0.33 & +0.06 \\
\midrule

\multirow{5}{*}{\textbf{Big 5}}
& Extraversion (658) & 0.31 & \textbf{0.35} & \textbf{0.33} & 0.34 & +0.04 \\
& Agreeableness (658) & 0.31 & 0.33 & 0.29 & 0.34 & +0.05 \\
& Conscientiousness (658) & 0.31 & 0.34 & \underline{0.29} & \underline{0.29} & +0.02 \\
& Neuroticism (658) & 0.32 & 0.34 & 0.28 & 0.34 & +0.06 \\
& Openness (658) & \textbf{0.32} & 0.34 & 0.29 & \textbf{0.34} & +0.05 \\
\rowcolor{gray!10}
& \textbf{Avg.} & 0.31 & 0.34 & 0.30 & 0.33 & +0.04 \\
\bottomrule
\rowcolor{yellow!5}
\multicolumn{2}{l}{\textbf{Overall Avg.}} & \textbf{0.32} & \textbf{0.37} & \textbf{0.28} & \textbf{0.34} & \textbf{+0.06} \\
\bottomrule
\end{tabular}
\vspace{0.1cm}
\scriptsize \\
% \textit{Similarity scores (0-1) pre- vs. post-training. \textbf{Bold}: highest per group; \underline{Underlined}: lowest. "Max $\Delta$" shows largest improvement.}
\vspace*{-1.5em}
\end{table}

\vspace*{-0.4em}
\subsection{Ablation Study on Groups of Attributes}
\vspace*{-0.5em}
We investigate the feature importance of demographics, personality, locus of control and in context Q/As in predicting the individual's opinion using our best-trained model. These experiments aim to analyze model performance degradation and identify which features contribute most to the model's ability to emulate human response. To assess feature contribution, we prompt LLama (trained on randomized partial profiles) with only one feature group at a time -- demographics, personality traits, locus of control, or in-context Q/As. We employ target masking to isolate specific features:
For demographics, we exclude gender, race/ethnicity, education, or religion. For personality, we mask either the locus of control, the Big Five traits, or both. Finally, random masking removes $k$ random features from a specified group (demographics, personality, or in-context Q/As) before inference. Some key observations include: \\ 
(1) Providing the model with {\em personality information achieves higher accuracy ($77.71\%$) than providing the model with only demographics features ($74.89\%$)}, (2) providing the {\em locus of control along with the personality is better than providing only personality or just locus}. (3) {\em In-context opinion indicators, as opposed to demographics and personality, are sufficient to predict an aspect of that individual's opinion}. In particular, providing the model with the individual's responses to the stimuli (e.g., harm perception, concern), without the free form or the initial opinions pre-stimuli, is sufficient to achieve the highest score of $95\%$ (w/ $\pm$2) and $78\%$ at exact. 

% Prompting the model with only personality information provides higher accuracy than providing the model with only demographics. 

\subsubsection{Demographics}
We evaluate the impact demographics has on the model's ability to emulate different individuals (Table \ref{tab:gemma12b_ablation}). We begin by prompting our model with only the demographics information, and establish a baseline of $30.64\%$ (exact) and $-74.80\%$ (w/ $\pm$2) across all question categories. Next, we investigate the effect of different demographic features such as age, gender and race. 
We observe that prompting the model with all the demographics information produces higher accuracy than when some of those demographic features are masked. However, not at all demographics are as important: When we randomly mask 7 field, our accuracy drops only by approximately 2\%.

\begin{table}[ht]
\centering
\caption{\textbf{Relative Impact of Demographic Factors on Prediction Accuracy}: We quantify the relative importance of demographic factors (gender, race/ethnicity, education, religion) in predicting responses to public health messages using our trained Llama 11B. Our study reveals: (1) education removal causes the largest performance drop ($-1.08/-1.29)$ in average $\pm$1/2 accuracy. (2) gender and race show minimal impact ($\pm$0.63/$\pm$0.15 $\Delta$), and (3) models degrade gracefully with random feature removal ($\leq 2.21$ drop with 7 fields masked). Results demonstrate that {\em education-level information is the most critical for accurate prediction, while other demographics contribute modestly.}}
\label{tab:gemma12b_ablation}
\resizebox{\textwidth}{!}{%
\small
\setlength{\tabcolsep}{5pt}
\renewcommand{\arraystretch}{1.15}
\begin{tabular}{lccccc|c}
\toprule
\textbf{Comparison} & \textbf{Concern} & \textbf{Emotion} & \textbf{Harm} & \textbf{Motivation} & \textbf{Openness} & \textbf{Avg.} \\
\midrule
\rowcolor{gray!8}
\textbf{Reference Model} & \multicolumn{6}{c}{\textit{All Demographics (Baseline)}} \\
\rowcolor{gray!3}
& 30.55 / 58.21 / 81.61 & 21.90 / 44.43 / 58.78 & 39.67 / 61.85 / 84.80 & 28.42 / 54.10 / 74.62 & 32.67 / 57.90 / 74.62 & 30.64 / 55.30 / 74.89 \\
\midrule
\rowcolor{gray!8}
\multicolumn{7}{l}{\textit{Single Demographic Removals ($\Delta$ from Reference)}} \\
\rowcolor{gray!3}
- Gender & \cellcolor{green!8}+0.45 / \cellcolor{green!8}+1.21 / \cellcolor{green!8}+0.30 & \cellcolor{orange!8}-0.21 / \cellcolor{orange!8}-0.40 / \cellcolor{orange!8}-0.12 & \cellcolor{orange!8}-0.16 / \cellcolor{orange!8}-0.45 / \cellcolor{orange!8}-0.61 & \cellcolor{red!8}-1.67 / \cellcolor{red!8}-2.12 / \cellcolor{green!8}+0.76 & \cellcolor{orange!8}-0.60 / \cellcolor{orange!8}-1.37 / \cellcolor{orange!8}-1.06 & \cellcolor{orange!8}-0.44 / \cellcolor{orange!8}-0.63 / \cellcolor{orange!8}-0.15 \\
\rowcolor{gray!3}
- Race & \cellcolor{green!8}+0.61 / \cellcolor{green!8}+0.76 / \cellcolor{green!8}+0.61 & \cellcolor{green!8}+0.23 / \cellcolor{green!8}+0.30 / \cellcolor{orange!8}-0.10 & \cellcolor{green!8}+0.15 / \cellcolor{orange!8}-0.60 / \cellcolor{orange!8}-0.76 & \cellcolor{green!8}+0.30 / \cellcolor{green!8}+1.22 / \cellcolor{orange!8}-0.00 & \cellcolor{red!8}-2.88 / \cellcolor{orange!8}-2.28 / \cellcolor{orange!8}-0.00 & \cellcolor{orange!8}-0.32 / \cellcolor{orange!8}-0.12 / \cellcolor{orange!8}-0.05 \\
\rowcolor{gray!3}
- Education & \cellcolor{orange!8}-0.31 / \cellcolor{orange!8}-0.00 / \cellcolor{red!8}-1.97 & \cellcolor{orange!8}-0.19 / \cellcolor{red!8}-1.30 / \cellcolor{red!8}-0.65 & \cellcolor{orange!8}-0.16 / \cellcolor{red!8}-1.67 / \cellcolor{red!8}-1.82 & \cellcolor{green!8}+0.30 / \cellcolor{orange!8}-0.00 / \cellcolor{red!8}-1.37 & \cellcolor{green!8}+0.61 / \cellcolor{orange!8}-2.43 / \cellcolor{orange!8}-0.61 & \cellcolor{green!8}+0.05 / \cellcolor{red!8}-1.08 / \cellcolor{red!8}-1.29 \\
\rowcolor{gray!3}
- Religion & \cellcolor{red!8}-1.52 / \cellcolor{green!8}+0.45 / \cellcolor{orange!8}-0.30 & \cellcolor{orange!8}-0.12 / \cellcolor{orange!8}-0.18 / \cellcolor{green!8}+0.30 & \cellcolor{red!8}-0.76 / \cellcolor{red!8}-1.52 / \cellcolor{orange!8}-0.61 & \cellcolor{green!8}+0.15 / \cellcolor{orange!8}-0.00 / \cellcolor{green!8}+0.61 & \cellcolor{orange!8}-0.15 / \cellcolor{orange!8}-0.76 / \cellcolor{orange!8}-0.00 & \cellcolor{orange!8}-0.48 / \cellcolor{orange!8}-0.40 / \cellcolor{orange!8}-0.00 \\
\midrule
\rowcolor{gray!8}
\multicolumn{7}{l}{\textit{Random Removals ($\Delta$ from Reference)}} \\
\rowcolor{gray!3}
-3 Fields & \cellcolor{orange!8}-0.61 / \cellcolor{green!8}+0.30 / \cellcolor{red!8}-1.52 & \cellcolor{green!8}+0.32 / \cellcolor{orange!8}-0.56 / \cellcolor{orange!8}-0.33 & \cellcolor{orange!8}-0.61 / \cellcolor{red!8}-2.12 / \cellcolor{red!8}-1.97 & \cellcolor{red!8}-1.67 / \cellcolor{red!8}-3.04 / \cellcolor{orange!8}-0.15 & \cellcolor{orange!8}-1.21 / \cellcolor{orange!8}-2.58 / \cellcolor{orange!8}-0.00 & \cellcolor{orange!8}-0.75 / \cellcolor{red!8}-1.60 / \cellcolor{orange!8}-0.80 \\
\rowcolor{gray!3}
-5 Fields & \cellcolor{green!8}+1.67 / \cellcolor{red!8}-1.22 / \cellcolor{red!8}-3.19 & \cellcolor{orange!8}-0.54 / \cellcolor{red!8}-0.88 / \cellcolor{red!8}-0.96 & \cellcolor{orange!8}-0.61 / \cellcolor{red!8}-2.58 / \cellcolor{red!8}-2.43 & \cellcolor{orange!8}-0.30 / \cellcolor{orange!8}-0.91 / \cellcolor{red!8}-1.37 & \cellcolor{orange!8}-1.67 / \cellcolor{red!8}-3.95 / \cellcolor{red!8}-1.98 & \cellcolor{orange!8}-0.29 / \cellcolor{red!8}-1.91 / \cellcolor{red!8}-1.99 \\
\rowcolor{gray!3}
-7 Fields & \cellcolor{red!8}-1.22 / \cellcolor{red!8}-1.83 / \cellcolor{red!8}-4.10 & \cellcolor{orange!8}-0.34 / \cellcolor{red!8}-1.91 / \cellcolor{red!8}-0.99 & \cellcolor{green!8}+1.82 / \cellcolor{green!8}+0.46 / \cellcolor{red!8}-1.82 & \cellcolor{red!8}-1.06 / \cellcolor{red!8}-3.49 / \cellcolor{red!8}-1.52 & \cellcolor{red!8}-3.34 / \cellcolor{red!8}-6.38 / \cellcolor{red!8}-2.58 & \cellcolor{red!8}-0.83 / \cellcolor{red!8}-2.63 / \cellcolor{red!8}-2.21 \\
\bottomrule
\end{tabular}
}
% \vspace{2mm}
\footnotesize \textit{Values show $\Delta$ from reference model (exact/±1/±2). \textcolor{green!50}{green} = improvement ($\Delta$ > +0.3), \textcolor{orange!50}{orange} = minor drop, \\ \textcolor{red!50}{red} = significant drop ($\Delta \leq$ -1.5).}
\end{table}

\subsubsection{Personality and Locus of Control}
In the second experiment , we investigate how personality traits influence predictive performance (Table \ref{tab:personality_ablation}). We operationalize personality through two complementary frameworks: the BFI-2 and the locus of control (LOC).
We evaluate the model using: 
1. Personality alone,
2. LOC alone,
3. Big 5 domains alone,
4. Big Five + LOC,
5. Full psychological profile (personality + LOC). We replicate the random masking procedure from experiment 1 to assess stability and quantify information loss when partial trait data is available. As seen with the random masking, our model's performance does not degrade, suggesting that partial personality information can be sufficient for the model to predict an individual's opinion.
\begin{table}[ht]
\centering
\caption{\textbf{Contribution of Personality Components to Prediction Accuracy}: We evaluate how different personality measures (Big Five traits, 15 facets, and Locus of Control) predict responses to public health messages. Our ablation study reveals that: (1) the 15 personality facets combined with Locus of Control yield highest accuracy (77.84\% average at ±2 tolerance), (2) full personality profiles marginally underperform this configuration (77.71\%), and (3) individual traits show varying predictive power, with facets being most informative. Results demonstrate personality's strong predictive value while highlighting the particular importance of facet-level traits. Accuracy reported as exact/±1/±2 matches."
}
\label{tab:personality_ablation}
\resizebox{\textwidth}{!}{%
\small
\setlength{\tabcolsep}{5pt}
\renewcommand{\arraystretch}{1.15}
\begin{tabular}{llccccc|c}
\toprule
\textbf{Comparison} & \textbf{Components} & \textbf{Concern} & \textbf{Emotion} & \textbf{Harm} & \textbf{Motivation} & \textbf{Openness} & \textbf{Avg.} \\
\midrule
\rowcolor{gray!8}
\textbf{Reference Model} & Full Personality + LOC & 30.09 / 62.01 / 83.74 & 24.51 / 45.99 / 63.79 & 44.38 / 63.37 / 79.94 & 29.64 / 59.42 / 80.40 & 33.89 / 58.81 / 80.70 & 32.50 / 57.92 / 77.71 \\
\midrule
\multicolumn{8}{l}{\textit{Component Contributions ($\Delta$ from Reference)}} \\
\rowcolor{gray!3}
- LOC & Full Personality & \cellcolor{green!8}+1.22 / \cellcolor{orange!8}-0.61 / \cellcolor{orange!8}-0.76 & \cellcolor{orange!8}-0.59 / \cellcolor{orange!8}-0.57 / \cellcolor{red!8}-2.76 & \cellcolor{orange!8}-0.31 / \cellcolor{orange!8}-0.30 / -0.00 & \cellcolor{green!8}+1.36 / \cellcolor{orange!8}-0.45 / \cellcolor{orange!8}-1.22 & \cellcolor{orange!8}-0.30 / \cellcolor{green!8}+1.68 / \cellcolor{orange!8}-1.67 & \cellcolor{green!8}+0.28 / \cellcolor{orange!8}-0.05 / \cellcolor{orange!8}-1.28 \\
\rowcolor{gray!3}
- Big5 & 15 Facets + LOC & \cellcolor{orange!8}-0.30 / \cellcolor{green!8}+0.00 / \cellcolor{orange!8}-0.31 & \cellcolor{orange!8}-0.53 / \cellcolor{green!8}+0.59 / \cellcolor{green!8}+0.63 & \cellcolor{orange!8}-0.16 / \cellcolor{green!8}+0.00 / \cellcolor{green!8}+0.15 & \cellcolor{orange!8}-0.92 / \cellcolor{orange!8}-1.21 / -0.00 & \cellcolor{red!8}-2.58 / \cellcolor{green!8}+1.22 / \cellcolor{green!8}+0.15 & \cellcolor{orange!8}-0.90 / \cellcolor{green!8}+0.12 / \cellcolor{green!8}+0.13 \\
\rowcolor{gray!3}
- Both & Only 15 Facets & \cellcolor{green!8}+0.91 / \cellcolor{orange!8}-1.68 / -0.00 & \cellcolor{orange!8}-0.90 / \cellcolor{orange!8}-0.80 / \cellcolor{red!8}-1.93 & \cellcolor{orange!8}-0.46 / \cellcolor{orange!8}-0.45 / -0.00 & \cellcolor{green!8}+0.91 / \cellcolor{orange!8}-1.06 / \cellcolor{orange!8}-0.46 & \cellcolor{orange!8}-1.06 / \cellcolor{green!8}+0.31 / \cellcolor{orange!8}-1.37 & \cellcolor{orange!8}-0.12 / \cellcolor{orange!8}-0.74 / \cellcolor{orange!8}-0.75 \\
\midrule
\multicolumn{8}{l}{\textit{Individual Component Performance ($\Delta$ from Reference)}} \\
\rowcolor{gray!3}
& Only Big5 & \cellcolor{orange!8}-1.15 / \cellcolor{red!8}-5.78 / \cellcolor{red!8}-3.34 & \cellcolor{red!8}-2.34 / \cellcolor{red!8}-3.43 / \cellcolor{red!8}-5.43 & \cellcolor{orange!8}-0.31 / \cellcolor{orange!8}-0.30 / -0.00 & \cellcolor{orange!8}-1.22 / \cellcolor{red!8}-5.01 / \cellcolor{red!8}-4.72 & \cellcolor{red!8}-4.25 / \cellcolor{red!8}-4.87 / \cellcolor{red!8}-4.10 & \cellcolor{orange!8}-1.62 / \cellcolor{red!8}-3.51 / \cellcolor{red!8}-3.52 \\
\rowcolor{gray!3}
& Only LOC & \cellcolor{red!8}-3.04 / \cellcolor{red!8}-7.30 / \cellcolor{red!8}-2.28 & \cellcolor{red!8}-4.21 / \cellcolor{red!8}-4.12 / \cellcolor{red!8}-5.55 & \cellcolor{orange!8}-0.16 / \cellcolor{orange!8}-0.15 / -0.00 & \cellcolor{red!8}-6.24 / \cellcolor{red!8}-8.20 / \cellcolor{orange!8}-1.68 & \cellcolor{red!8}-4.86 / \cellcolor{red!8}-5.01 / \cellcolor{red!8}-6.23 & \cellcolor{red!8}-3.70 / \cellcolor{red!8}-4.96 / \cellcolor{red!8}-3.14 \\
\midrule
\multicolumn{8}{l}{\textit{Random Ablations ($\Delta$ from Reference)}} \\
\rowcolor{gray!3}
-3 Traits & Random subset & \cellcolor{green!8}+0.32 / \cellcolor{orange!8}-0.55 / \cellcolor{orange!8}-1.22 & \cellcolor{orange!8}-0.69 / \cellcolor{orange!8}-0.80 / \cellcolor{red!8}-2.08 & \cellcolor{orange!8}-0.16 / \cellcolor{orange!8}-0.15 / -0.00 & \cellcolor{green!8}+1.06 / \cellcolor{orange!8}-1.37 / \cellcolor{orange!8}-1.22 & \cellcolor{green!8}+0.46 / \cellcolor{green!8}+0.61 / \cellcolor{orange!8}-1.52 & \cellcolor{green!8}+0.32 / \cellcolor{orange!8}-0.46 / \cellcolor{orange!8}-1.20 \\
\rowcolor{gray!3}
-5 Traits & Random subset & \cellcolor{green!8}+1.41 / \cellcolor{orange!8}-0.92 / \cellcolor{orange!8}-0.76 & \cellcolor{orange!8}-0.31 / \cellcolor{orange!8}-1.11 / \cellcolor{red!8}-2.25 & \cellcolor{orange!8}-0.46 / \cellcolor{orange!8}-0.45 / -0.00 & \cellcolor{green!8}+0.60 / \cellcolor{orange!8}-0.30 / \cellcolor{orange!8}-0.61 & \cellcolor{orange!8}-1.06 / \cellcolor{green!8}+0.76 / \cellcolor{orange!8}-0.91 & \cellcolor{green!8}+0.12 / \cellcolor{orange!8}-0.40 / \cellcolor{orange!8}-0.90 \\
\rowcolor{gray!3}
-7 Traits & Random subset & \cellcolor{green!8}+0.61 / \cellcolor{orange!8}-1.83 / \cellcolor{orange!8}-0.91 & \cellcolor{red!8}-1.30 / \cellcolor{orange!8}-0.95 / \cellcolor{red!8}-2.08 & \cellcolor{orange!8}-0.31 / \cellcolor{orange!8}-0.30 / -0.00 & \cellcolor{green!8}+1.67 / \cellcolor{orange!8}-0.30 / \cellcolor{orange!8}-0.92 & \cellcolor{orange!8}-0.91 / \cellcolor{orange!8}-0.45 / \cellcolor{orange!8}-1.06 & \cellcolor{orange!8}-0.05 / \cellcolor{orange!8}-0.77 / \cellcolor{orange!8}-0.99 \\
\bottomrule
\end{tabular}%
}
% \vspace{2mm}
\footnotesize \textit{Values show $\Delta$ from reference model (exact/±1/±2). \textcolor{green!50}{green} = improvement, \textcolor{orange!50}{orange} = moderate drop (0-3\%), \\ \textcolor{red!50}{red} = large drop (>3\%). LOC = Locus of Control.}
\end{table}

\subsubsection{In-Context Questions and Answers}
In context questions and answers provide the model hints on how the participant responded to other questions asked. For example, if the model is trying to predict the participant's harm perception, then we ask, is providing the emotional responses to that topic helpful? We evaluate the impact of in-context information on model accuracy (Table \ref{tab:response_ablation}). When we provide the model only with the in-context demonstrations of their reactions to the stimuli, it becomes trivial for the model to complete the missing piece, achieving $78.13\%$ at exact and over $95\%$ with $\pm$2 tolerance. When we provide the model with only the initial opinions (prior to the stimuli), our performance drops to $32.53$ at exact and $80.06\%$ with $\pm$2. The free form responses are also not sufficient and cause a significant drop in performance if they are the only available feature ($29.96\%$ at exact and $69.48$ with $\pm$2). Since the in-context responses to the stimuli are most indicative of one's response, we mask 3-7 of them for the random masking experiment.
\begin{table}[ht]
\vspace*{-1em}
\centering
\caption{\textbf{Impact of In-Context Responses on Prediction Accuracy}: We evaluate how different response components (initial opinions, free-form text, and structured Q/A responses) contribute to predicting public health message reception. Using Llama 11B, we demonstrate that structured Q/A responses alone achieve comparable performance $(\Delta +0.68/+0.10/+0.16$) to the full model, while initial opinions and free-form text show limited predictive value individually ($\Delta -44.32/-26.61/-15.66$ and -47.49/-38.63/-26.24 respectively). Removing all response components causes severe performance degradation (-44.32 average exact match), highlighting their critical role in accurate prediction. Results are reported as exact/±1/±2 match accuracy across five psychological dimensions.}
\label{tab:response_ablation}
\resizebox{\textwidth}{!}{%
\small
\setlength{\tabcolsep}{5pt}
\renewcommand{\arraystretch}{1.15}
\begin{tabular}{llccccc|c}
\toprule
\textbf{Comparison} & \textbf{Components} & \textbf{Concern} & \textbf{Emotion} & \textbf{Harm} & \textbf{Motivation} & \textbf{Openness} & \textbf{Avg.} \\
\midrule
\rowcolor{gray!8}
\textbf{Reference Model} & All components & 98.63 / 99.70 / 100.00 & 99.71 / 99.83 / 99.89 & 49.54 / 76.29 / 92.10 & 98.78 / 98.94 / 99.09 & 40.58 / 73.10 / 87.54 & 77.45 / 89.57 / 95.72 \\
\midrule
\multicolumn{8}{l}{\textit{Component Contributions ($\Delta$ from Reference)}} \\
\rowcolor{gray!3}
- Initial Opinions & All except initial & +0.15 / -0.31 / -0.15 & +0.00 / +0.00 / +0.00 & \cellcolor{green!8}+2.13 / \cellcolor{green!8}+3.04 / \cellcolor{green!8}+1.21 & -0.76 / -0.31 / -0.15 & \cellcolor{green!8}+2.13 / -2.13 / -0.76 & \cellcolor{green!8}+0.73 / \cellcolor{green!8}+0.06 / +0.03 \\
\rowcolor{gray!3}
- Free Form & All except free form & \cellcolor{green!8}+1.07 / \cellcolor{green!8}+0.30 / +0.00  & -0.01 / -0.04 / -0.04 & \cellcolor{green!8}+1.37 / \cellcolor{green!8}+2.59 / \cellcolor{green!8}+1.67 & -0.30 / -0.16 / -0.15 & \cellcolor{green!8}+0.76 / \cellcolor{green!8}+0.15 / -0.15 & \cellcolor{green!8}+0.57 / \cellcolor{green!8}+0.57 / +0.27 \\
\rowcolor{gray!3}
- Responses & Only responses & \cellcolor{green!8}+1.07 / +0.00 / +0.00 & \cellcolor{green!8}+0.06 / \cellcolor{green!8}+0.04 / \cellcolor{green!8}+0.01 & \cellcolor{green!8}+1.98 / \cellcolor{green!8}+3.19 / \cellcolor{green!8}+1.37 & -1.67 / -0.76 / -0.46 & \cellcolor{green!8}+1.97 / -1.98 / -0.15 & \cellcolor{green!8}+0.68 / \cellcolor{green!8}+0.10 / \cellcolor{green!8}+0.16 \\
\midrule
\multicolumn{8}{l}{\textit{Individual Component Performance ($\Delta$ from Reference)}} \\
\rowcolor{gray!3}
& Only initial & \cellcolor{red!8}-64.74 / \cellcolor{red!8}-29.34 / \cellcolor{red!8}-13.07 & \cellcolor{red!8}-78.15 / \cellcolor{red!8}-55.71 / \cellcolor{red!8}-41.36 & -5.92 / -8.81 / -2.59 & \cellcolor{red!8}-66.71 / \cellcolor{red!8}-33.89 / \cellcolor{red!8}-17.48 & \cellcolor{orange!8}-6.08 / \cellcolor{orange!8}-5.32 / \cellcolor{orange!8}-3.80 & \cellcolor{red!8}-44.32 / \cellcolor{red!8}-26.61 / \cellcolor{red!8}-15.66 \\
\rowcolor{gray!3}
& Only free form & \cellcolor{red!8}-72.03 / \cellcolor{red!8}-47.12 / \cellcolor{red!8}-25.38 & \cellcolor{red!8}-79.41 / \cellcolor{red!8}-59.86 / \cellcolor{red!8}-45.97 & -5.32 / -12.92 / -11.55 & \cellcolor{red!8}-70.21 / \cellcolor{red!8}-49.70 / \cellcolor{red!8}-28.42 & \cellcolor{red!8}-10.49 / \cellcolor{red!8}-23.56 / \cellcolor{red!8}-19.91 & \cellcolor{red!8}-47.49 / \cellcolor{red!8}-38.63 / \cellcolor{red!8}-26.24 \\
\midrule
\multicolumn{8}{l}{\textit{Random Ablations ($\Delta$ from Reference)}} \\
\rowcolor{gray!3}
-3 Fields & Random subset & \cellcolor{red!8}-16.72 / \cellcolor{orange!8}-6.84 / \cellcolor{orange!8}-2.58 & \cellcolor{red!8}-17.21 / \cellcolor{red!8}-8.96 / \cellcolor{orange!8}-4.64 & -0.45 / \cellcolor{green!8}+1.37 / \cellcolor{green!8}+1.37 & \cellcolor{red!8}-21.27 / \cellcolor{red!8}-11.86 / \cellcolor{red!8}-6.08 & -2.13 / \cellcolor{green!8}+19.76 / -1.37 & \cellcolor{red!8}-11.56 / \cellcolor{red!8}-6.14 / \cellcolor{red!8}-2.66 \\
\rowcolor{gray!3}
-5 Fields & Random subset & \cellcolor{red!8}-27.96 / \cellcolor{red!8}-12.92 / \cellcolor{red!8}-5.17 & \cellcolor{red!8}-29.54 / \cellcolor{red!8}-15.46 / \cellcolor{red!8}-8.43 & -2.43 / -1.67 / +0.76 & \cellcolor{red!8}-33.89 / \cellcolor{red!8}-19.15 / \cellcolor{red!8}-8.97 & \cellcolor{red!8}-3.04 / \cellcolor{red!8}-5.93 / \cellcolor{orange!8}-2.28 & \cellcolor{red!8}-19.37 / \cellcolor{red!8}-11.02 / \cellcolor{red!8}-4.81 \\
\rowcolor{gray!3}
-7 Fields & Random subset & \cellcolor{red!8}-39.21 / \cellcolor{red!8}-19.91 / \cellcolor{red!8}-7.90 & \cellcolor{red!8}-40.93 / \cellcolor{red!8}-23.04 / \cellcolor{red!8}-13.04 & -0.91 / -3.04 / -3.35 & \cellcolor{red!8}-45.44 / \cellcolor{red!8}-27.06 / \cellcolor{red!8}-12.77 & \cellcolor{red!8}-7.75 / \cellcolor{red!8}-8.21 / \cellcolor{red!8}-3.50 & \cellcolor{red!8}-26.85 / \cellcolor{red!8}-16.25 / \cellcolor{red!8}-8.11 \\
\bottomrule
\end{tabular}%
}
% \vspace{2mm}
\footnotesize \textit{Values show $\Delta$ from reference model (exact/±1/±2). Color coding: \textcolor{green!50}{green} = improvement, \textcolor{orange!50}{orange} = moderate drop (0-10\%), \textcolor{red!50}{red} = large drop (>10\%).}
% \vspace*{-1em}
\end{table}

\subsubsection{Opinion Prediction to Visual Stimuli}
Our experiments in Table \ref{tab:image-ablation} reveal a nuanced relationship between visual stimuli and opinion prediction:
Images contribute critically to exact-match accuracy, with removal causing a $36.32\%$ drop in motivation and $30.85\%$ in concern. Average exact-match accuracy falls from 77.58 → 57.25\% (-20.33), proving visuals to be strong affective anchors for categorical judgments. In minimal-input conditions (Personality + Demographics + Locus of Control, image removal improves $\pm$1,2 accuracy ($+5.46/+3.80)$, implying that the textual signal dominates nuanced opinion spectra. This supports a dual-process model of AI opinion prediction: System 1 (Fast): Image-driven, affective processing dominates initial classification. System 2 (Slow): Deliberative analysis of stable traits (P+D+LOC) enables fine-grained prediction.

\begin{table}[ht]
\vspace*{-0.5em}
\centering
\caption{\textbf{Impact of Visual and Contextual Inputs on Opinion Prediction}}
\label{tab:image-ablation}
\vspace*{-0.5em}
\resizebox{\textwidth}{!}{%
\small
\setlength{\tabcolsep}{5pt}
\renewcommand{\arraystretch}{1.15}
\begin{tabular}{llccccc|c}
\toprule
\textbf{Comparison} & \textbf{Inputs Used} & \textbf{Concern} & \textbf{Emotion} & \textbf{Harm} & \textbf{Motivation} & \textbf{Openness} & \textbf{Avg.} \\
\midrule
\rowcolor{gray!8}
\multicolumn{8}{l}{\textit{Visual Impact Analysis (Image Removal)}} \\
\rowcolor{gray!3}
& All fields (w/ image) & 98.63 / 99.54 / 100.00 & 99.73 / 99.83 / 99.89 & 49.24 / 74.32 / 90.58 & 99.09 / 99.24 / 99.39 & 41.19 / 71.73 / 87.08 & 77.58 / 88.93 / 95.39 \\
\rowcolor{gray!3}
& All fields - image & 67.78 / 88.75 / 95.90 & 80.64 / 92.55 / 95.71 & 40.88 / 68.09 / 93.92 & 62.77 / 85.56 / 94.07 & 34.19 / 69.00 / 84.65 & 57.25 / 80.79 / 92.85 \\
\cmidrule{2-8}
\rowcolor{gray!8}
\textbf{Change} & \textbf{Image contribution} & \cellcolor{red!10}-30.85 / \cellcolor{red!10}-10.79 / -4.10 & \cellcolor{red!10}-19.09 / \cellcolor{orange!10}-7.28 / -4.18 & -8.36 / -6.23 / \cellcolor{green!10}+3.34 & \cellcolor{red!10}-36.32 / \cellcolor{red!10}-13.68 / -5.32 & -7.00 / -2.73 / -2.43 & \cellcolor{red!10}-20.33 / \cellcolor{red!10}-8.14 / -2.54 \\
\midrule
\rowcolor{gray!8}
\multicolumn{8}{l}{\textit{Contextual Impact Analysis (Responses + Image Removal)}} \\
\rowcolor{gray!3}
& P+D+LOC (w/ image) & 35.41 / 63.98 / 84.35 & 25.96 / 47.23 / 63.69 & 42.71 / 62.46 / 81.46 & 31.31 / 62.77 / 79.64 & 32.07 / 61.40 / 81.46 & 33.49 / 59.57 / 78.12 \\
\rowcolor{gray!3}
& P+D+LOC - both & 36.17 / 72.19 / 86.63 & 23.94 / 52.83 / 65.35 & 39.06 / 65.05 / 92.25 & 30.85 / 67.93 / 81.91 & 33.28 / 67.17 / 83.43 & 32.66 / 65.03 / 81.92 \\
\cmidrule{2-8}
\rowcolor{gray!8}
\textbf{Change} & \textbf{Combined effect} & \cellcolor{green!10}+0.76 / \cellcolor{green!20}+8.21 / +2.28 & -2.02 / \cellcolor{green!10}+5.60 / +1.66 & -3.65 / +2.59 / \cellcolor{green!20}+10.79 & -0.46 / \cellcolor{green!10}+5.16 / +2.27 & \cellcolor{green!10}+1.21 / \cellcolor{green!10}+5.77 / +1.97 & -0.83 / \cellcolor{green!10}+5.46 / +3.80 \\
\bottomrule
\end{tabular}%
}
% \vspace{1mm}
\footnotesize \textit{LOC = Locus of Control. \textcolor{green!50}{green} = improvement, \textcolor{orange!50}{orange} = moderate drop (0-10\%), \textcolor{red!50}{red} = large drop (>20\%).}
% \vspace*{-1.5em}
\end{table}

% \vspace*{-2em}
\section{Conclusion}
% \vspace*{-1em}
This paper introduces PHORECAST, a novel multi-modal dataset that enables modeling of human reactions to public health campaigns using demographics and psychographic factors and personality traits. With training using PHORECAST, we capture fine-grained individual traits, inclusive of demographics, personality, and contextual behaviors, and demonstrate the utility of this dataset in predicting responses to health messaging. Our results highlight both the shortcomings of existing LLM/VLM models and the promise of individualized reasoning in shaping better human-aligned public health interventions. PHORECAST provides a foundation for future research in human-centric AI systems that aim to navigate the complexity and diversity of real-world decision-making.

% \vspace*{-0.25em}
\noindent
{\bf Discussion, Limitations and Future Directions: }
This works present several promising avenues for future research. First, this dataset comprises primarily of English-speaking participants, which may bias the findings and limit global applicability. Cultural, linguistic, and geopolitical factors critically shape attitudes towards public health issues, and future efforts must broaden the participant base to support more globally representative insights. Second, while our models incorporate rich contextual features, they remain static with respect to time. Health beliefs are not fixed; they evolve with social, political, and personal contexts. Modeling temporal dynamics is essential to anticipate these shifts and to evaluate the long-term impact of interventions. \\

% \vspace*{-0.25em}
{\bf Contraindications}
PHORECAST is not a global dataset. As such, PHORECAST should not be used to infer cross-cultural, multilingual, or non-U.S. population responses to health communication. Researchers seeking to generalize beyond U.S.-based, English-speaking populations should treat PHORECAST results as hypothesis-generating only, and pursue follow-up studies with more representative data.

The PHORECAST dataset has comparatively low representation from certain demographic groups, including racial groups such as Asians and Hispanics, religious groups such as Agnostics and Buddhists, and older adults (particularly those over 65). Addressing these demographic gaps will be crucial for enhancing the inclusivity and generalizability of future analyses.

\newpage

\bibliography{neurips.bib}
\bibliographystyle{plainnat}

\appendix
\clearpage
\appendix
% \section*{Supplementary Material}
\addcontentsline{toc}{section}{Supplementary Material}

% \vspace*{\fill}
\begin{center}
    \textbf{\Huge Supplementary Material}
\end{center}
\vspace{4em}
\hrulefill
\vspace{1em}
\begin{center}
    \textbf{\Large Table of Contents}
\end{center}
\vspace{1em}
\hrulefill
\vspace{1em}

\noindent\textbf{Part 1 - The PHORECAST Dataset}
\begin{itemize}
    % \item 1. Dataset Details \hfill \pageref{supp:part1_technical_appendices}
        % \begin{itemize}
    \item A. Dataset Access \hfill \pageref{supp:Access}
    \item B. Recruitment Process \hfill \pageref{sup:recruitment}
    \item C. Curation Rationale \hfill \pageref{supp:curation-rationale}
    \item  D. Integration with Existing Public Health Communication Frameworks \hfill \pageref{supp:integration}
    \item E. Ethical Consideration \hfill
    \pageref{supp:ethical-cons}
    \item F. Terms of Use \hfill
    \pageref{supp:terms}
    \item G. Data Rights Compliance and Issue Reporting \hfill \pageref{data-rights}
    \item H. Informed Consent \hfill \pageref{supp:consent}
    \item I. Survey Details \hfill \pageref{supp:survey}
    \item J. Data Processing \hfill \pageref{supp:data-processing}
    \item K. Data Analysis \hfill \pageref{supp:data-analysis}
    \item L. Media Analysis \hfill \pageref{supp:media-an}
\end{itemize}

\vspace{1em}
\noindent\textbf{Part 2 - Training Details and Evaluation Metrics}
\begin{itemize}
    \item L. Training Details \hfill \pageref{supp:training} 
    \item M. Dataset Preparation \hfill
            \pageref{supp:part1_data_prep}
    \item N. Free Form Metrics \hfill 
    \pageref{supp:free-form}
    \item  O. Generalization \hfill \pageref{supp:generalization}
        % \begin{itemize}
        %     \item 3.1 SDE \hfill \pageref{supp:free-form}
        %     \item 3.2 Discriminator Accuracy \hfill \pageref{free-form-acc}
        % \end{itemize}
    \item P. Qualitative Examples \hfill \pageref{supp:qual}
\end{itemize}

% \vspace{1em}
\noindent\textbf{Part 3 - Generalizing Response Predictions to Unseen Communication Strategies}
\hfill \pageref{unseen-comm}

\vspace{1em}

\noindent\textbf{Part 4 - Future Practical Use Case: Communication Strategy Recommendation}
\hfill \pageref{comm-rec}

\vspace*{\fill}
\clearpage
% comm-rec

\section{The PHORECAST Dataset}
\label{supp:part1_technical_appendices}
\subsection{Access}
\label{supp:Access}
The data can be accessed on HuggingFace at \hyperlink{PHORECAST}{https://huggingface.co/datasets/tomg-group-umd/PHORECAST}. We also provide scripts for data preparation and transforming it into training and validation splits on \hyperlink{Github}{github.com/rifaaQ/PHORECAST}. In doing so, each row represents a single prediction task for a particular individual (one of the 12 Likert scale items or free form response).

\subsection{Recruitment Process}
\label{sup:recruitment}
We recruit participants through the \hyperlink{SurveyCircle}{https://www.surveycircle.com/en/} platform and via the social media platform LinkedIn. Each participant is incentivized with a \$10 Tango gift card. Participants are provided with details of potential risks of participating in the survey and their ability to voluntarily skip questions or end the survey before completion. All data was collected between Jan 5th, 2025 to Jan 9th, 2025.
% The project is funded by the T-01 Teaching and Learning Innovation Grant from the Teaching and Learning Transformation Center at the University of Maryland, College Park.

\subsection{Curation Rationale}
\label{supp:curation-rationale}
The PHORECAST dataset aims to map real human profiles (demographics, personality, and locus of control) to their responses / reactions from interacting with various public health campaigns. The primary purpose is for academic research to study how different people interact with stimuli and simulate how and why different communities respond differently to visuals. The results will be used to build an AI simulator that can mimic real world communities. 
\subsection{Integration with Existing Public Health Communication Frameworks}
\label{supp:integration}
Our approach is designed to complement existing public health communication frameworks by offering a scalable, low-risk method for pre-testing messages prior to real-world deployment. Currently, message evaluation relies heavily on human subjects through focus groups or randomized trials — ad-hoc approaches that are often resource-intensive, time-consuming, and difficult to implement in time-sensitive or emergency contexts.

By simulating audience responses using LLM-generated communities, our method offers practitioners a novel way to assess potential message effectiveness and refine content in advance. This novel approach for public-health messaging has the potential to reduce reliance on real-time, post-dissemination adjustments, which can be both costly and disruptive, potentially biased and less representative given the limited sampled community groups and resource-constraints.

In practice, this new approach, empowered by LLM and broader online community surveys, supports:
\begin{enumerate}
    \item Rapid iteration and scenario testing within existing campaign development workflows 
    \item Enhanced training for public health students and practitioners, analogous to surgical simulators in medicine, allowing them to practice message design and delivery in a controlled, harm-free environment
    \item Improved trust and efficiency in community engagement, by minimizing the need for repeated in-person testing, particularly with vulnerable populations
\end{enumerate}
While not a replacement for field validation, this newly designed AI tool aims to serve as an intermediary step that enhances message development and helps operationalize core principles from established frameworks, such as the Health Belief Model, Social Cognitive Theory, and the Theory of Planned Behavior.

\subsection{Ethical Considerations}
\label{supp:ethical-cons}
This study is approved by the Institutional Review Board (IRB) at the University of Maryland, College Park, under an exemption category. All participants provide digital informed consent prior to beginning the survey. The consent form clearly outlines potential risks and benefits of participation, and informs participants of their right to withdraw at any time and to decline to answer any questions. To maintain anonymity and separate responses from personal identifying information, participants are directed to a separate form for incentive compensation. All collected data are de-identified before analysis.
\subsection{Terms of Use}
\label{supp:terms}
\paragraph{Purpose}
The dataset can only be used for educational and research purposes or to develop and evaluate AI models.
\vspace{-0.3cm}
\paragraph{Restrictions}
This dataset is a public-use dataset as defined by the Data Procedures Manual by NCES (\url{https://nces.ed.gov/}). All individually identifiable information has been removed to protect the confidentiality of participants and no license is needed to access the dataset.
\paragraph{Deanonymization}
\vspace{-0.3cm}
The users are prohibited from de-anonymizing the individuals represented in the dataset. 
\paragraph{Content Warning}
\vspace{-0.3cm}
The dataset may include text, images and videos that could be considered unsafe or offensive for some individuals. The users must use appropriate measures to filter content when used for educational or training purposes to adhere to the ethical and safety standards.
\paragraph{Endorsement and Liability}
\vspace{-0.3cm}
The authors of the paper, the dataset creators, funders and the affiliated institution do not endorse the views and opinions expressed in the data and are not liable to damages resulting from the use of the dataset.

\subsection{Data Rights Compliance and Issue Reporting}
\label{data-rights}
We are committed to complying with data protection rights. If any individual whose data is included in the PHORECAST dataset wishes to have their data removed, we provide a straightforward process for issue reporting and resolution. Concerned parties are encouraged to contact the authors directly via making a formal issue reporting on our GitHub page at \url{https://github.com/rifaaQ/PHORECAST}. Upon receiving a request, we will engage with the individual to verify their identity and proceed to remove the relevant entries from the dataset. We commit to addressing and resolving such requests within 30 days of verification.

\subsection{Informed Consent}
\label{supp:consent}
\paragraph{Eligibility}
Thank you for expressing interest in helping us build an AI Community Simulator. Before we get started, please answer the following eligibility questions:\\
Are you 18 years of age or older?
$\bigcirc$ Yes 
$\bigcirc$ No \\
Do you currently reside in the United States? 
$\bigcirc$ Yes 
$\bigcirc$ No \\
If “no” to any one of these questions:
Thank you for taking the time to express interest in our study. For more information
about the work we do, please visit: 
\hyperlink{PH}{https://sph.umd.edu/about/office-public-health-
practice-community-engagement}. If eligible, the participants are directed to the informed consent:
\vspace{-0.2cm}
\paragraph{Purpose of the Study}
This research project aims to develop an Al-powered community simulator modeled after real-world communities (e.g., Prince George's County) to enhance public health training and practice. This study seeks to understand how individuals' demographic identities and personality traits interact and react to public health messages. The results will be used to develop a prototype of an Al-powered community simulator to test public health trainees' development of public health messages.
\vspace{-0.2cm}
\paragraph{Procedures}
You will be asked to answer questions about your social and physical identity and personality traits. Then, you will be asked to review and react to various public health communication messages. The survey will take approximately 30 minutes to complete. Once you complete the survey, you will be asked to enter your contact information separately to receive your compensation.
\vspace{-0.2cm}
\paragraph{Potential Risks and Discomforts}
There are minimal risks or inconveniences from participating in this research study. The length of time required to take this survey (30 minutes) may be inconvenient for some; however the survey has been designed to limit this possibility. If you feel uncomfortable answering some questions, you have the right to skip any questions you do not want to answer.
\vspace{-0.2cm}
\paragraph{Potential Benefits}
While there are no direct benefits to you for participating in this study, your involvement will significantly benefit the development of training opportunities for public health trainees to increase the effectiveness of public health messaging to improve health outcomes.
\vspace{-0.2cm}
\paragraph{Confidentiality}
Any potential loss of confidentiality will be minimized by storing data in a password-protected computer.
If we write a report or article about this research project, your identity will be protected to the maximum extent possible. Your information may be shared with representatives of the University of Maryland, College Park, or governmental authorities if you or someone else is in danger or if we are required to do so by law. You will be taken to a separate form to enter your contact information for compensation to ensure that your personal information will not be linked to your responses.
\vspace{-0.2cm}
\paragraph{Compensation}
By participating in this study, you will receive a \$10 gift card. You will be responsible for any taxes assessed on the compensation. A separate email containing your compensation will be sent to you within 30 days of completing the survey
\vspace{-0.2cm}
\paragraph{Right to Withdraw and Questions}
Your participation in this study is completely voluntary. You may choose not to take part at all. If you decide to participate in this study, you may stop participating at any time. If you decide not to participate in this study or if you stop participating at any time, you will not be penalized or lose any benefits to which you otherwise qualify.
If you decide to stop taking part in the study, have any questions, concerns, or complaints, or if you need to report an injury related to the research, please contact the investigators:
Dr. Tracy Zeeger
School of Public Health, Room 1234T
\textit{tzeeger@umd.edu}
301-405-3453
Dr. Sylvette La Touche- Howard
School of Public Health, Room 2242
\textit{latouche@umd.edu}
301-405-8161
\vspace{-0.2cm}
\paragraph{Participant Rights}
If you have questions about your rights as a research participant or wish to report a research-related injury, please contact:
University of Maryland College Park
Institutional Review Board Office
1204 Marie Mount Hall
College Park, Maryland, 20742
E-mail: irb@umd.edu
Telephone: 301-405-0678
For more information regarding participant rights, please visit:
\hyperlink{IRB}{https://research.umd.edu/irb-research-participants}
This research has been reviewed according to the University of Maryland, College Park IRB procedures for research involving human subjects.
\vspace{-0.2cm}
\paragraph{Statement of Consent}
Your signature indicates that you are at least 18 years of age; you have read this consent form or have had it read to you; your questions have been answered to your satisfaction and you voluntarily agree to participate in this research study. You will receive a copy of this signed consent form.
By checking the box below, you indicate that you are at least 18 years of age; you have read this consent form or have had it read to you; your questions have been answered to your satisfaction and you voluntarily agree to participate in this research study. If you agree to participate, please check the box:
$\bigcirc$ I agree 
\vspace{-0.1cm}
\subsection{Survey}
\label{supp:survey}
\begin{figure}
    \centering
    \includegraphics[trim={2cm 8.5cm 2cm 10.5cm},clip,width=1\linewidth]{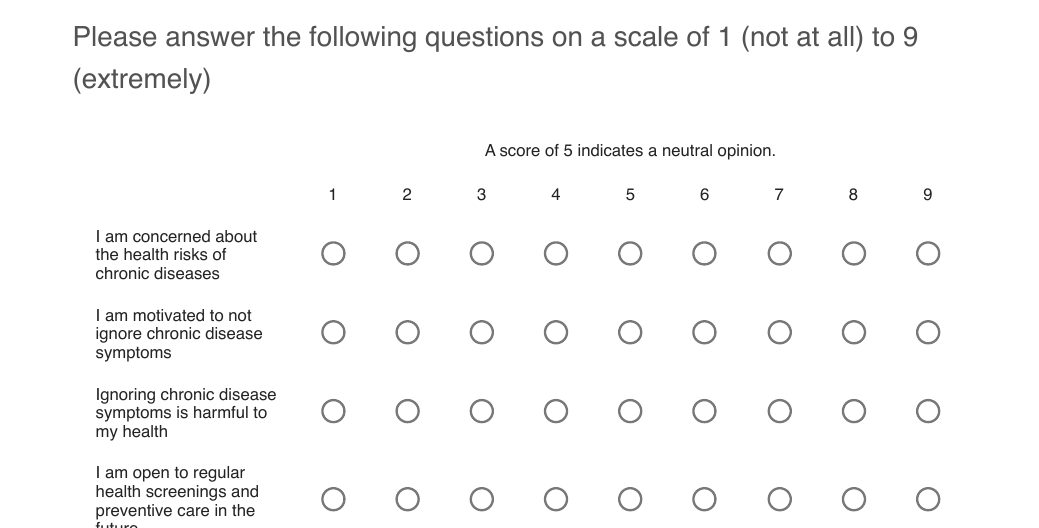}
    \includegraphics[trim={2cm 8.5cm 2cm 10.5cm},clip,width=1\linewidth]{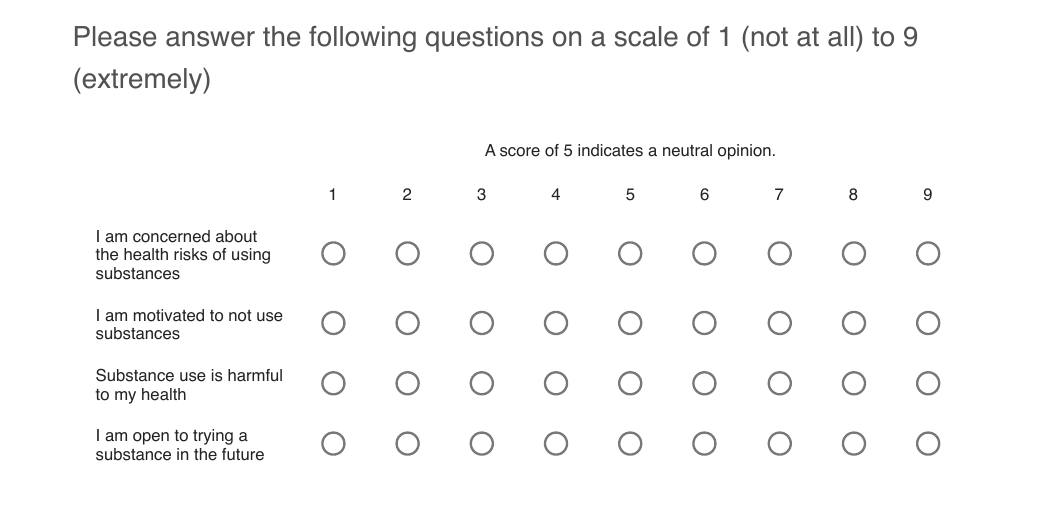}
    \caption{Example of baseline opinions we ask participants prior to viewing any public health campaigns. Each participants answers four baseline questions on five health topics, such as Chronic Diseases, Substance Use, Smoking/COPD, Nutrition etc}
    \label{fig:baseline-ex}
\end{figure}

\begin{figure}
    \centering
    \includegraphics[trim={1cm 1cm 3cm 6cm},clip,width=1\linewidth]{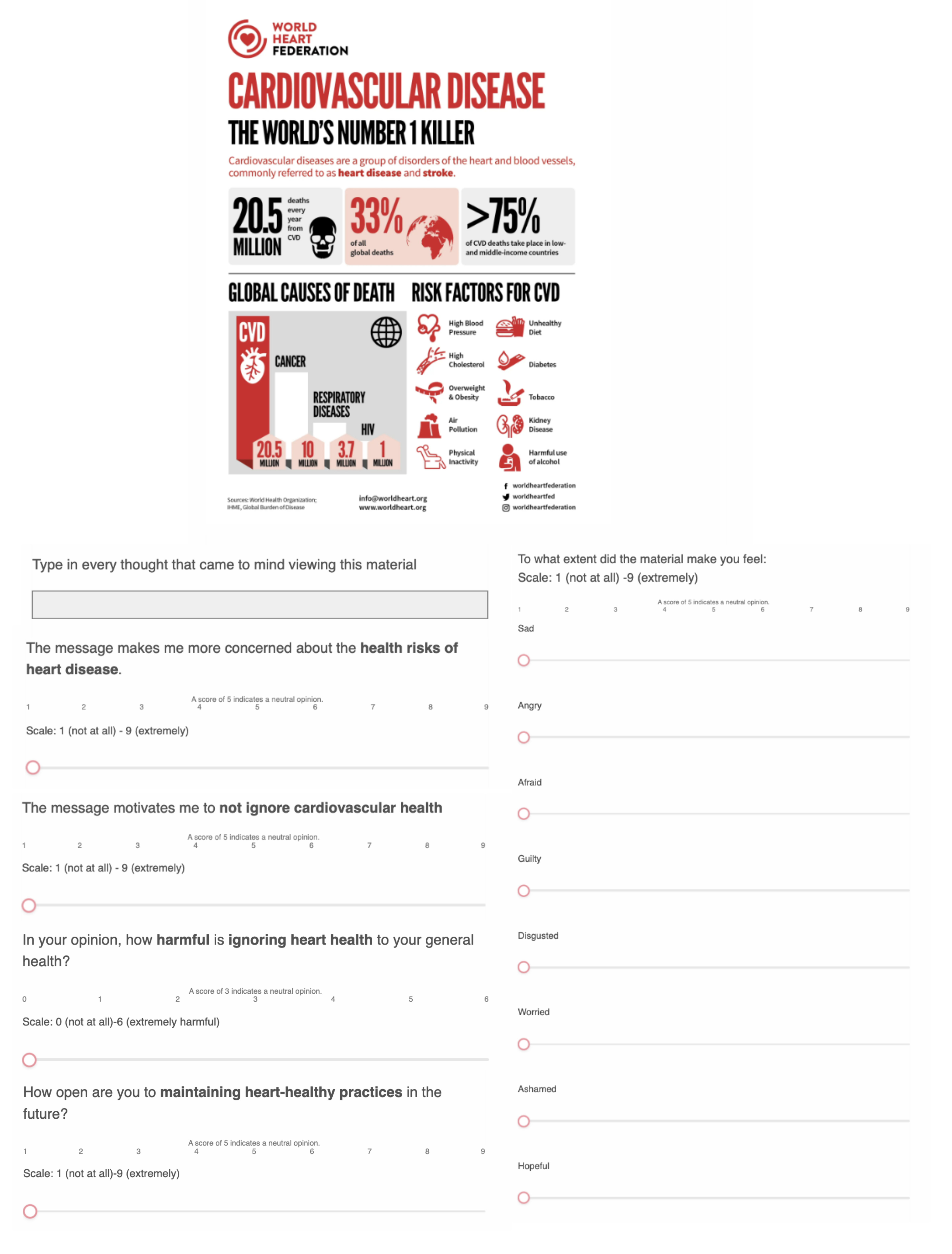}
    \caption{An example of the public health campaign survey shown to participants. This snippet illustrates the diverse question types, including free-form responses, 4-point Likert scales (gauging concern, motivation, harm perception, and openness), and emotional assessments. Note: Display order here does not reflect the actual survey flow.}
    \label{fig:survey-ex}
\end{figure}

We use the Big Five Inventory (BFI-2) by \cite{Soto2017BFI2} along with their scoring methods\footnote{For more information, please visit \url{http://www.colby.edu/psych/personality-lab/}}. We collect demographics information and personality traits by having participants self-report and answer the questions. Participants answer baseline questions prior to viewing any campaigns on five different health topics. We show an example of the questions asked for the topic of Chronic Diseases and Substance Use in Fig.\ref{fig:baseline-ex}. Subsequently, participants engaged with five distinct campaigns related to the same baseline topics, and provided their opinions and reactions through a series of questions. We show an example of this in Fig.\ref{fig:survey-ex}. 

% \begin{figure}
%     \centering
%     \includegraphics[trim={1cm 0cm 1cm 0cm},clip,width=0.4\linewidth]{media/supplementary/personality_.pdf}
%     % \includegraphics[trim={0cm 10cm 0.2cm 4cm},clip,width=0.4\linewidth]{media/supplementary/personality_key.pdf}
%     \caption{The Big Personality Questionnaire and the scales, directly copied from \cite{Soto2017BFI2}.}
%     \label{fig:big5}
% \end{figure}
\subsection{Campaign Selection Process}
To build the campaign repository, we curated open-source health communication materials from nonprofit organizations, peer-reviewed publications, government agencies, and other entities focused on public health behavior change. Inclusion criteria required that messages (1) addressed a health behavior or outcome and (2) included a visual component (e.g., print or video campaigns). We intentionally sampled a diverse range of campaigns to capture messages targeting different age groups and demographic segments.

\textbf{Annotation Process:} 
Each campaign was annotated along three dimensions:

\textit{Target Population}: Classified into predefined age-based categories (Children $\leq$11, Adolescents 12–17, Young Adults 18–24, Adults 25–44, Adults 45–64, and Older Adults $\geq$ 65). When messages applied to multiple groups, overlapping categories were selected.

\textit{Message Type}: Based on the dominant communication strategy—informative, persuasive-efficacy, or persuasive-threat—categories commonly used in public health communication and grounded in established health behavior theories.

\textit{Health Behavior \& Outcome:} Specific health behaviors and linked outcomes were identified for each message.

The initial coding was conducted by trained research assistants under the guidance of the principal investigators. All annotations underwent a secondary review by the investigators to ensure coding consistency. Any discrepancies were resolved through discussion until consensus was reached.

\subsection{Data Processing}
\label{supp:data-processing}
Due to a discrepancy between compensation forms and completed surveys, we implemented several measures to ensure data integrity. Incomplete surveys and submissions from duplicate IP addresses were removed. Additionally, we identified and excluded four participants whose free-form responses exhibited characteristics consistent with generative model output, indicating potential fabrication. 

\subsection{Data Analysis}
\label{supp:data-analysis}
\begin{figure}
    \centering
    \includegraphics[trim={0.0cm 0.0cm 0.2cm 0.8cm},clip,width=0.95\linewidth]{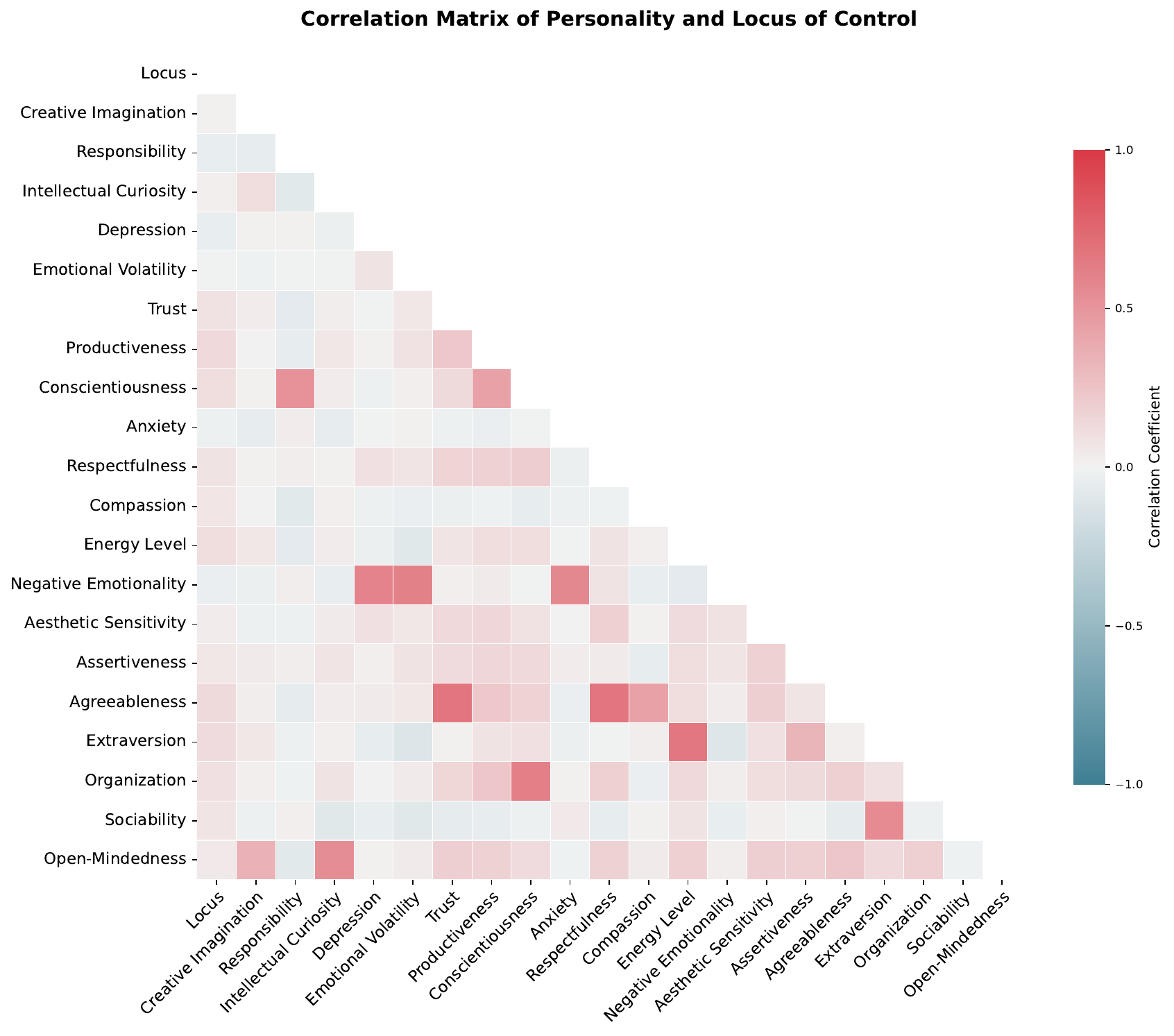}
    \caption{Covariance matrix of personality features. Agreeableness, Trust, Productiveness, Respectfulness and Compassion are highly correlated. Emotional Volatility, Anxiety, Depression and Negative Emotionality are highly correlated. The Locus of Control is more highly correlated with Productiveness, Agreeableness, Extraversion, Organization, Energy Level than it is with Creative Imagination, Responsibility, Depression and Emotional Volatility. }
    \label{fig:corr-pers}
\end{figure}

\begin{figure}
    \centering
    \includegraphics[width=0.9\linewidth]{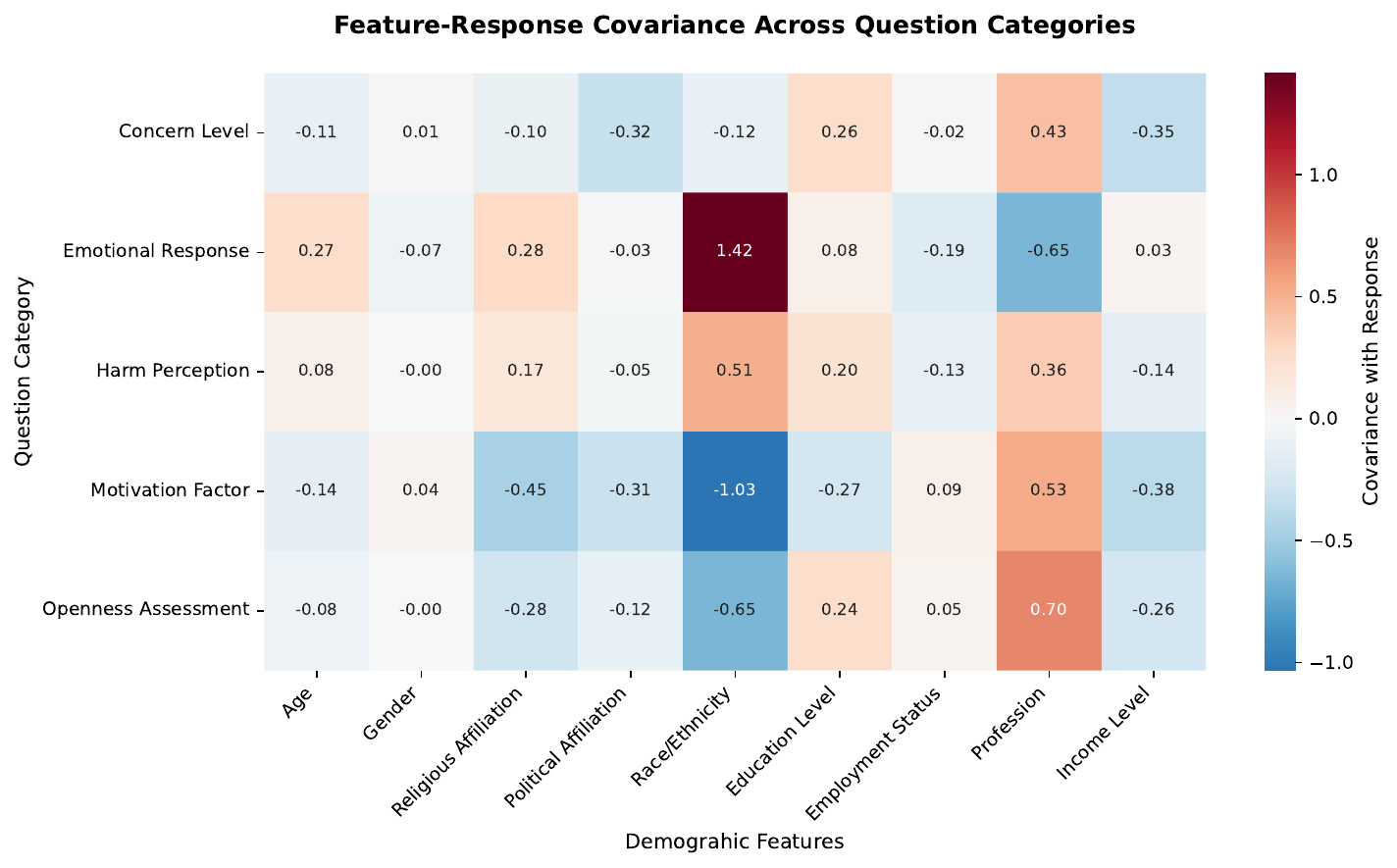}
    \includegraphics[trim={0cm 0cm 0cm 1cm},clip,width=0.9\linewidth]{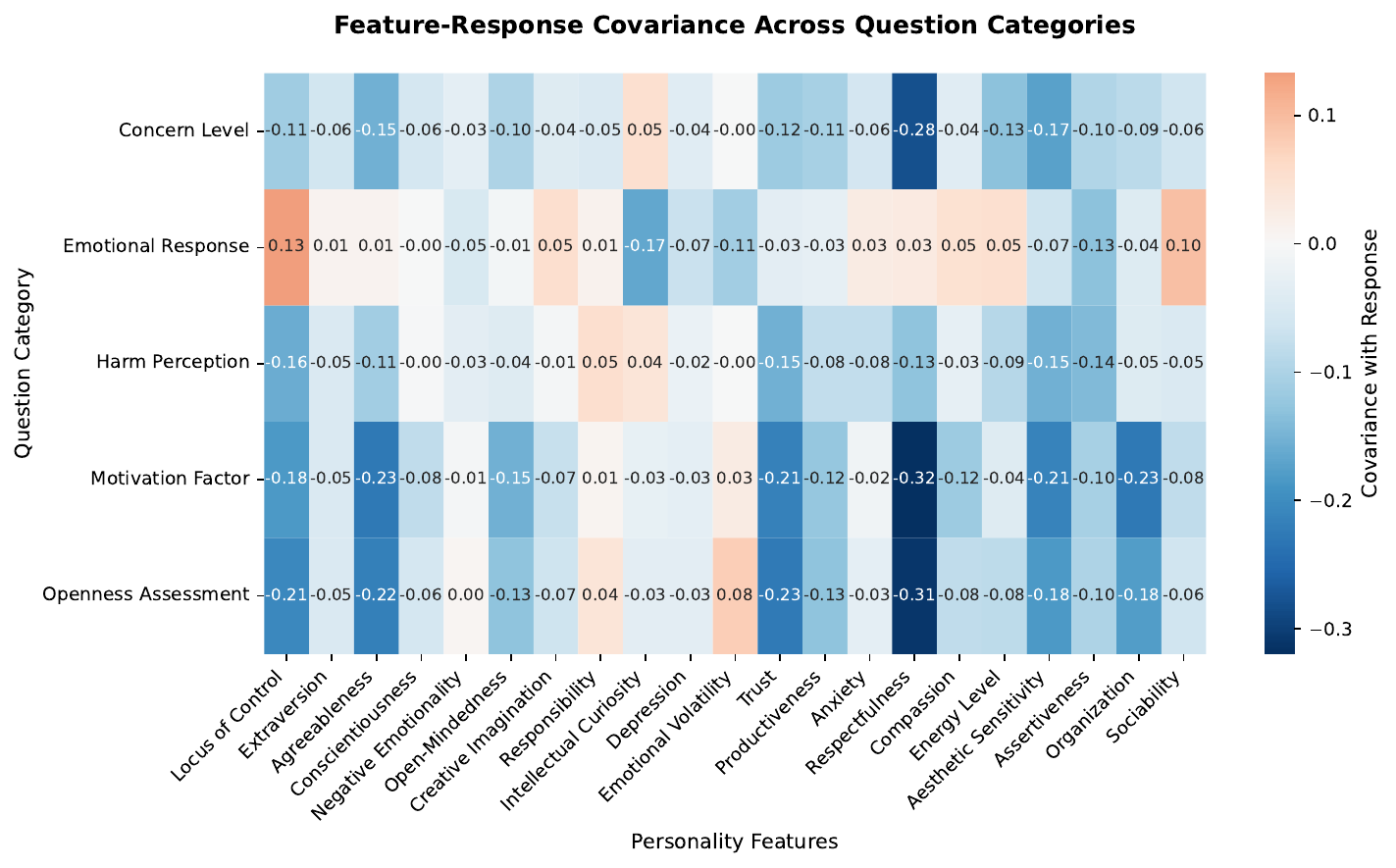}
    \caption{Demographics \& Personality Features and Response covariance across question categories. Race/Ethnicity shows strong correlations with emotional responses (cov = 1.42) and harm perception (cov = 0.51), but minimal influence on motivation or openness. Profession is a dominant factor across multiple dimensions, showing high covariance with openness (cov = 0.70), motivation (cov = 0.53), concern level (cov = 0.43), and harm perception. Within personality traits, emotional volatility and sociability are most strongly associated with emotional responses and openness. Locus of control exhibits the highest covariance among personality traits with emotional responses (cov = 0.13)}
    \label{fig:covariance-across-features}
\end{figure}

We look into the covariance matrix of our features to analyze their relationships. We observe high correlation between  different traits such as Respectfulness and Agreeableness, Extraversion and Energy Level, Negative Emotionality, Depression and Emotional Volatility. Please refer to Fig.\ref{fig:corr-pers} for the full analysis. We also analyze the correlation between demographics or personality and the responses. We observe that the locus of control and the sociability level of the individual greatly impact their emotional responses (Fig.\ref{fig:covariance-across-features}). Furthermore, an individual's race/ethnicity greatly impacts their emotional response, followed by their profession and education level. Those features also impact the harm perception, concern level, motivation and openness levels. 

\begin{figure}[h]
    \centering
    \includegraphics[trim={0.3cm 0.9cm 0.2cm 0.2cm},clip,width=0.49\linewidth]{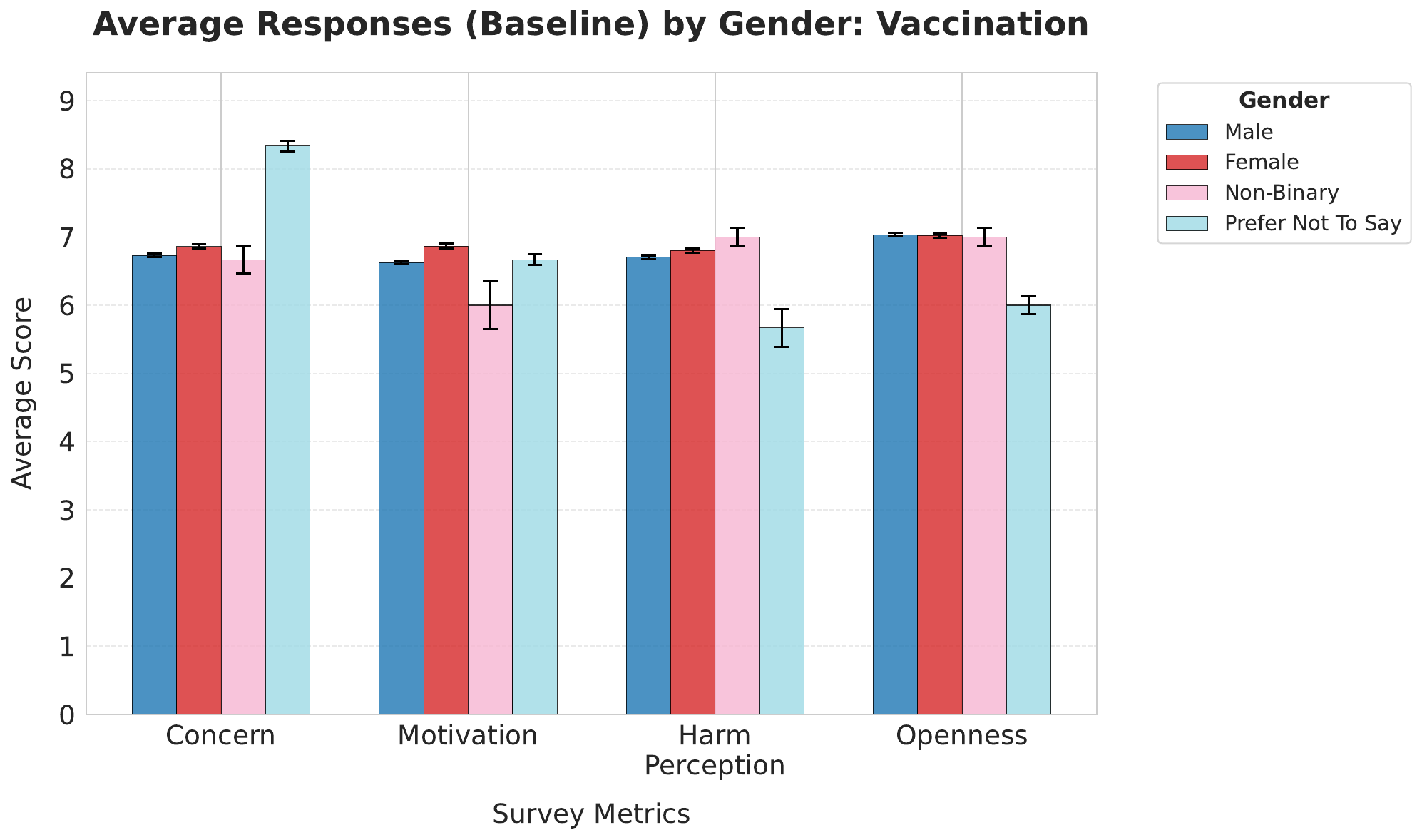}
    \includegraphics[trim={0.3cm 0.9cm 0.2cm 0.2cm},clip,width=0.49\linewidth]{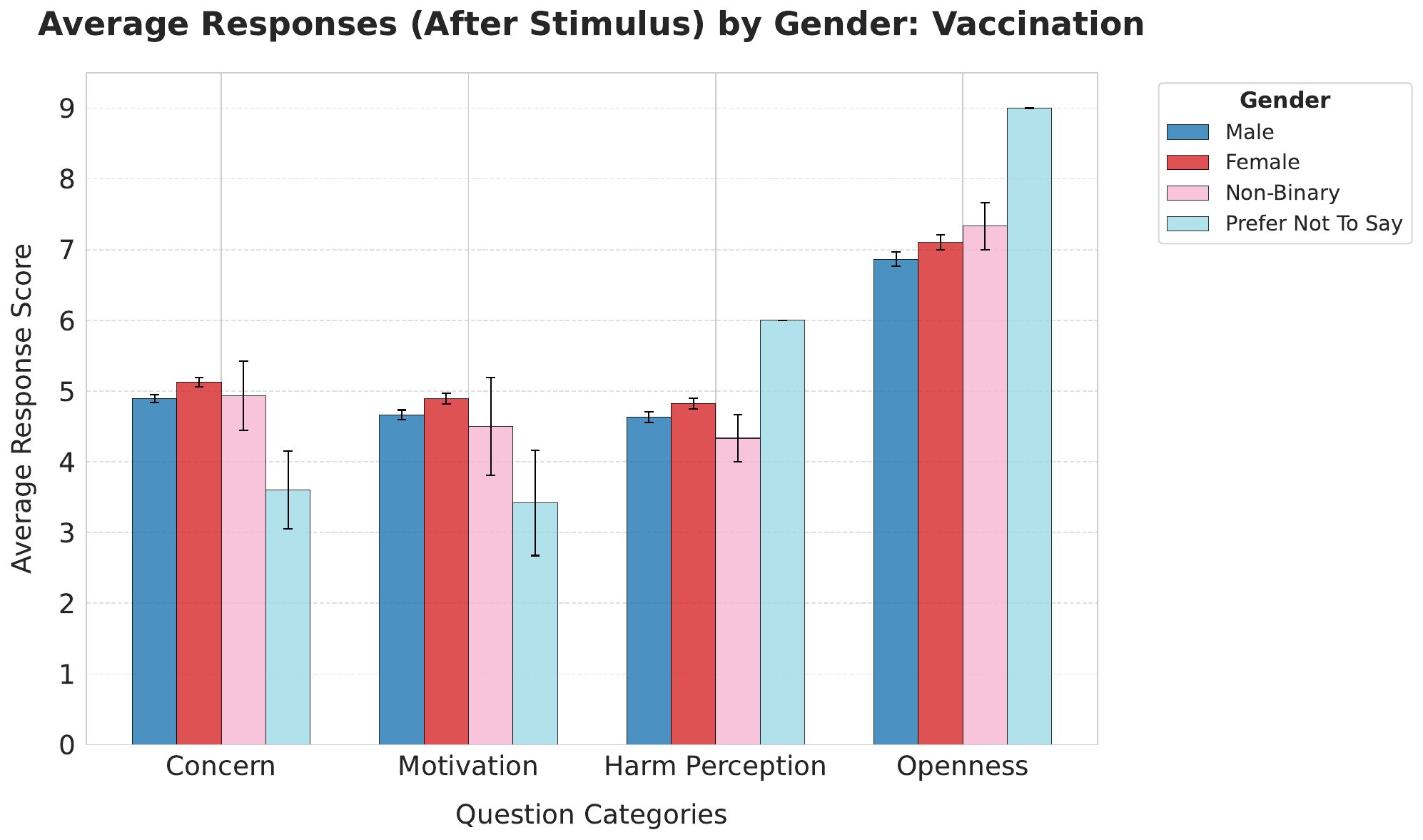}
    \includegraphics[trim={0.3cm 0.9cm 0.2cm 0.2cm},clip,width=0.49\linewidth]{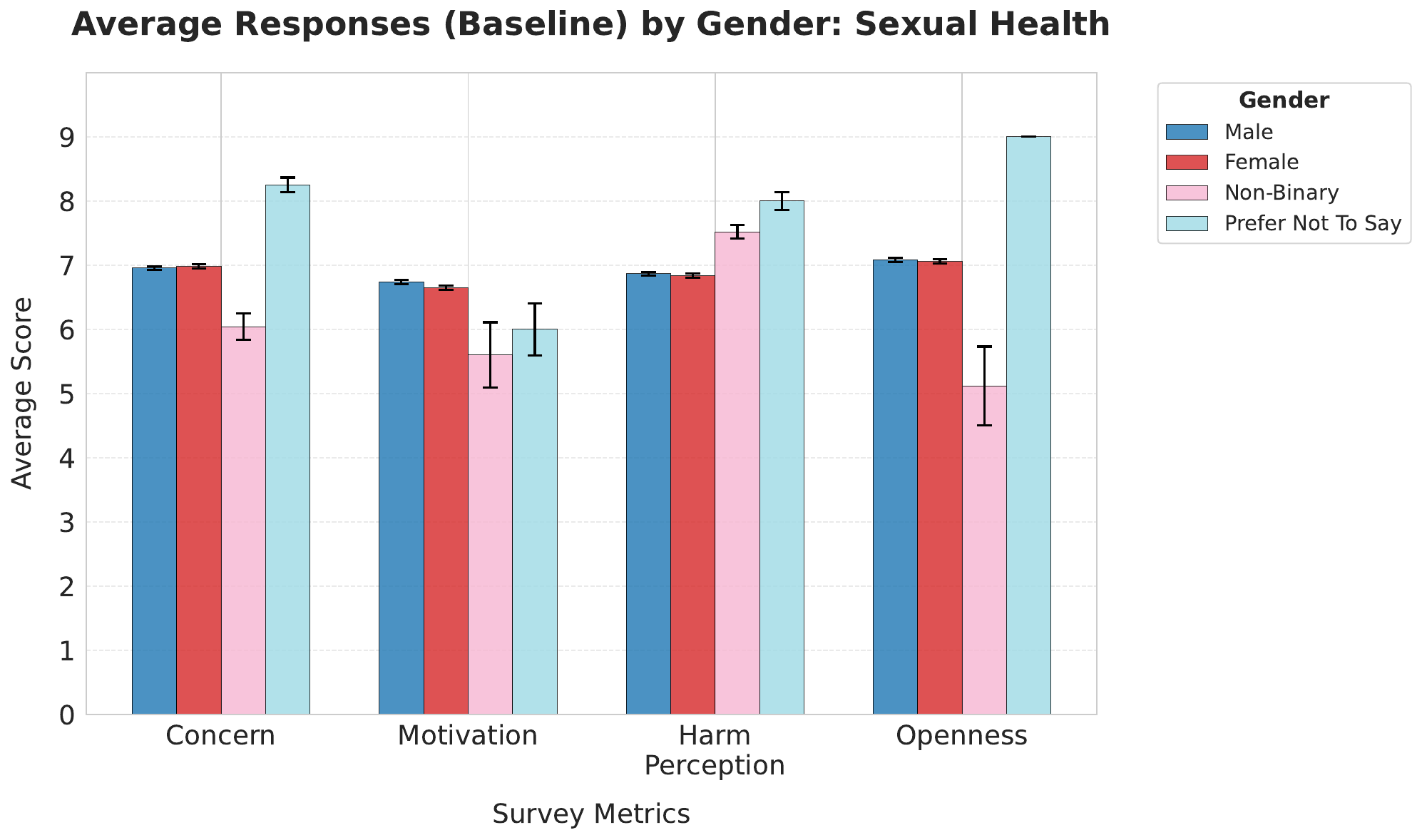}
    \includegraphics[trim={0.3cm 0.9cm 0.2cm 0.2cm},clip,width=0.49\linewidth]{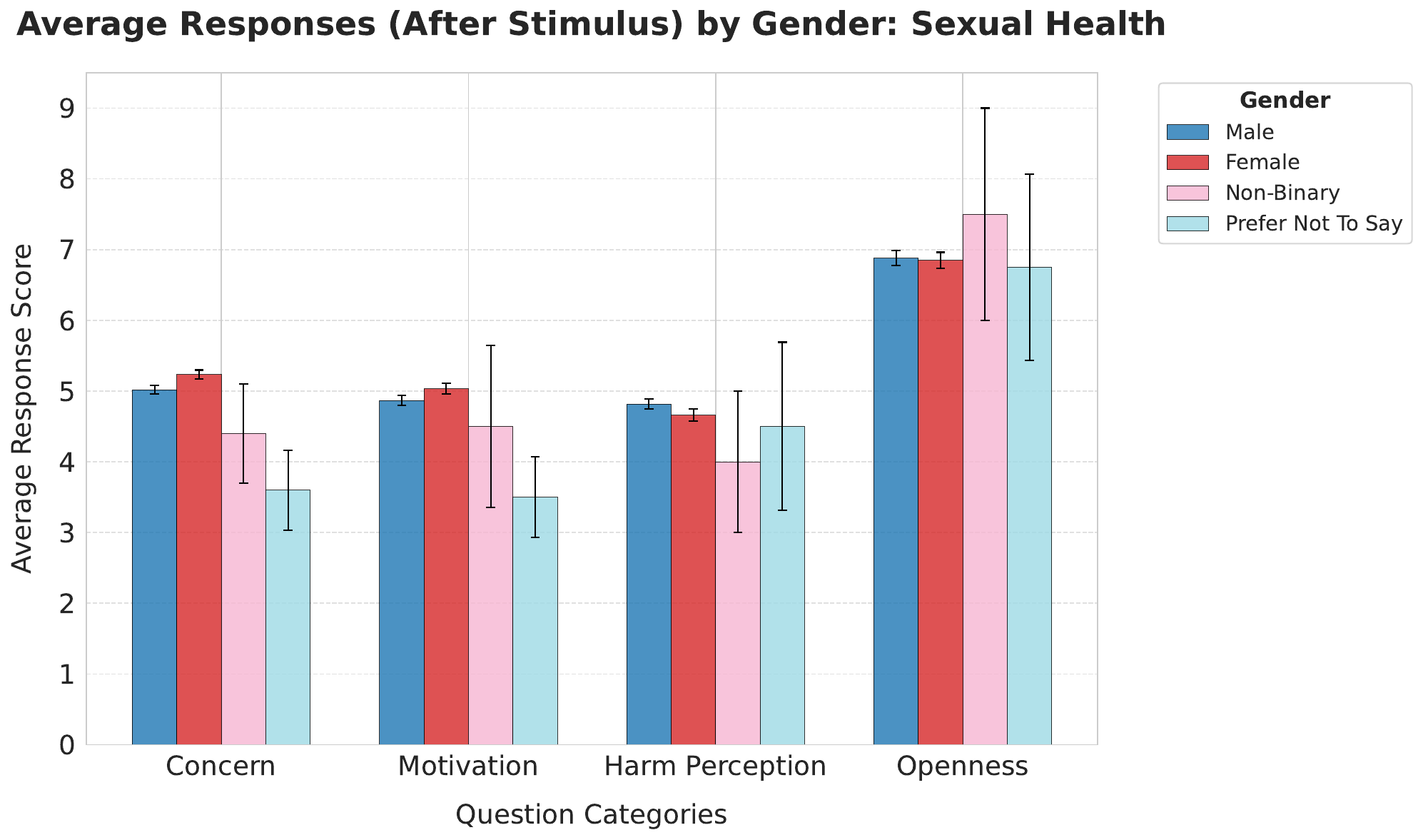}
    \includegraphics[trim={0.3cm 0.9cm 0.2cm 0.2cm},clip,width=0.49\linewidth]{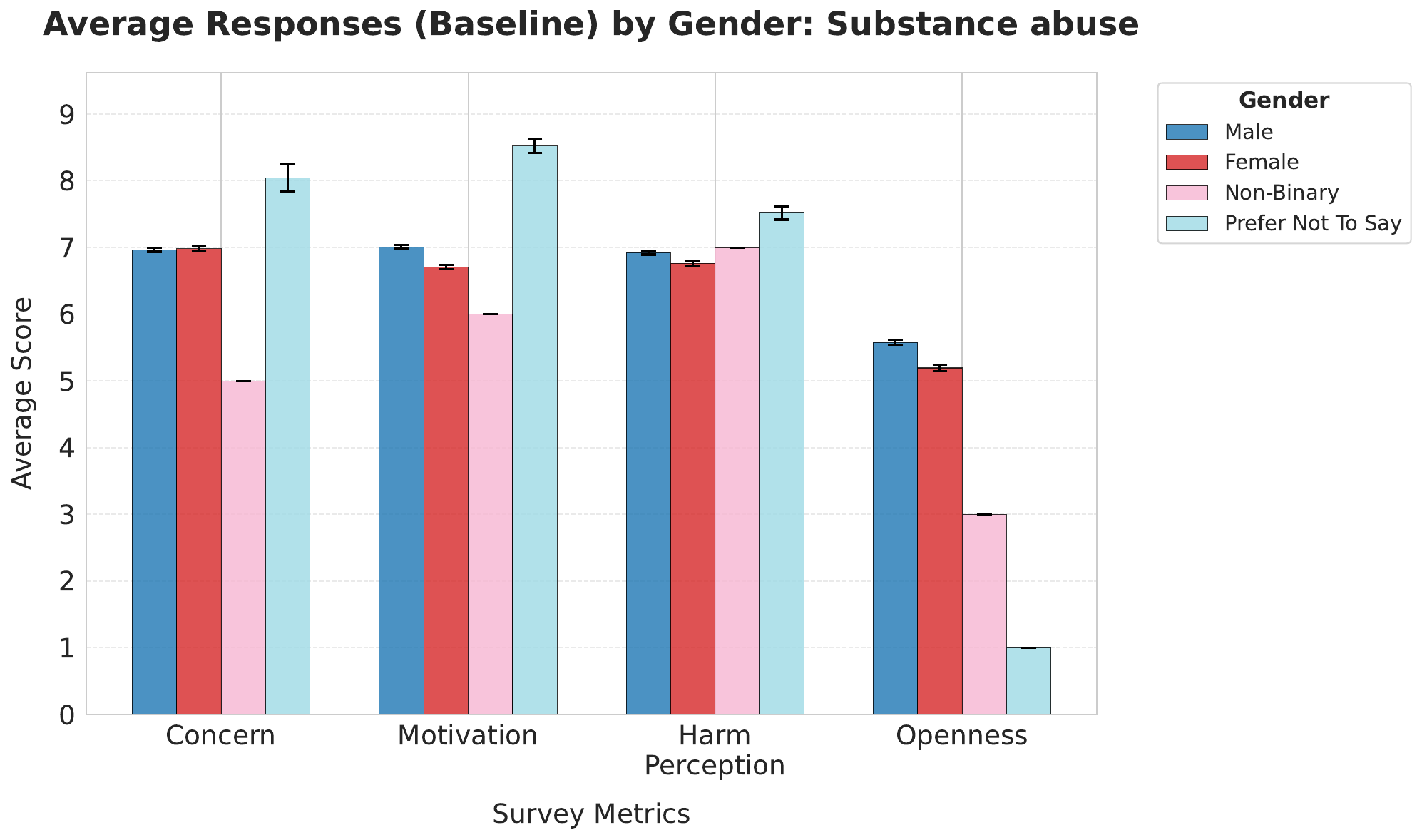}
    \includegraphics[trim={0.3cm 0.9cm 0.2cm 0.2cm},clip,width=0.49\linewidth]{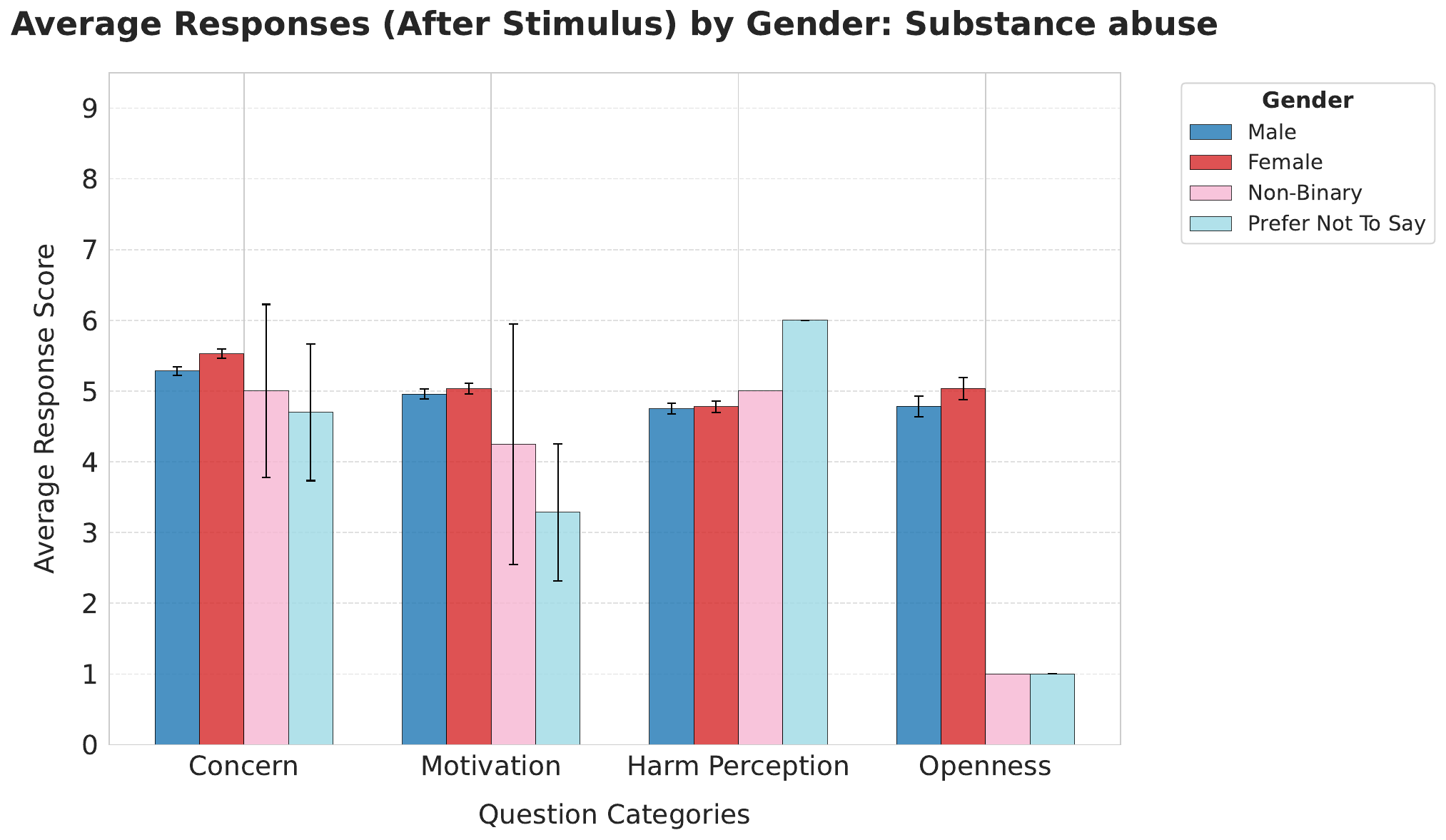}
    \includegraphics[trim={0.3cm 0.9cm 0.2cm 0.2cm},clip,width=0.49\linewidth]{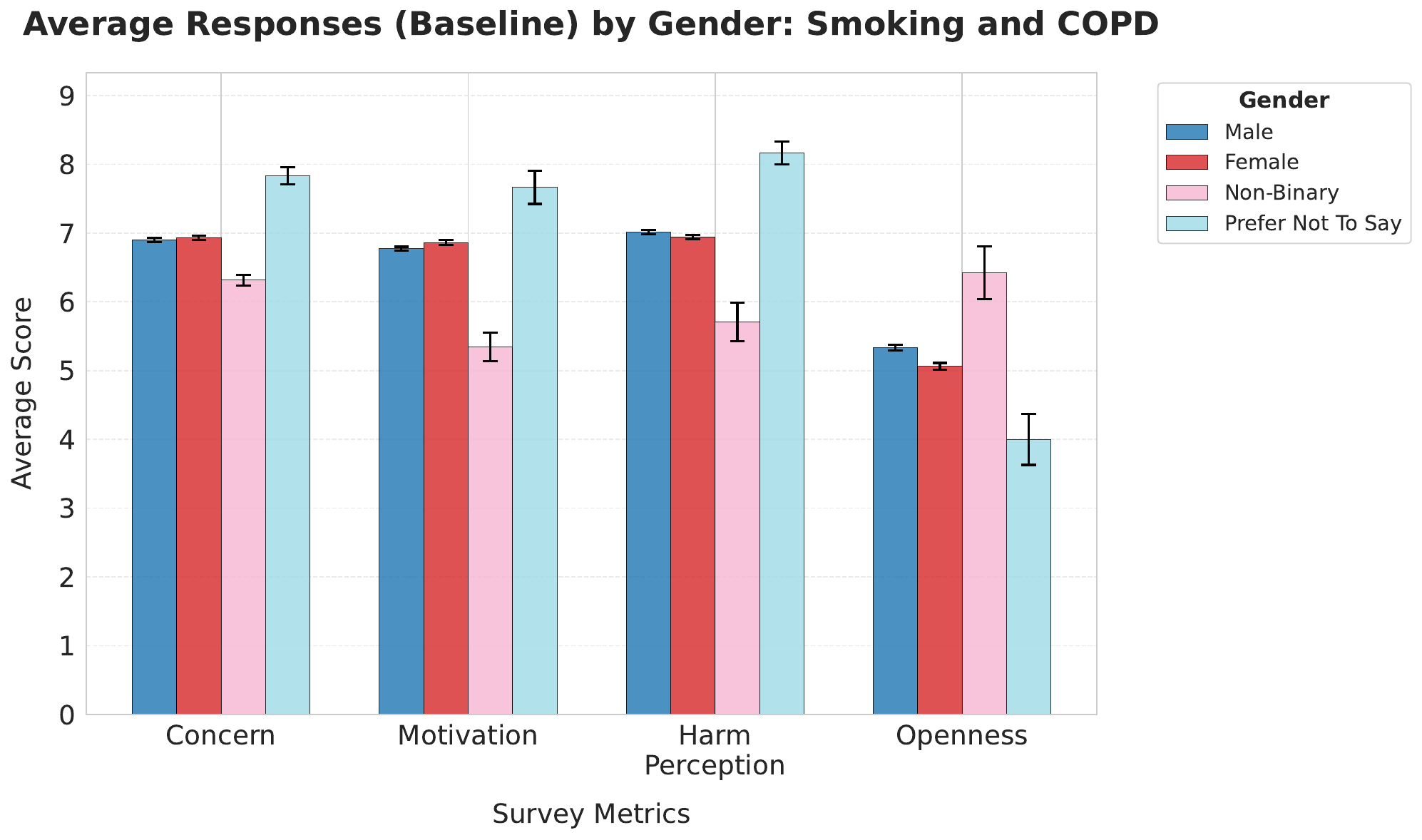}
    \includegraphics[trim={0.3cm 0.9cm 0.2cm 0.2cm},clip,width=0.49\linewidth]{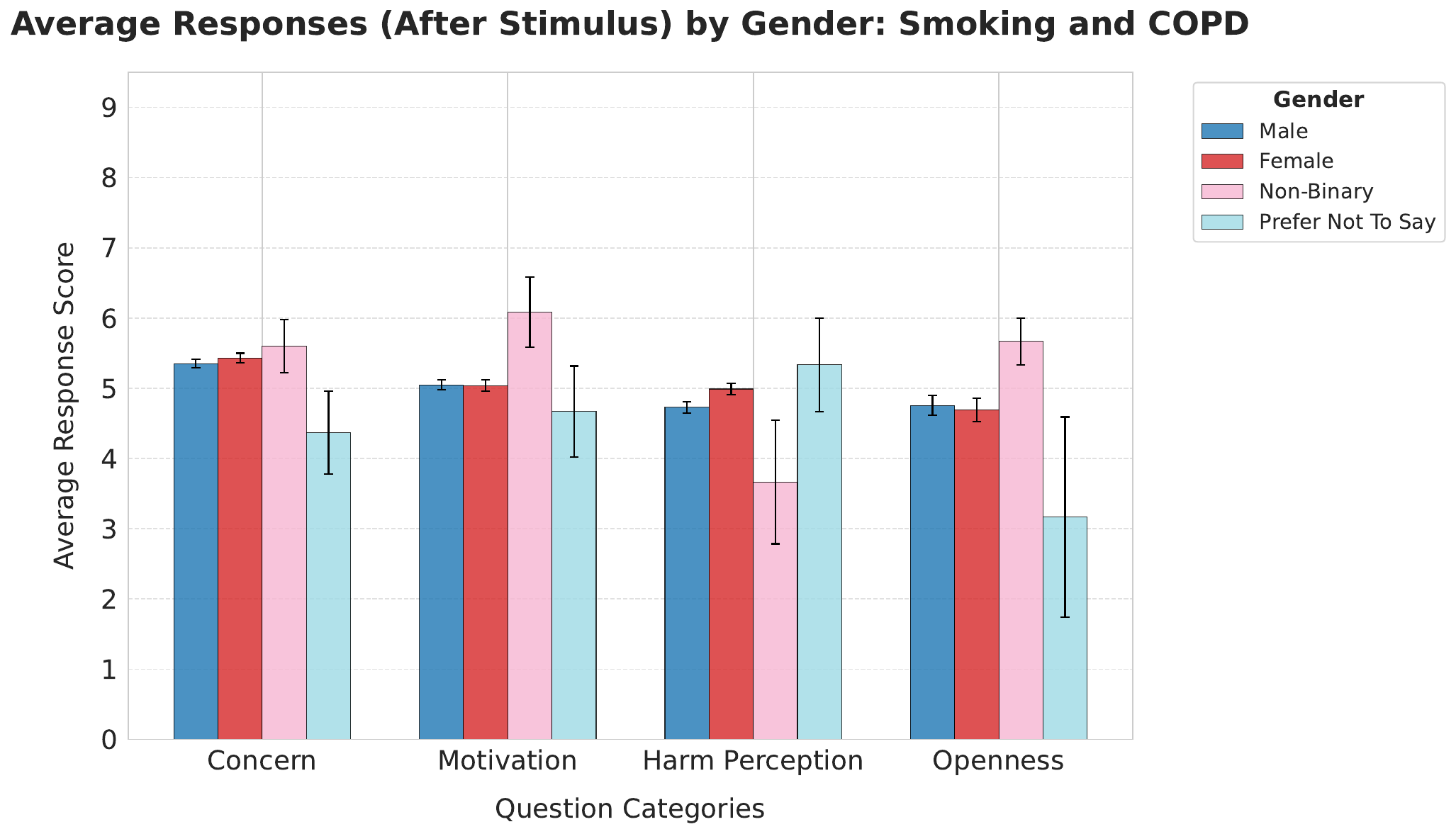}
    \includegraphics[trim={0.3cm 0.9cm 0.2cm 0.2cm},clip,width=0.49\linewidth]{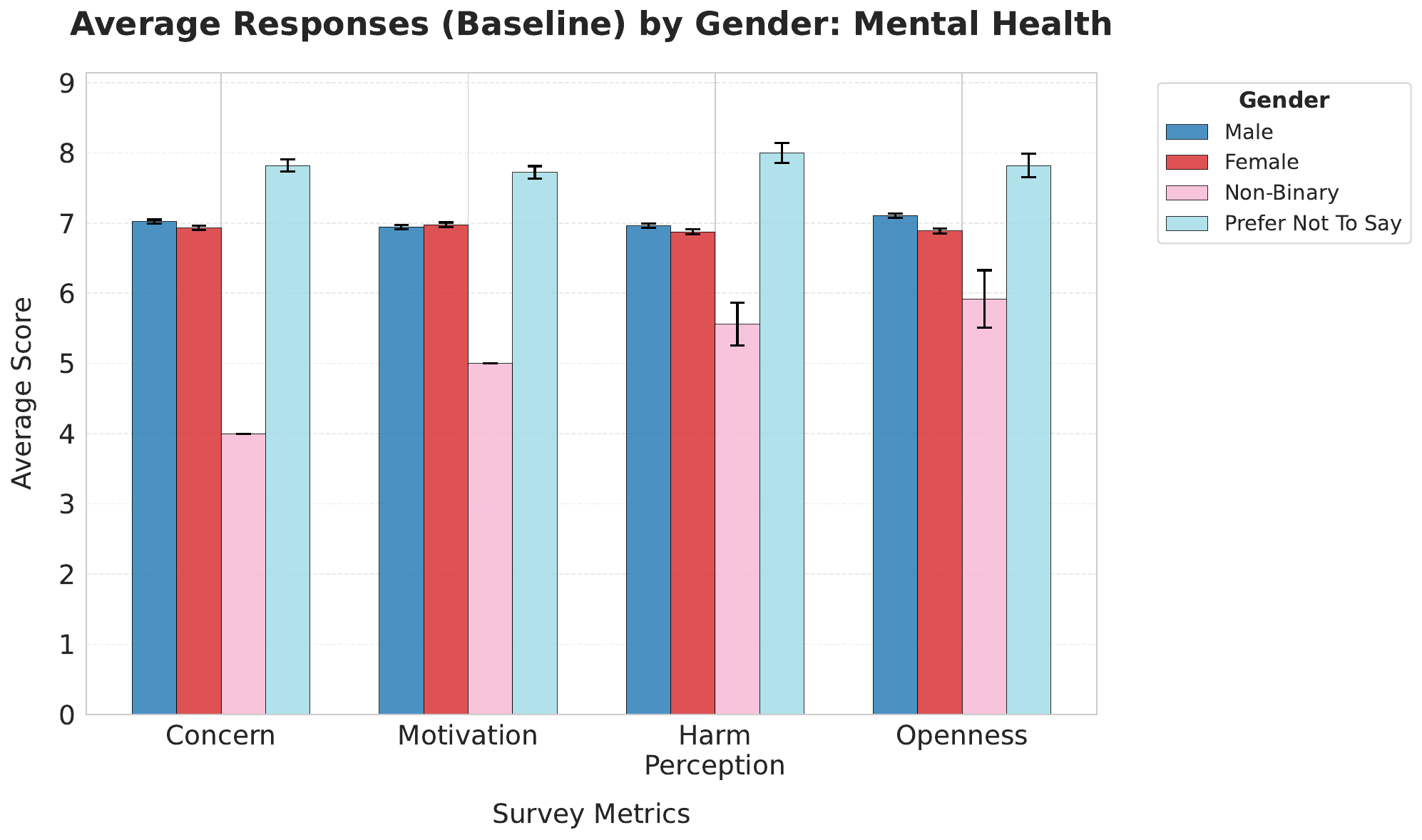}
    \includegraphics[trim={0.3cm 0.9cm 0.2cm 0.2cm},clip,width=0.49\linewidth]{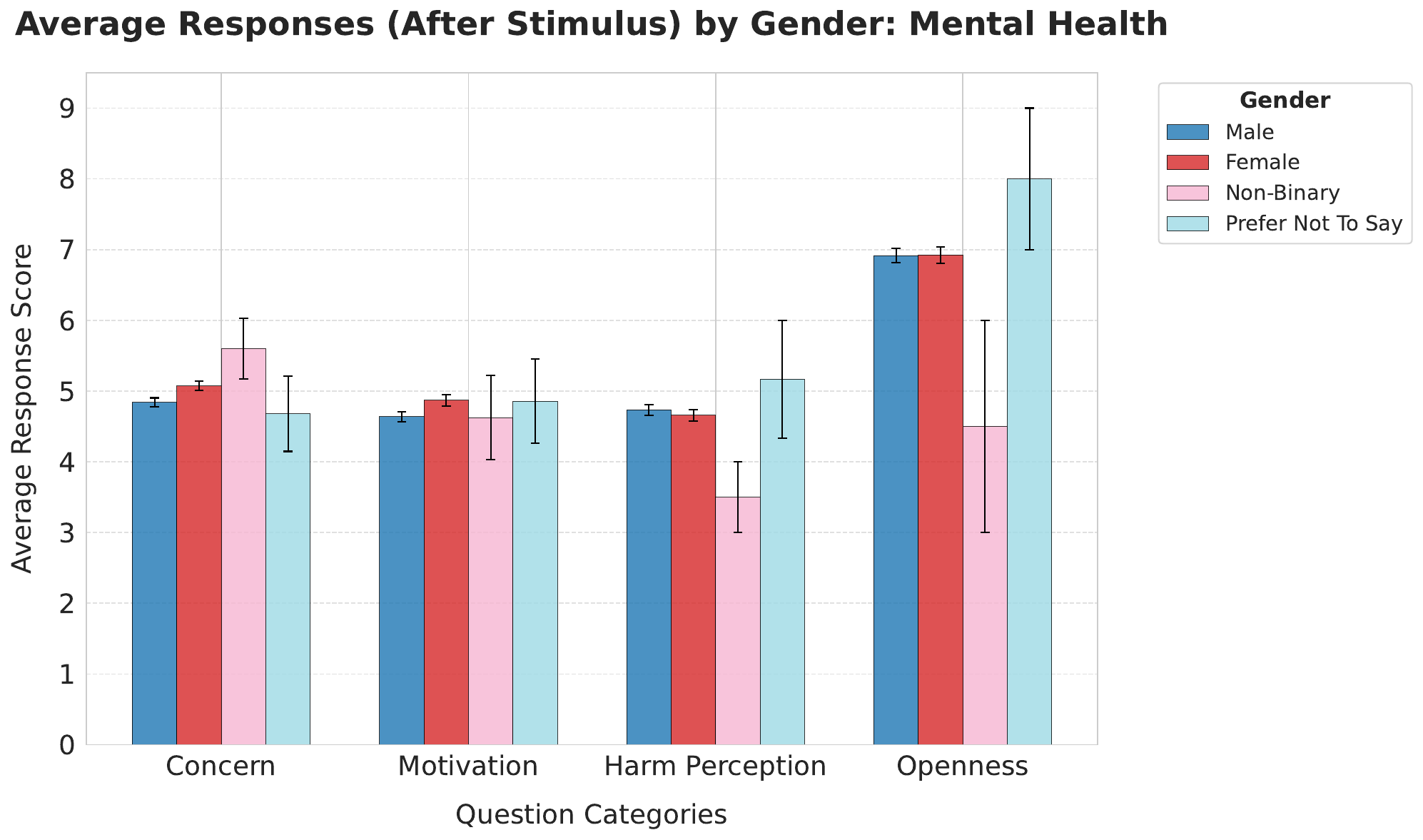}
    \caption{\textbf{Average Responses by Gender:}: While baseline levels of concern, motivation, harm perception, and openness were generally similar between men and women, women exhibited a greater change in responses following exposure to marketing stimuli. Non-binary and other genders show elevated perception of sexual health and HIV risks.}
    \label{fig:gender-diff}
\end{figure}

\begin{figure}
    \centering
    \includegraphics[trim={0.3cm 0.9cm 0.2cm 0.2cm},clip,width=0.49\linewidth]{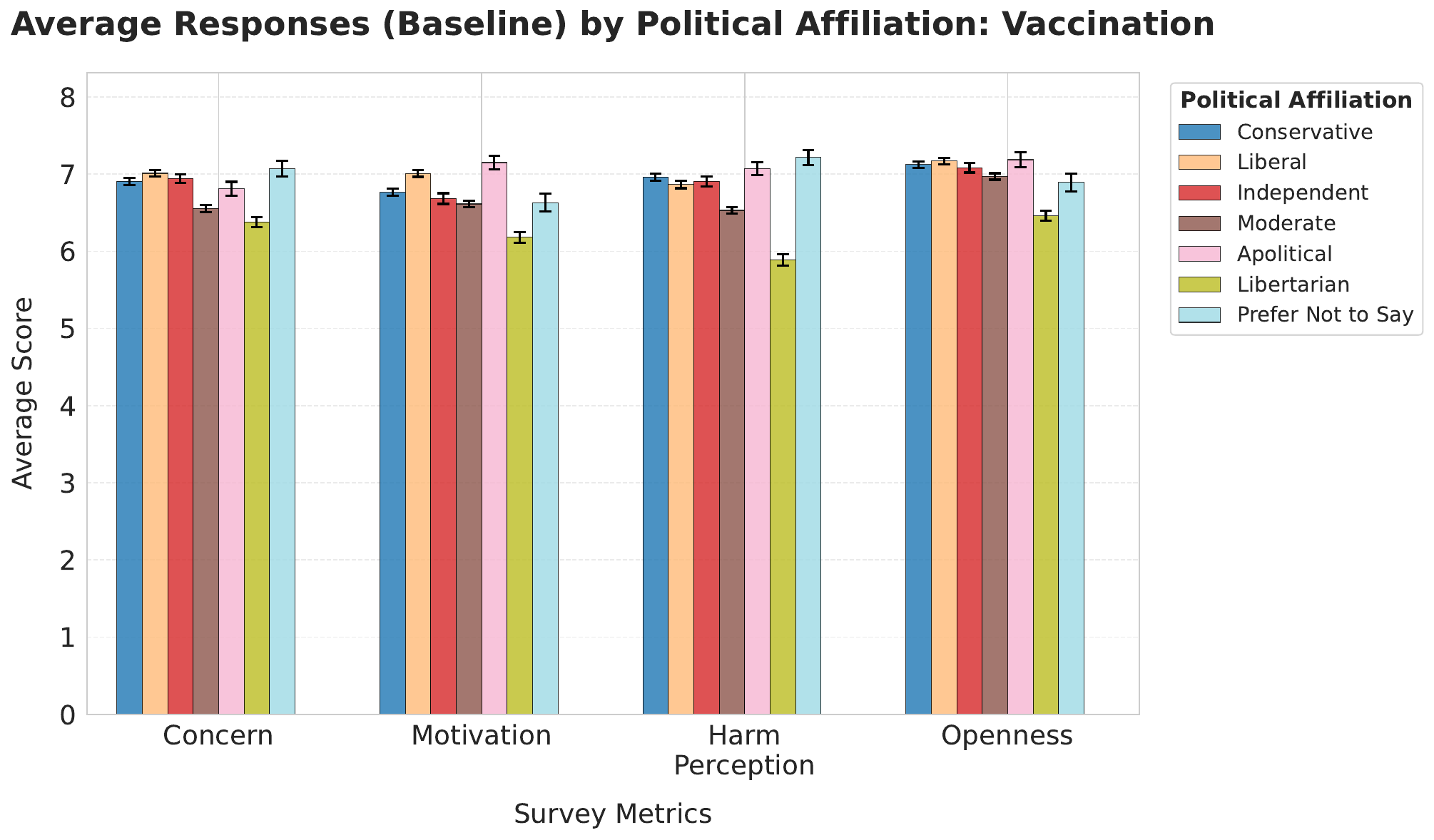}
    \includegraphics[trim={0.3cm 0.9cm 0.2cm 0.2cm},clip,width=0.49\linewidth]{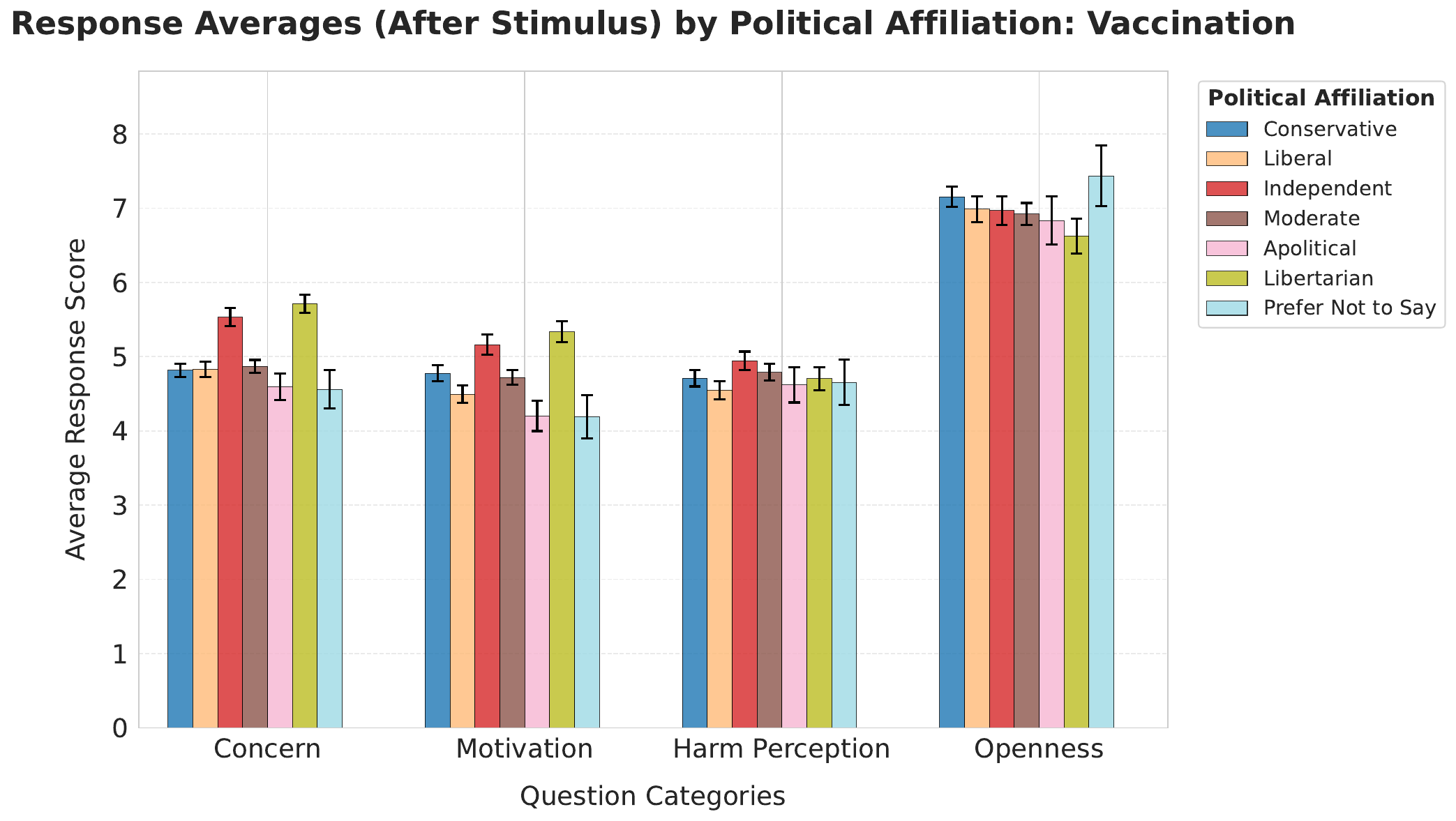}
    \includegraphics[trim={0.3cm 0.9cm 0.2cm 0.2cm},clip,width=0.49\linewidth]{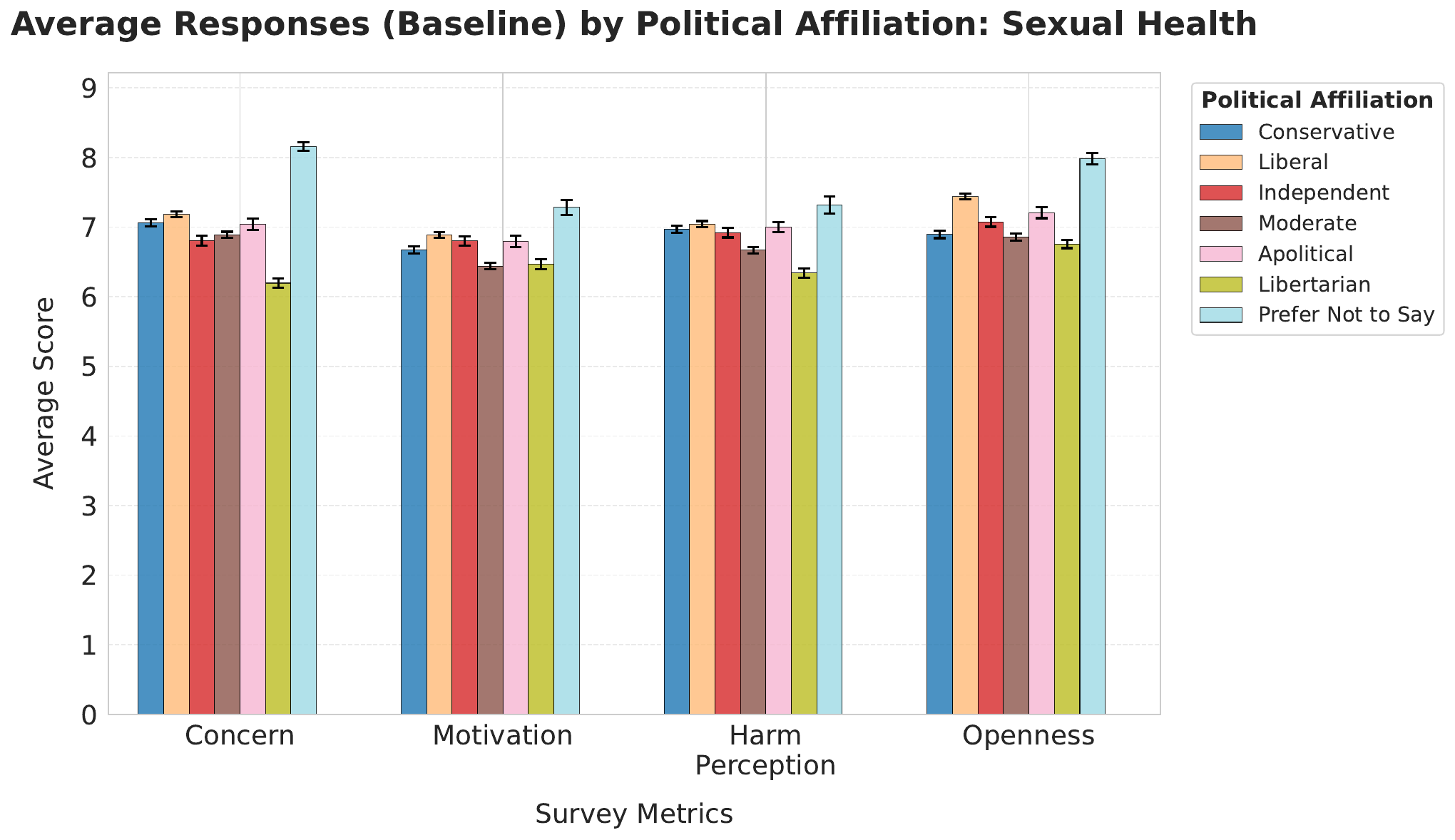}
    \includegraphics[trim={0.3cm 0.9cm 0.2cm 0.2cm},clip,width=0.49\linewidth]{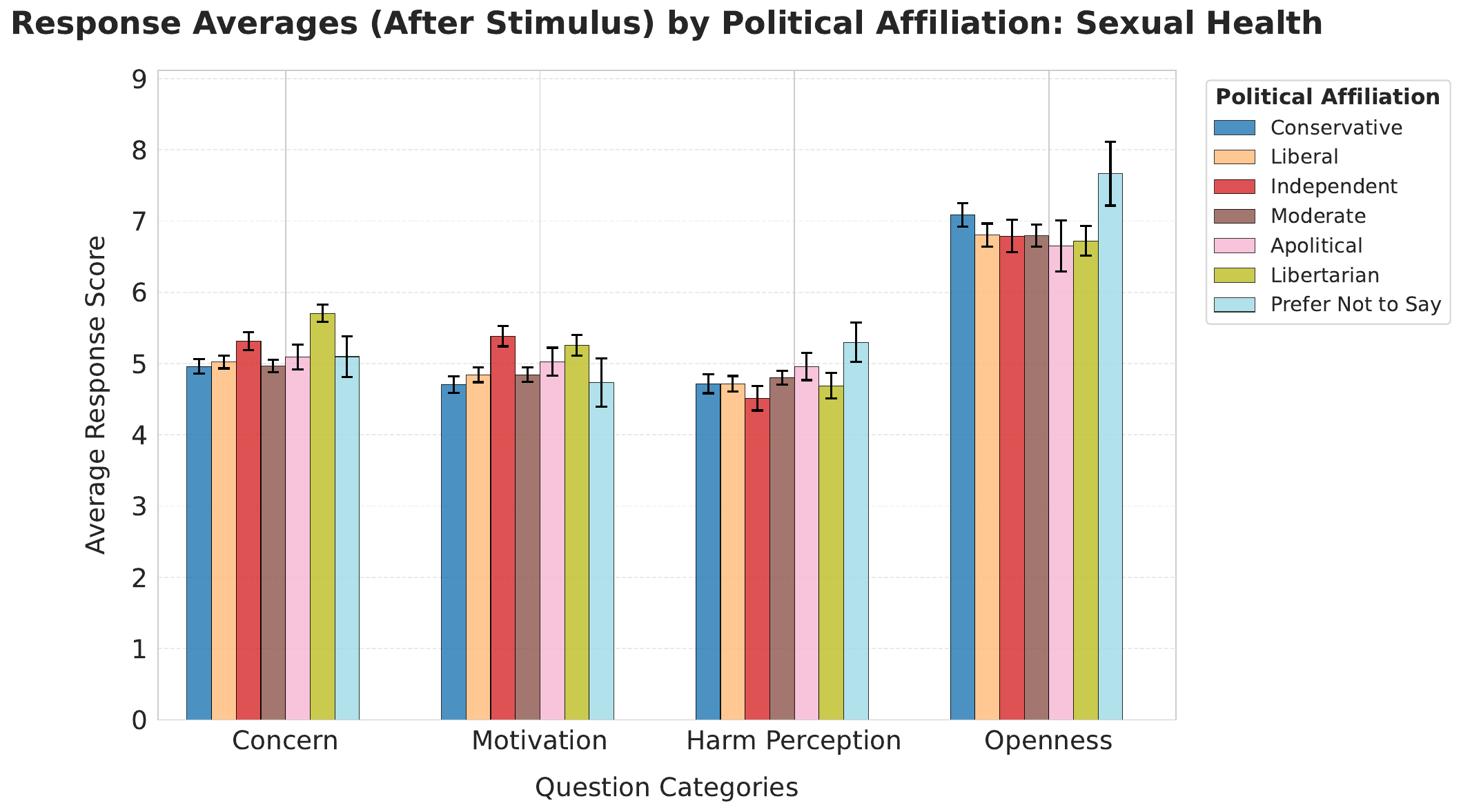}
    \includegraphics[trim={0.3cm 0.9cm 0.2cm 0.2cm},clip,width=0.49\linewidth]{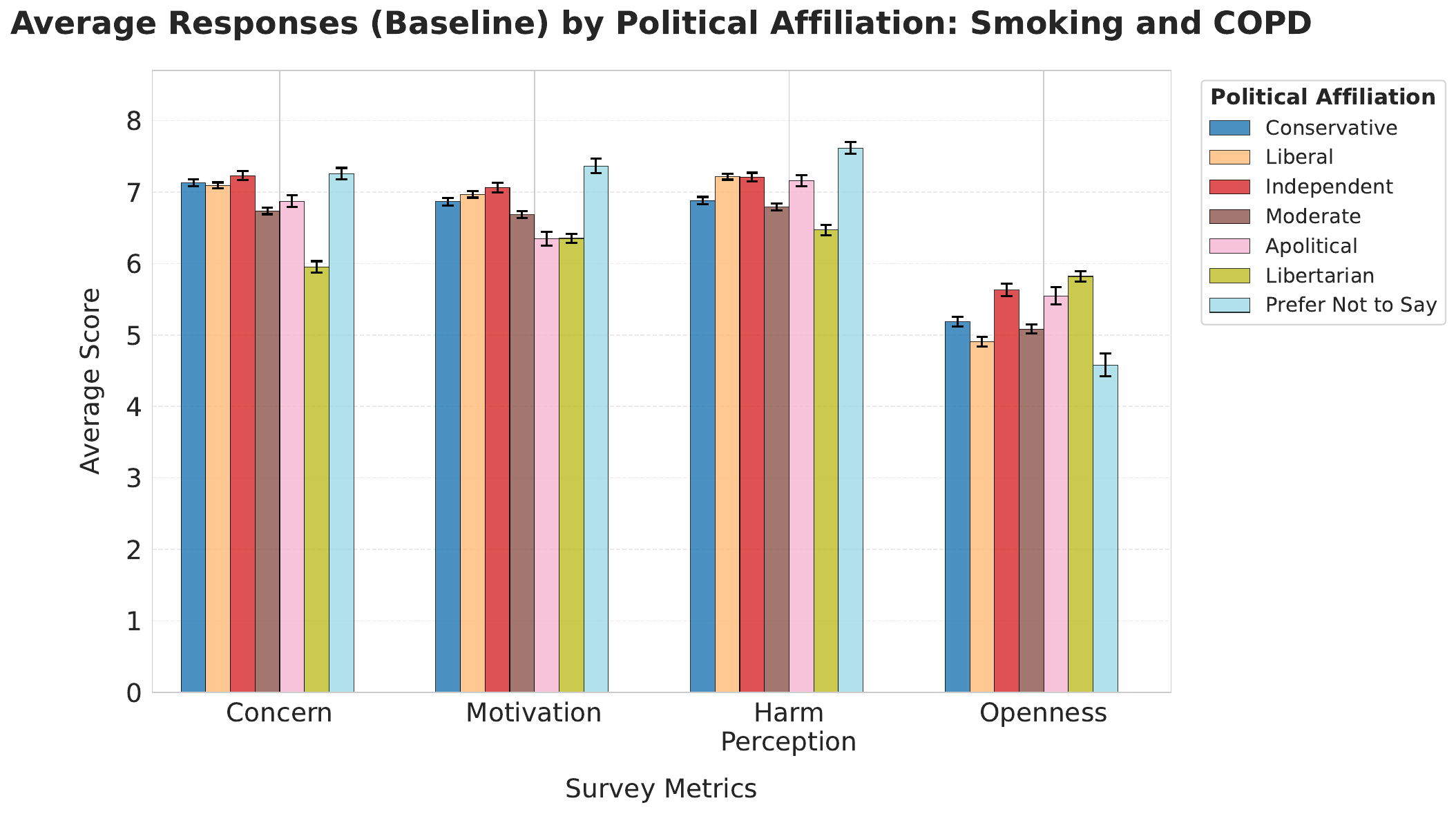}
    \includegraphics[trim={0.3cm 0.9cm 0.2cm 0.2cm},clip,width=0.49\linewidth]{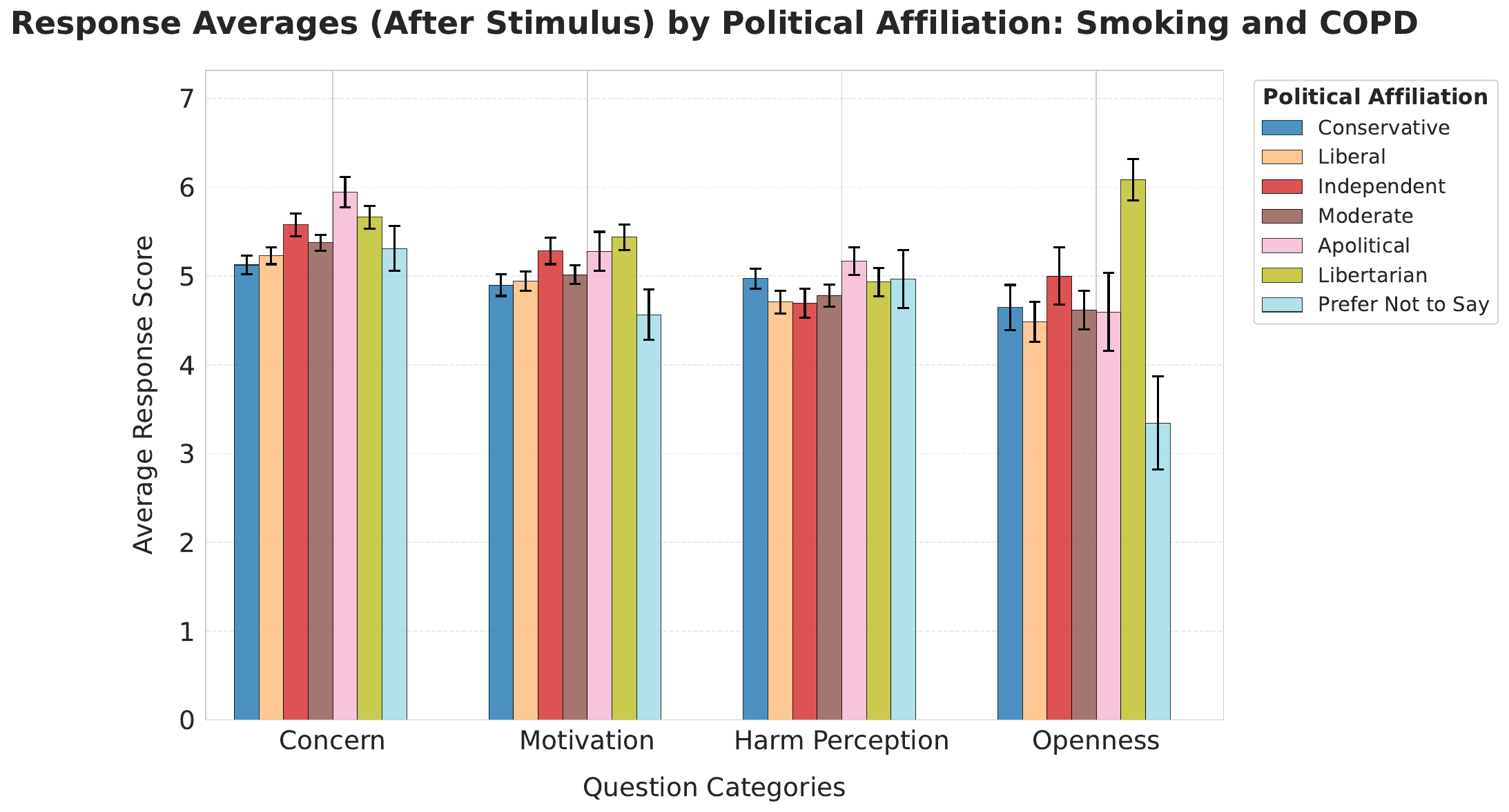}
    \includegraphics[trim={0.3cm 0.9cm 0.2cm 0.2cm},clip,width=0.49\linewidth]{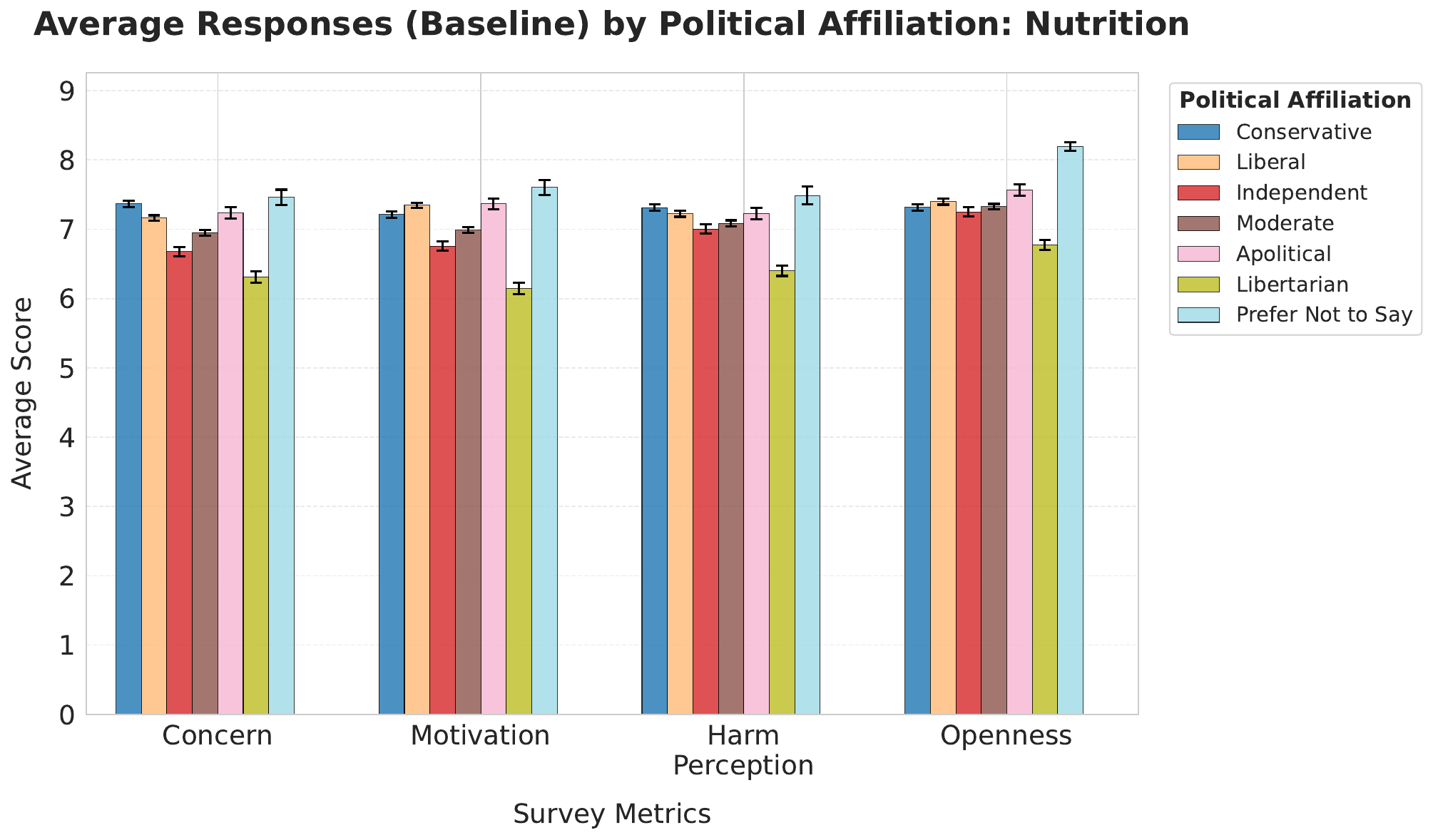}
    \includegraphics[trim={0.3cm 0.9cm 0.2cm 0.2cm},clip,width=0.49\linewidth]{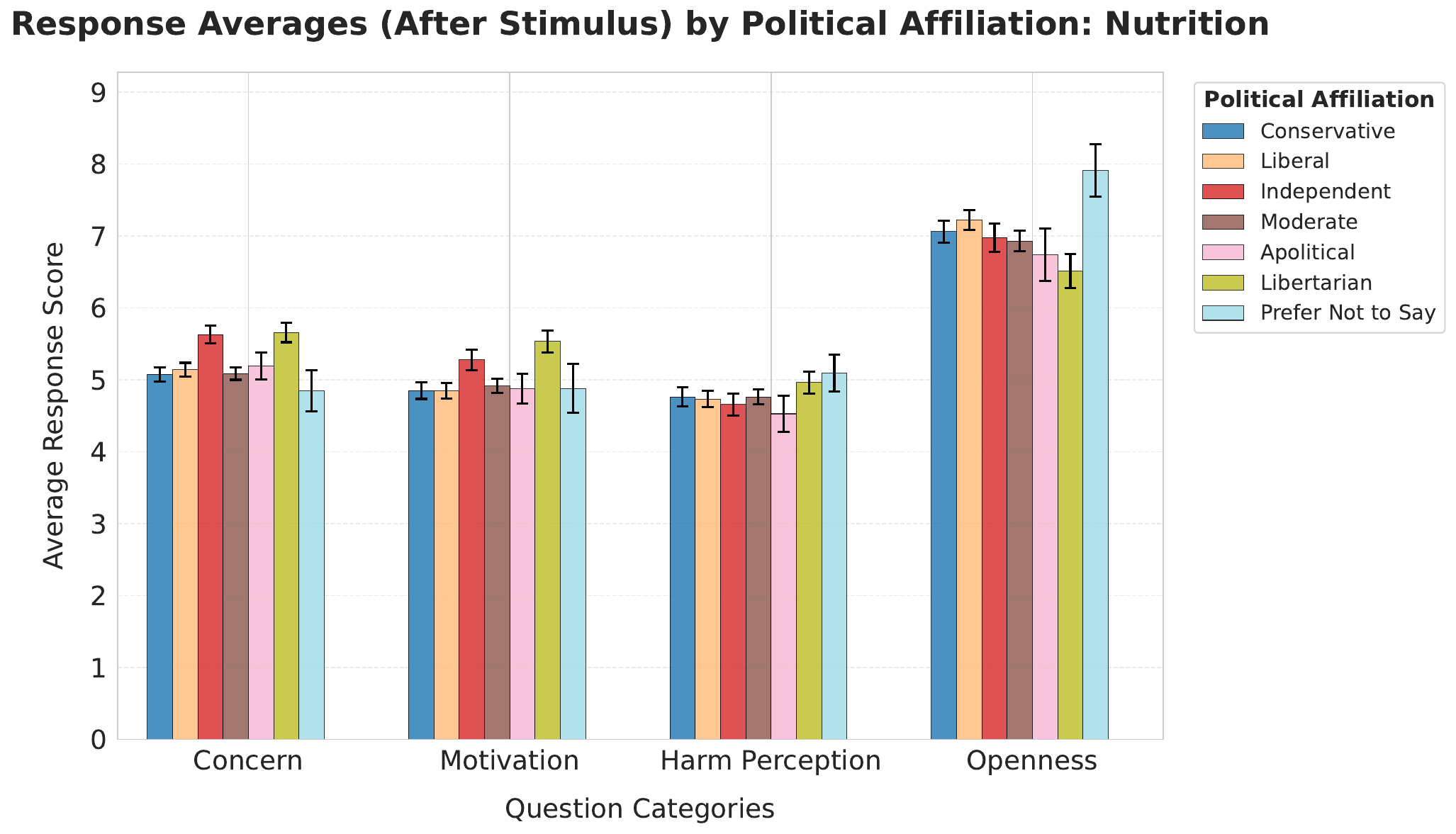}
    \includegraphics[trim={0.3cm 0.9cm 0.2cm 0.2cm},clip,width=0.49\linewidth]{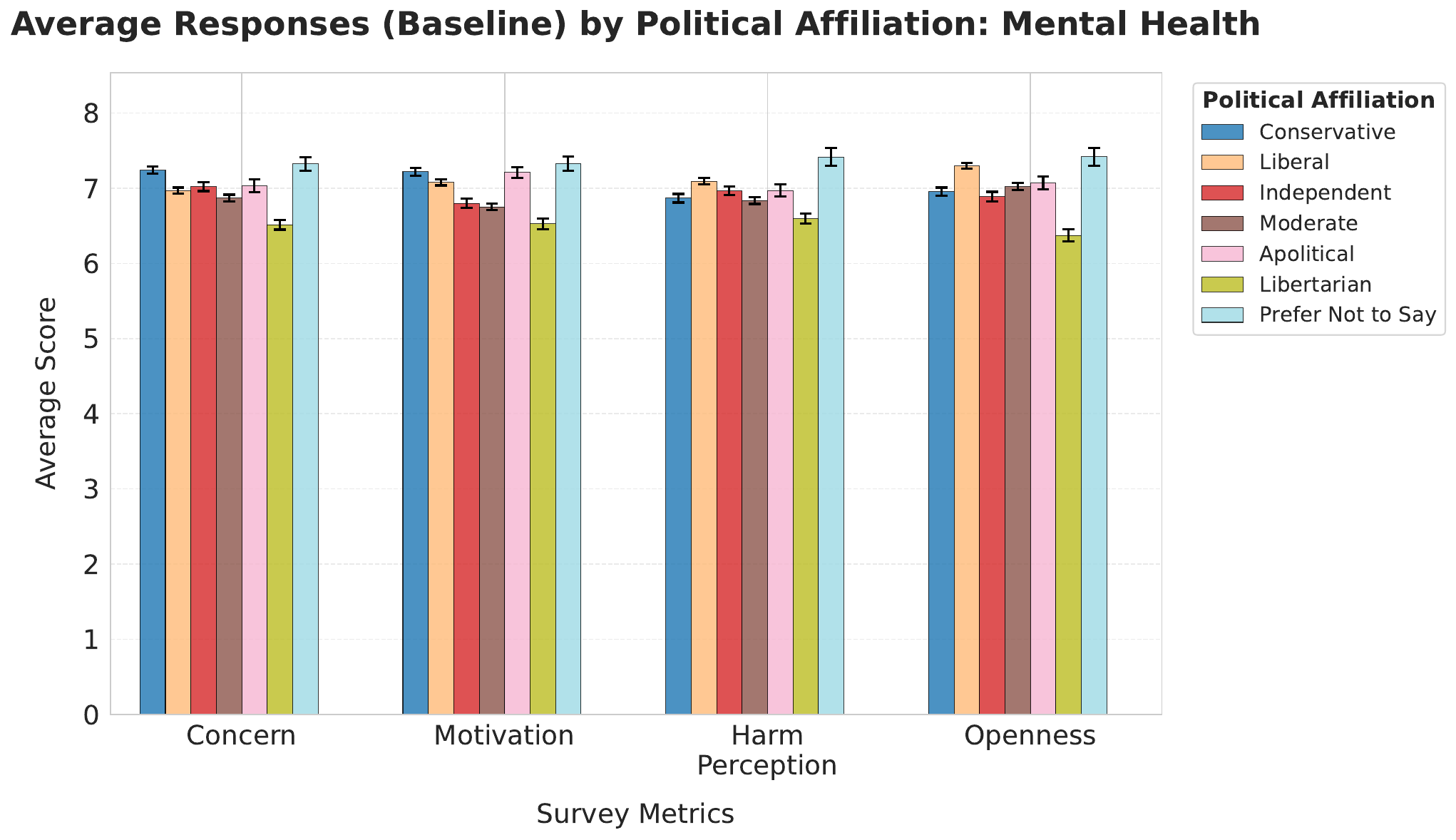}
    \includegraphics[trim={0.3cm 0.9cm 0.2cm 0.2cm},clip,width=0.49\linewidth]{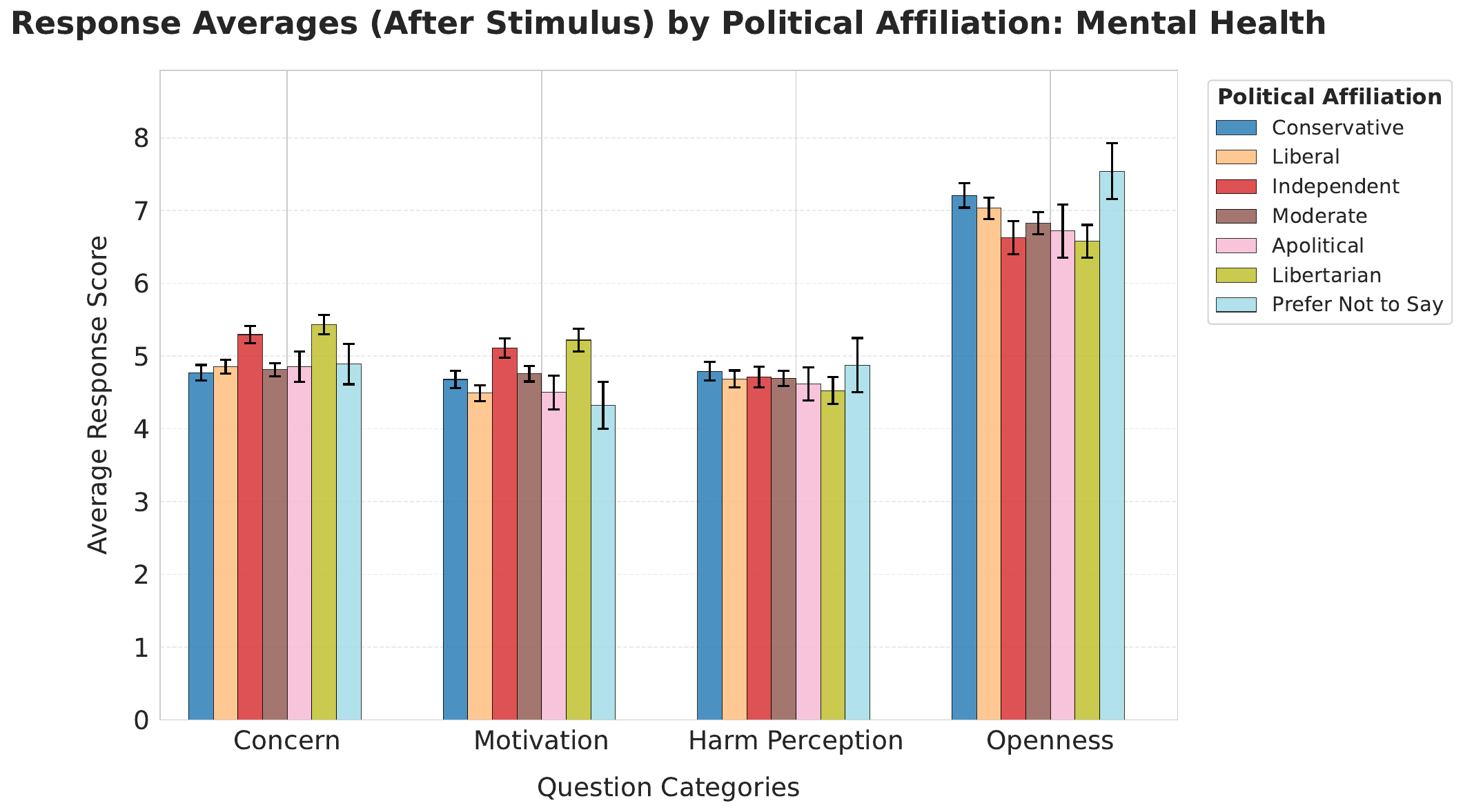}
    \caption{\textbf{Average Responses By Political Affiliation:} We analyze the opinions of different affiliated groups. We observe that Libertarians appear to have the least amount of concern, motivation and harm perception regarding various health topics. At the same time, the campaigns seem to have a bigger effect on them. Yet, they seem to be the most open to smoke. Conservatives seem to be less concerned about the risks of skipping vaccination, mental health and nutrition.}
    \label{fig:polit-resp}
\end{figure}

\begin{figure}
    \centering
    \includegraphics[trim={0.3cm 0.3cm 0.2cm 0.0cm},clip,width=0.49\linewidth]{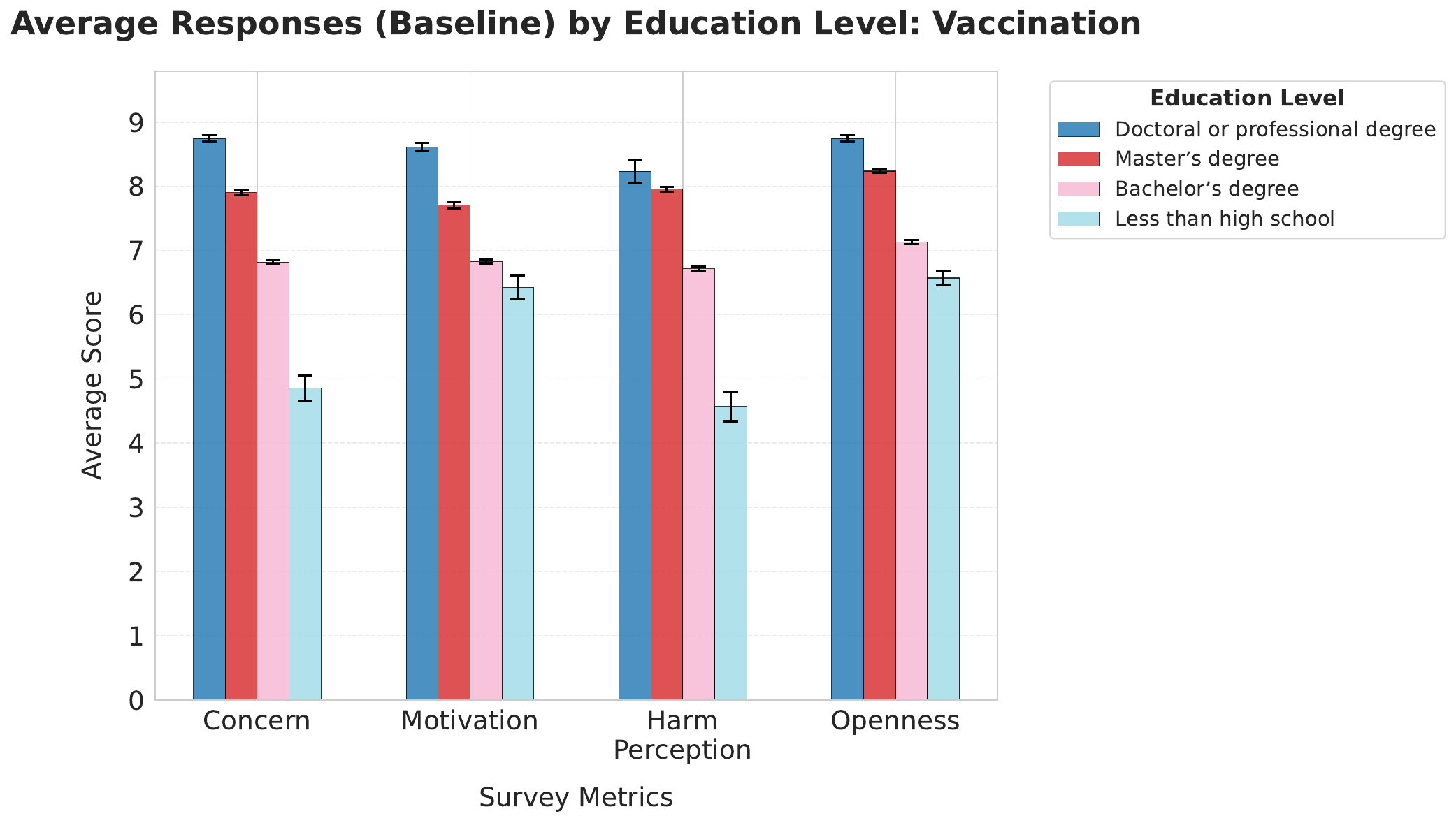}
    \includegraphics[trim={0.3cm 0.3cm 0.2cm 0.0cm},clip,width=0.49\linewidth]{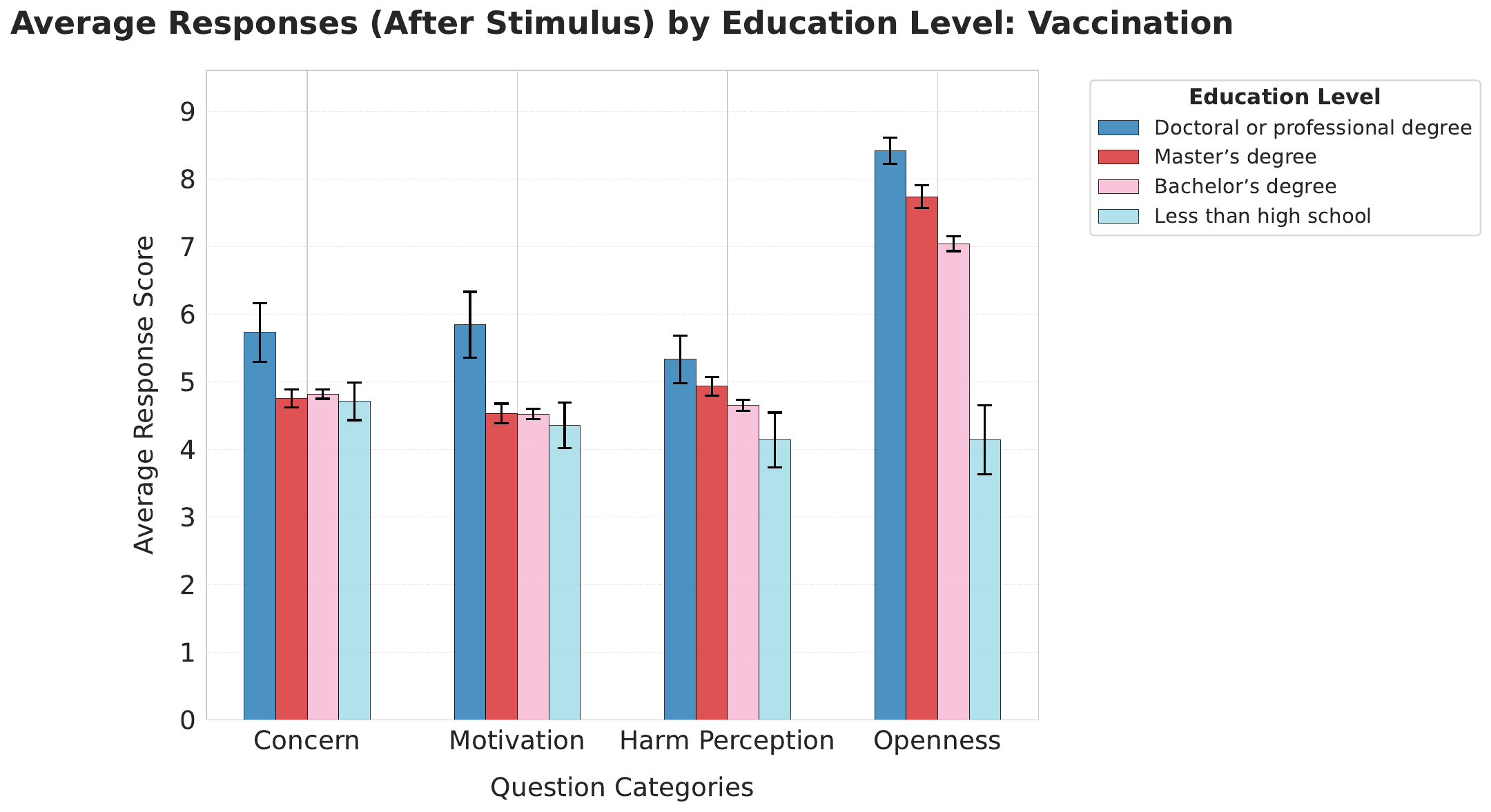}
    \includegraphics[trim={0.3cm 0.3cm 0.2cm 0.0cm},clip,width=0.49\linewidth]{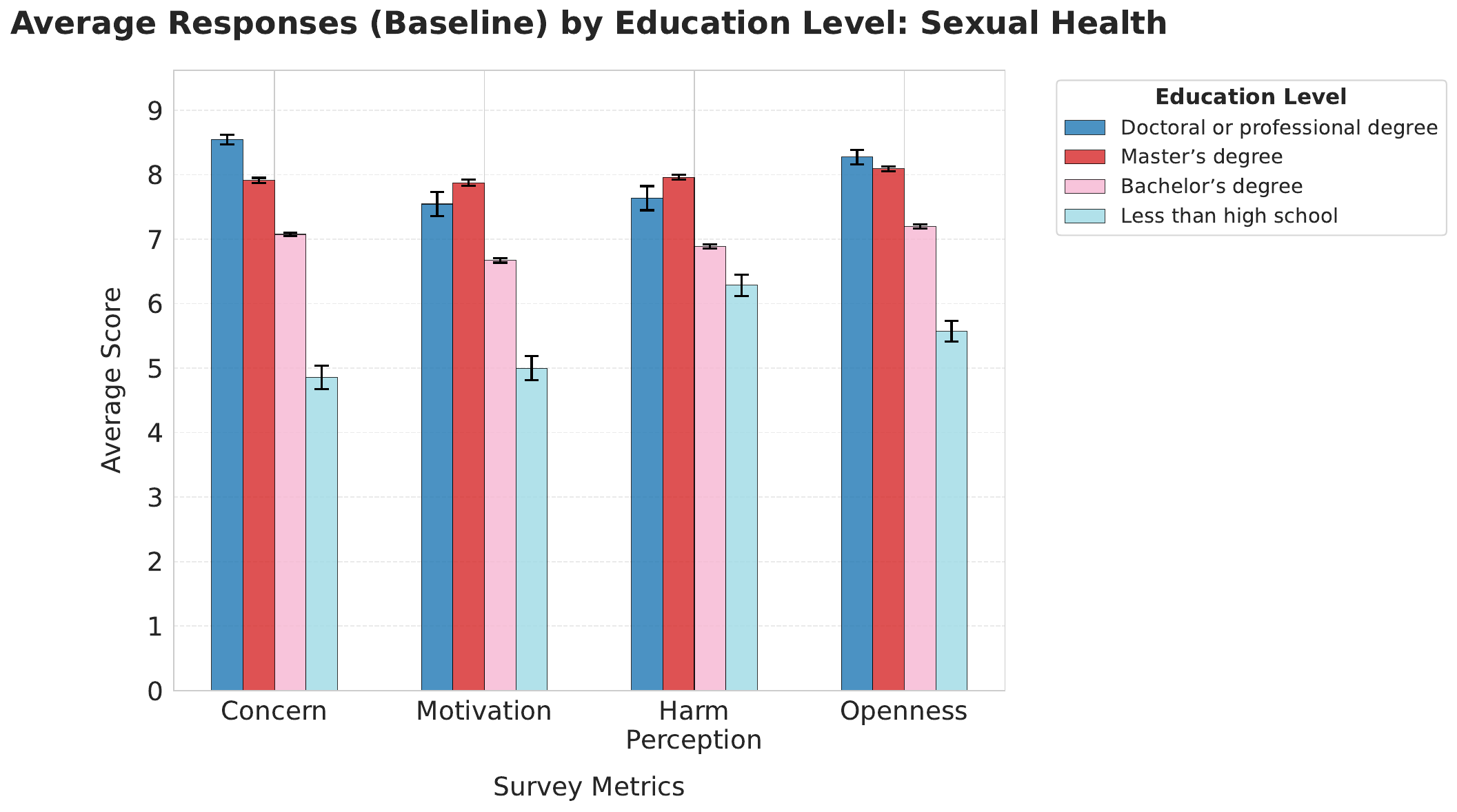}
    \includegraphics[trim={0.3cm 0.3cm 0.2cm 0.0cm},clip,width=0.49\linewidth]{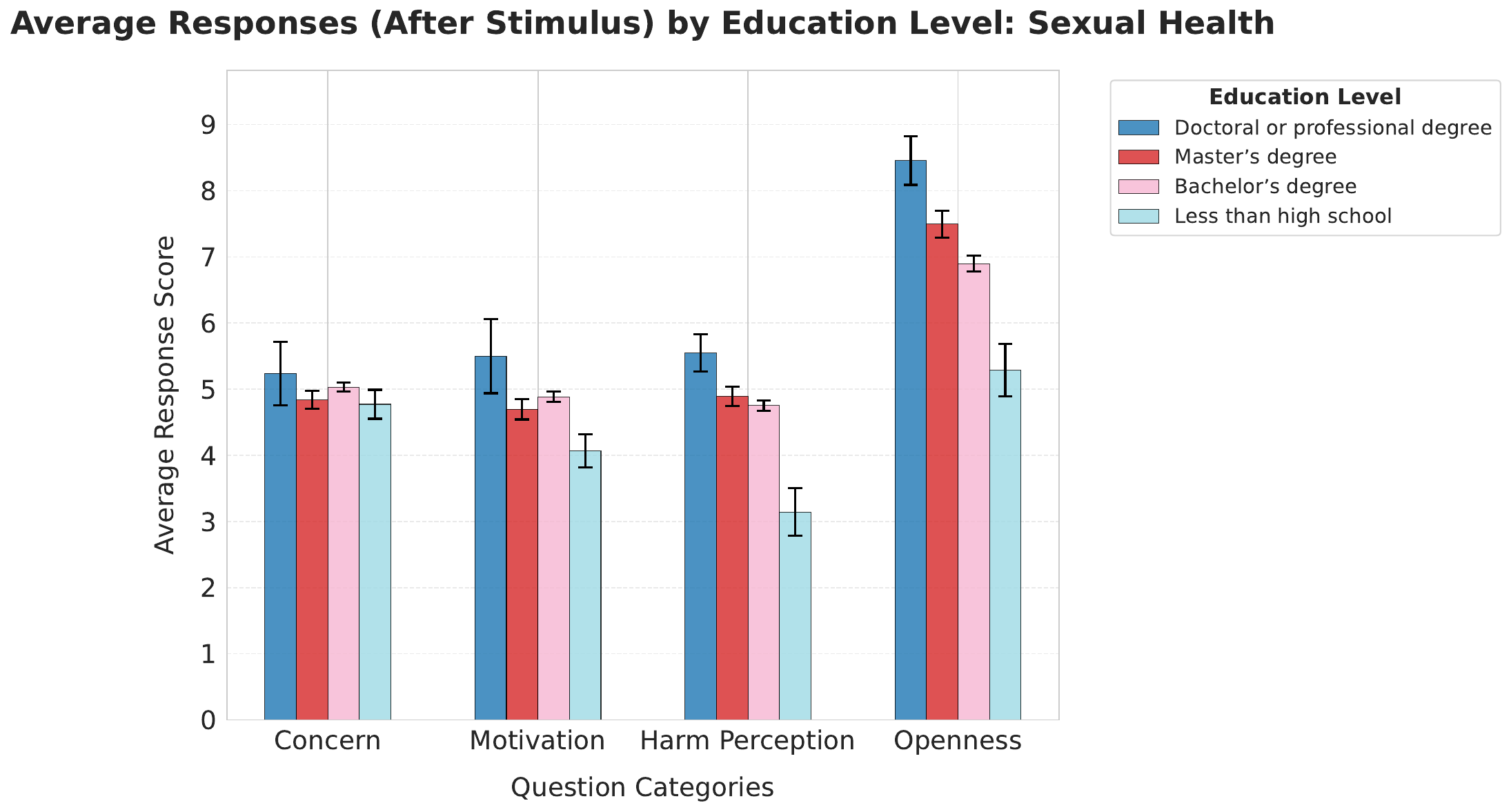}
    \includegraphics[trim={0.3cm 0.3cm 0.2cm 0.0cm},clip,width=0.49\linewidth]{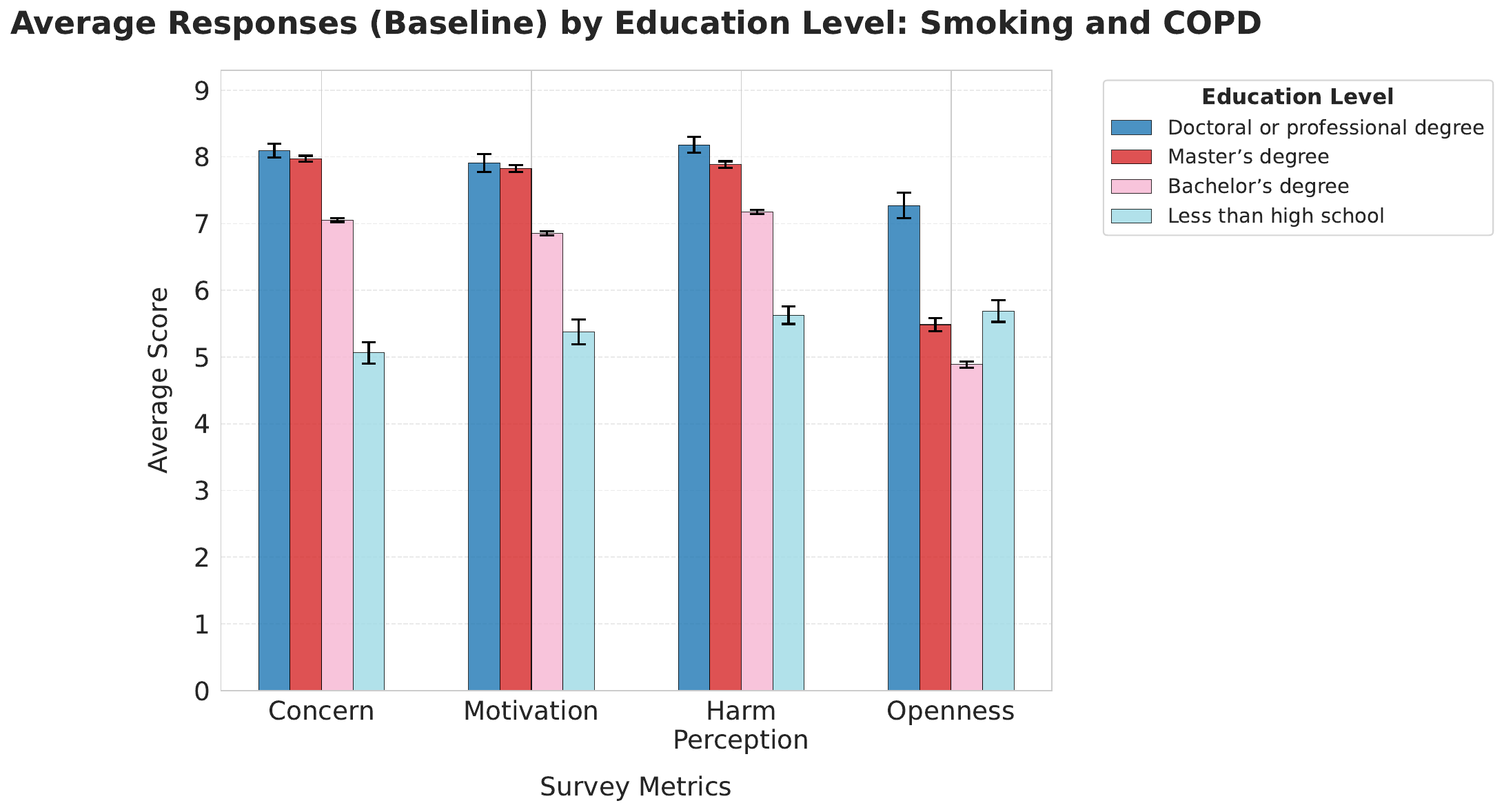}
    \includegraphics[trim={0.3cm 0.3cm 0.2cm 0.0cm},clip,width=0.49\linewidth]{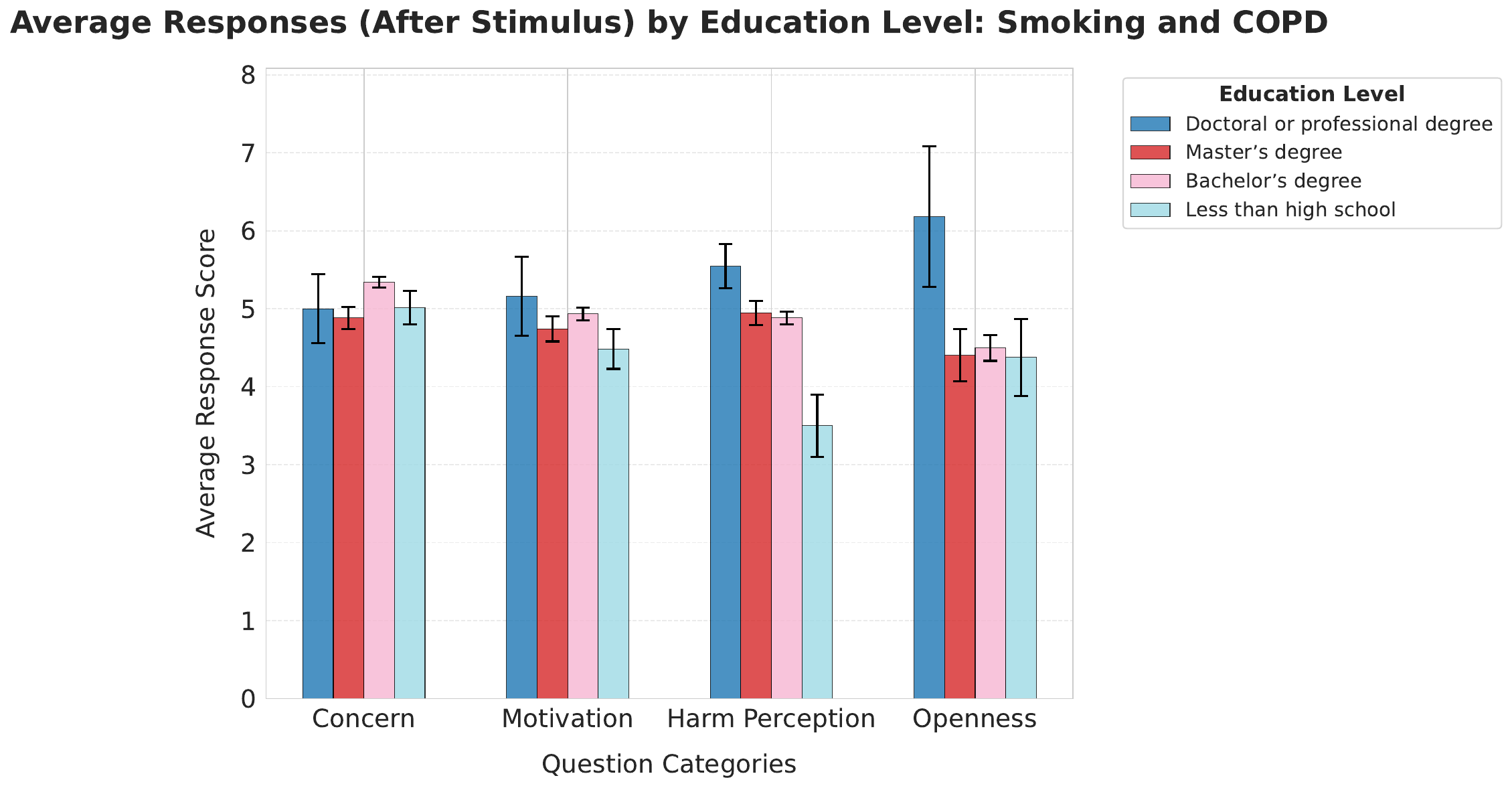}
    \includegraphics[trim={0.3cm 0.3cm 0.2cm 0.0cm},clip,width=0.49\linewidth]{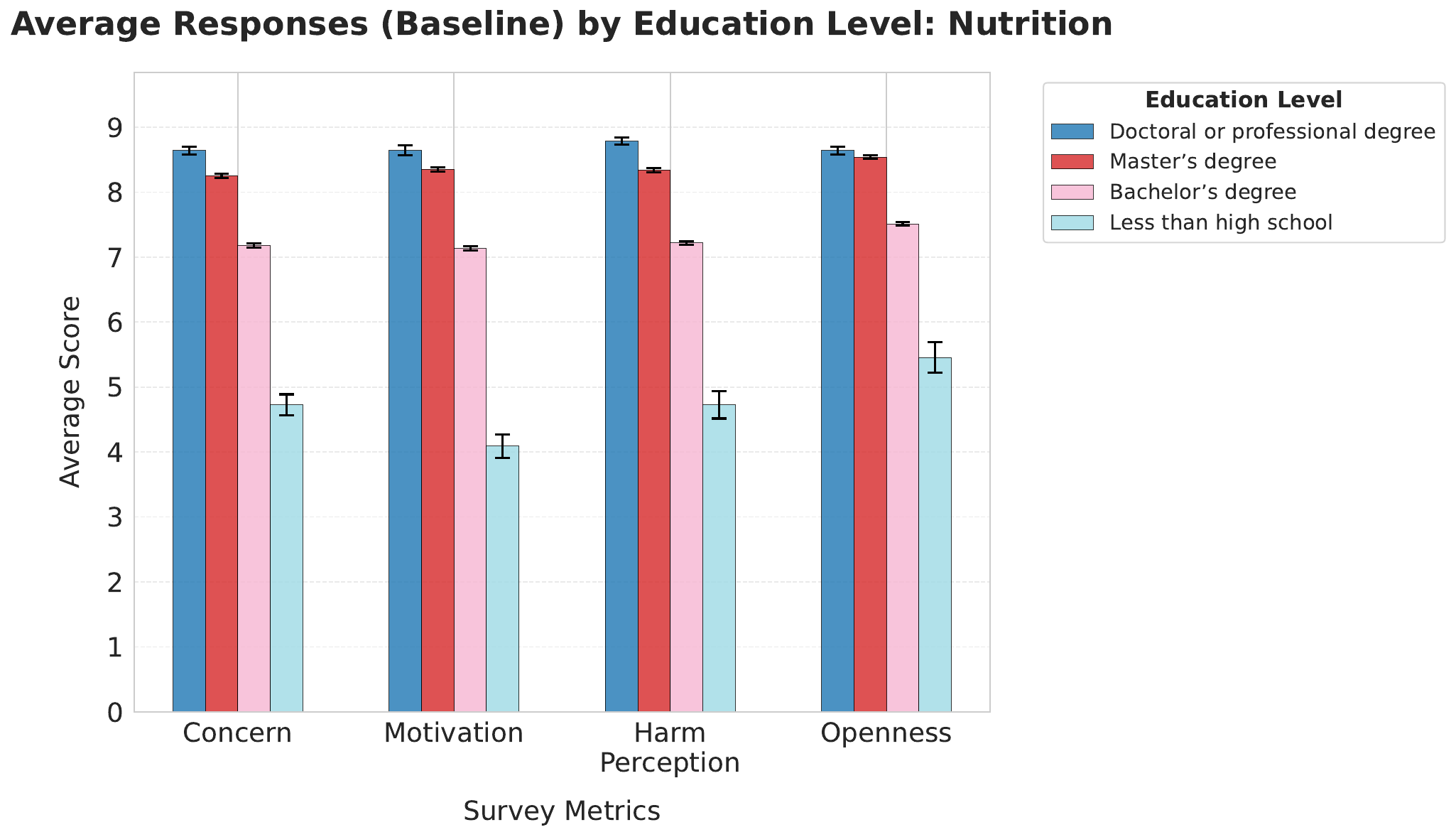}
    \includegraphics[trim={0.3cm 0.3cm 0.2cm 0.0cm},clip,width=0.49\linewidth]{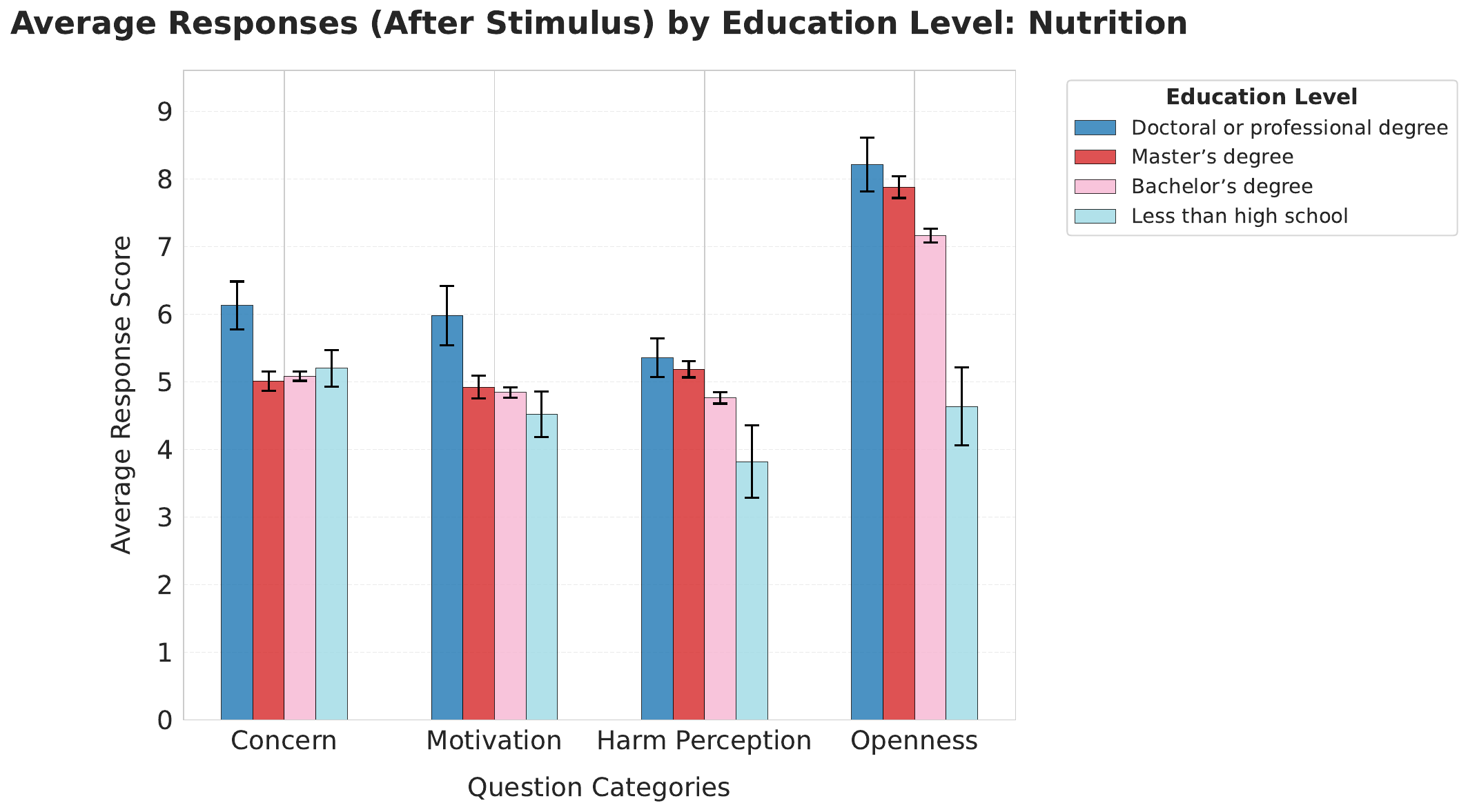}
    \includegraphics[trim={0.3cm 0.3cm 0.2cm 0.0cm},clip,width=0.49\linewidth]{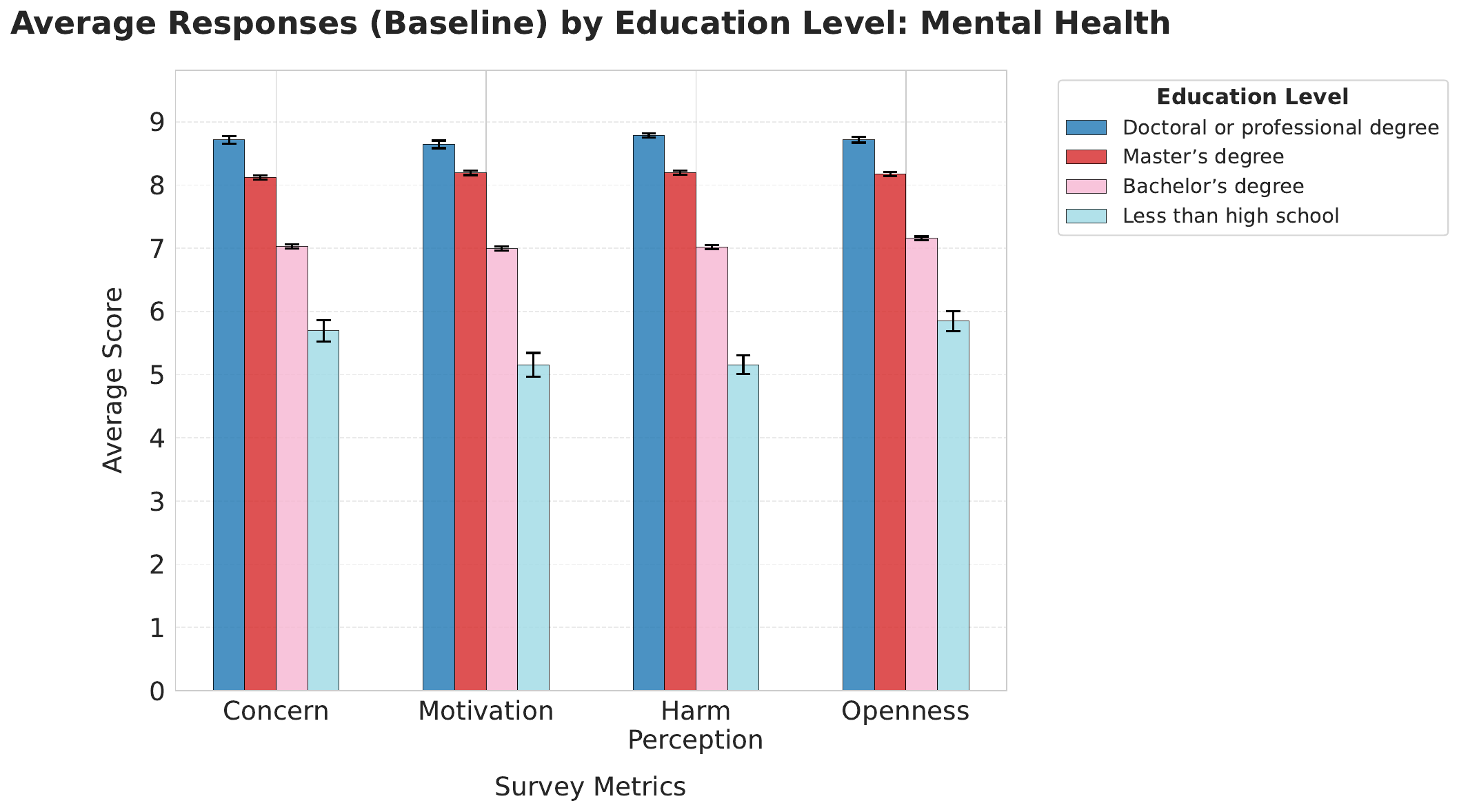}
    \includegraphics[trim={0.3cm 0.3cm 0.2cm 0.0cm},clip,width=0.49\linewidth]{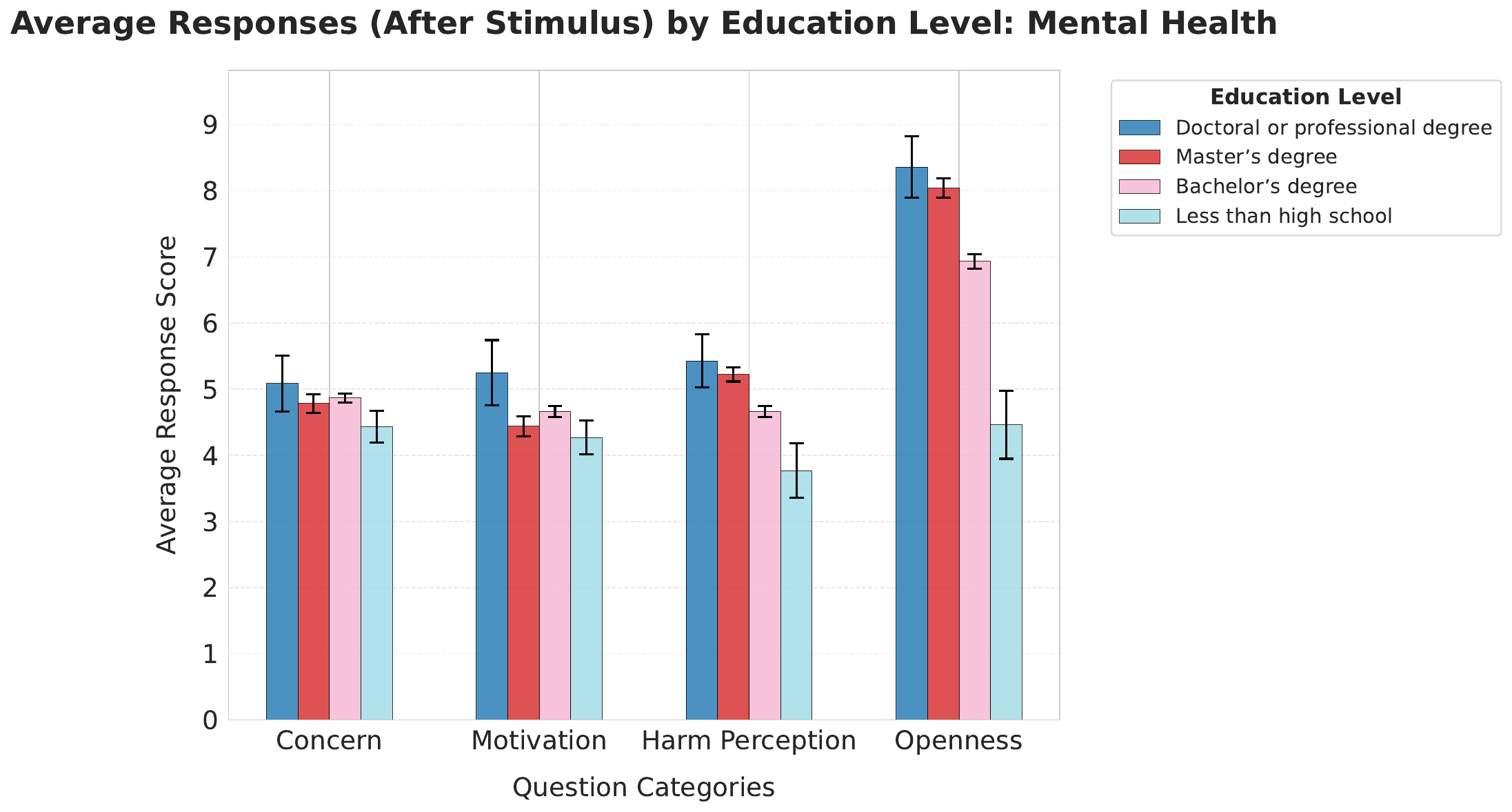}

    \caption{\textbf{Average Responses By Education Levels:} Concern, Motivation, Harm Perception and Openness of individuals with different education levels prior and after viewing any marketing content.  People with a higher degree such as masters and doctoral tend to be more concerned about health concerns, and more motivated or open to practice healthy behaviors. Interestingly, individuals with less than high school are less open to smoking than other educational groups, but less concerned about the health risks related to it and less motivated to not smoke or abuse substances.}
    \label{fig:education-resp}
\end{figure}

We analyze the distributional differences in responses among different groups including gender (Fig.\ref{fig:gender-diff}), political affiliation (Fig.\ref{fig:polit-resp}),  and education levels (Fig.\ref{fig:education-resp}). We make some key observations:

Women and men generally exhibit similar opinions prior to interacting with any campaigns. Post-intervention, women report slighter higher levels of concern across most categories. Notably, sexual health interventions appeared to elicit the strongest responses among non-binary participants. 

Our dataset further indicates significant variations in how different political affiliations perceive and respond to health-related topics. For instance, while conservatives exhibit high initial concern regarding mental health and nutrition, independents demonstrate increased concern scores following the stimulus. A unique pattern emerges among libertarians, who show the highest openness to smoking, yet the lowest propensity for vaccination or dietary care. Conversely, both libertarians and independents demonstrate the highest levels of concern and motivation regarding timely vaccination. 

Analysis of health-related behaviors and perceptions by education level reveal that individuals with higher educational attainment consistently report elevated concern for various health topics. Notably, for all categories except nutrition, the disparity in concern across education levels diminishes post-intervention, suggesting potential for more effective health interventions among lower education groups. 

\begin{figure}
    \centering
    \includegraphics[trim={0.3cm 0.85cm 0.2cm 0.0cm},clip,width=0.40\linewidth]{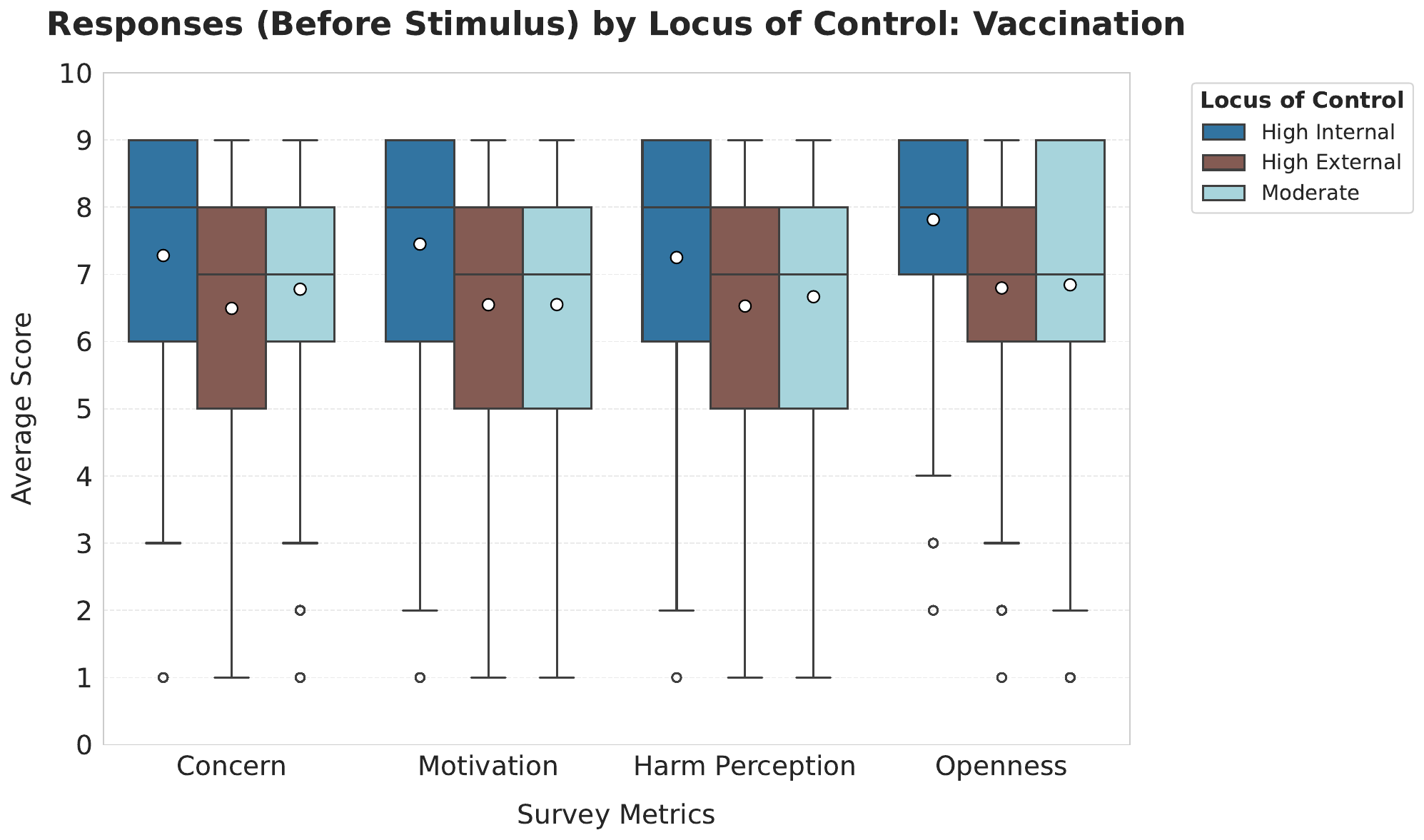}
    \includegraphics[trim={0.3cm 0.85cm 0.2cm 0.0cm},clip,width=0.40\linewidth]{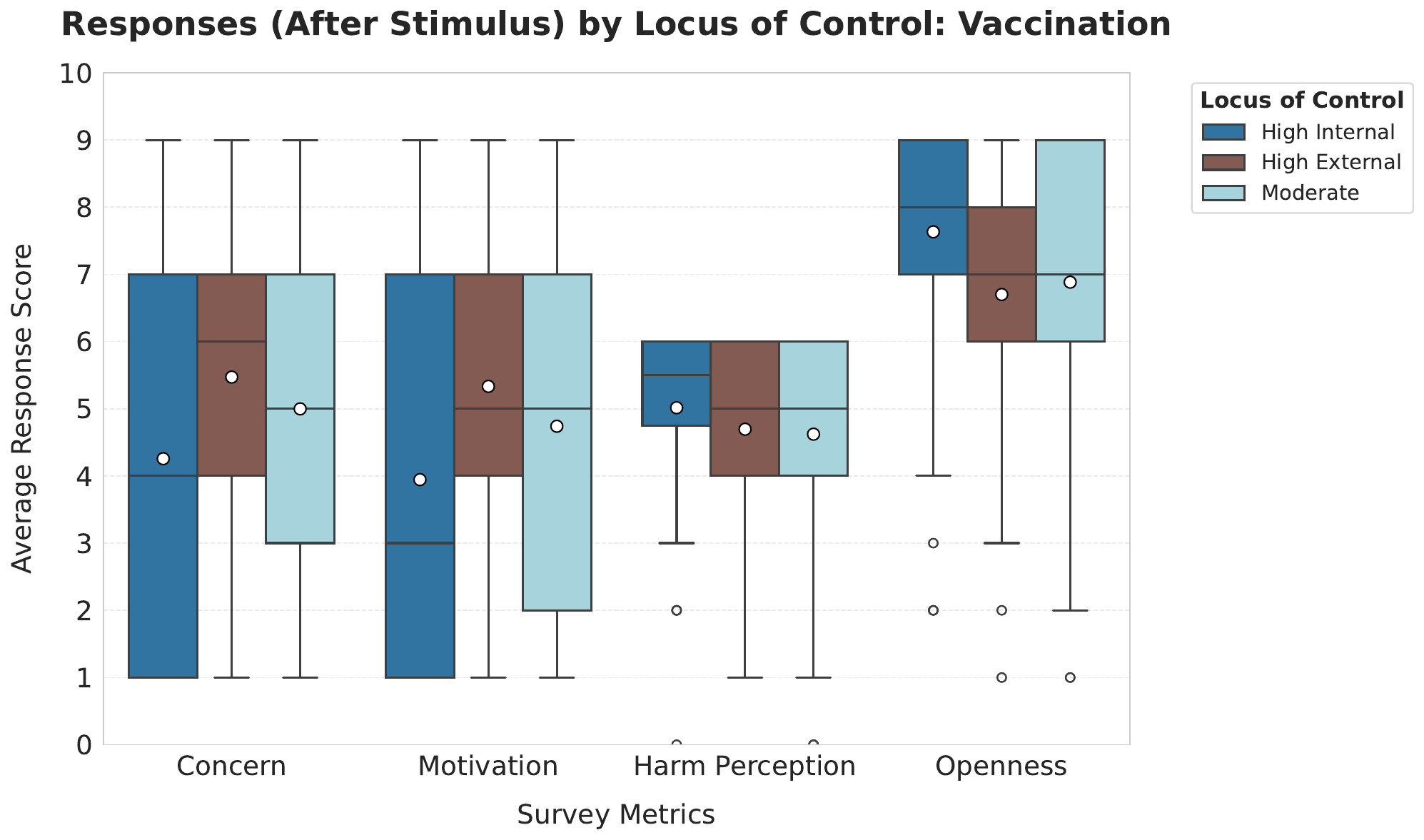}
    \includegraphics[trim={0.3cm 0.85cm 0.2cm 0.0cm},clip,width=0.40\linewidth]{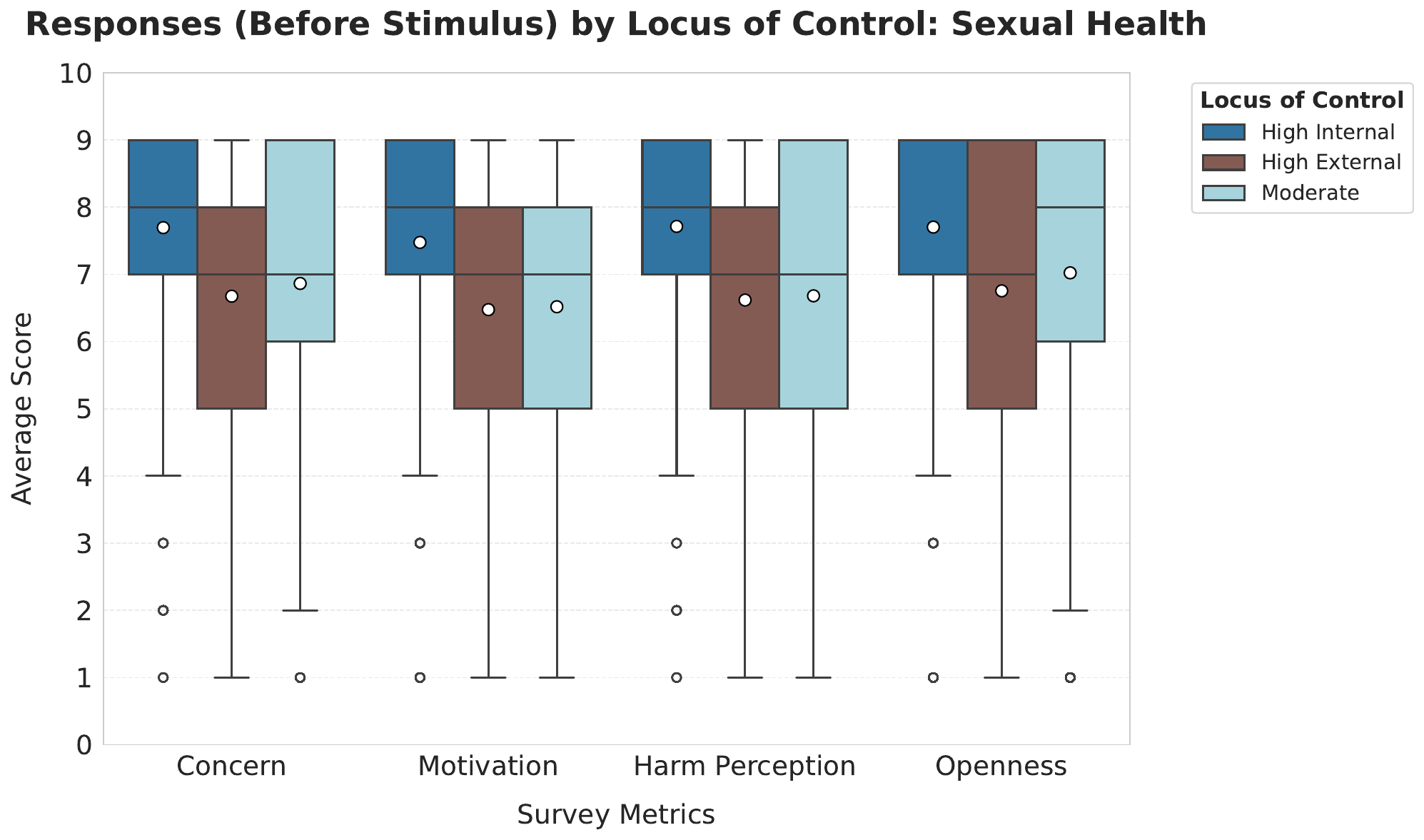}
    \includegraphics[trim={0.3cm 0.85cm 0.2cm 0.0cm},clip,width=0.40\linewidth]{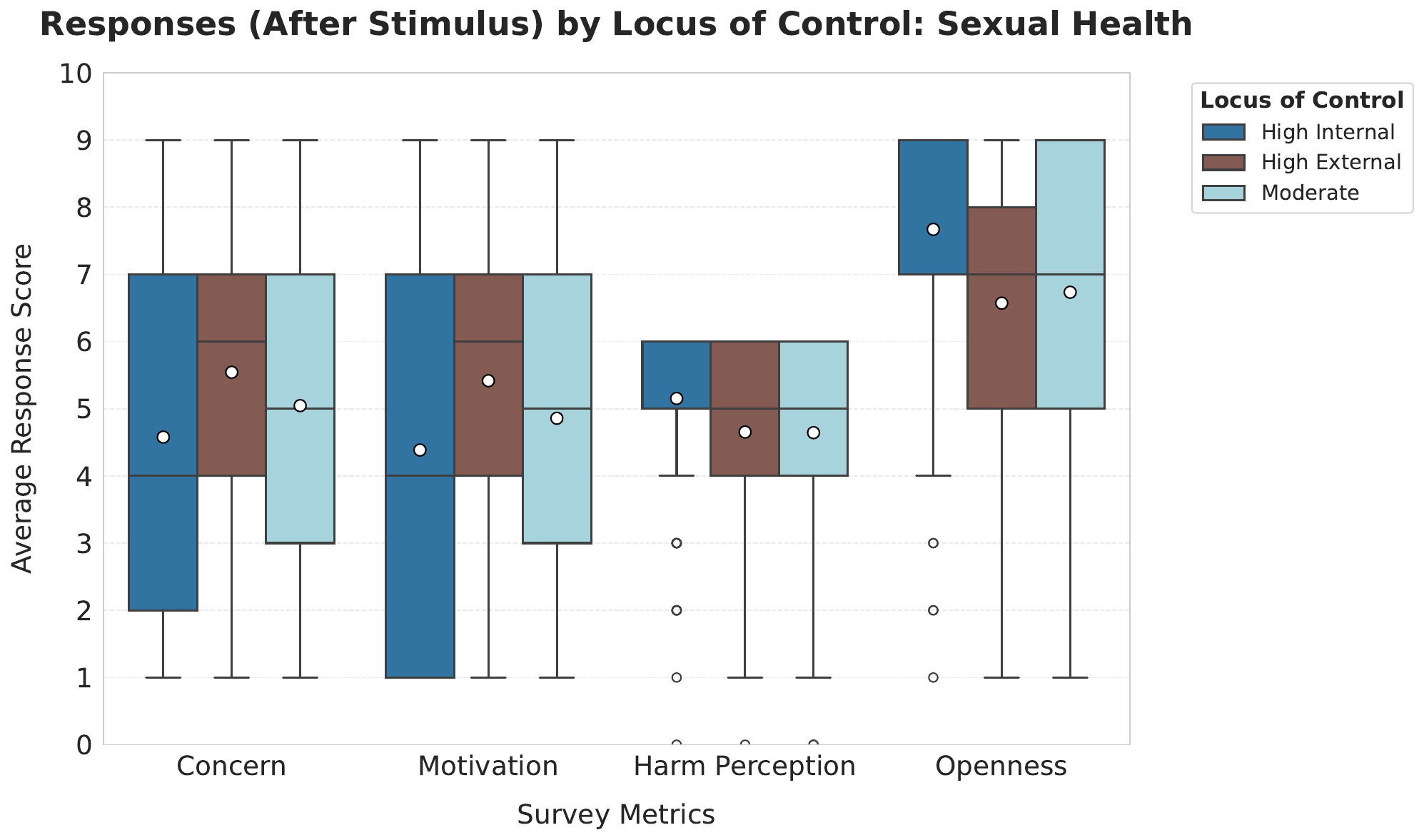}
    \includegraphics[trim={0.3cm 0.85cm 0.2cm 0.0cm},clip,width=0.40\linewidth]{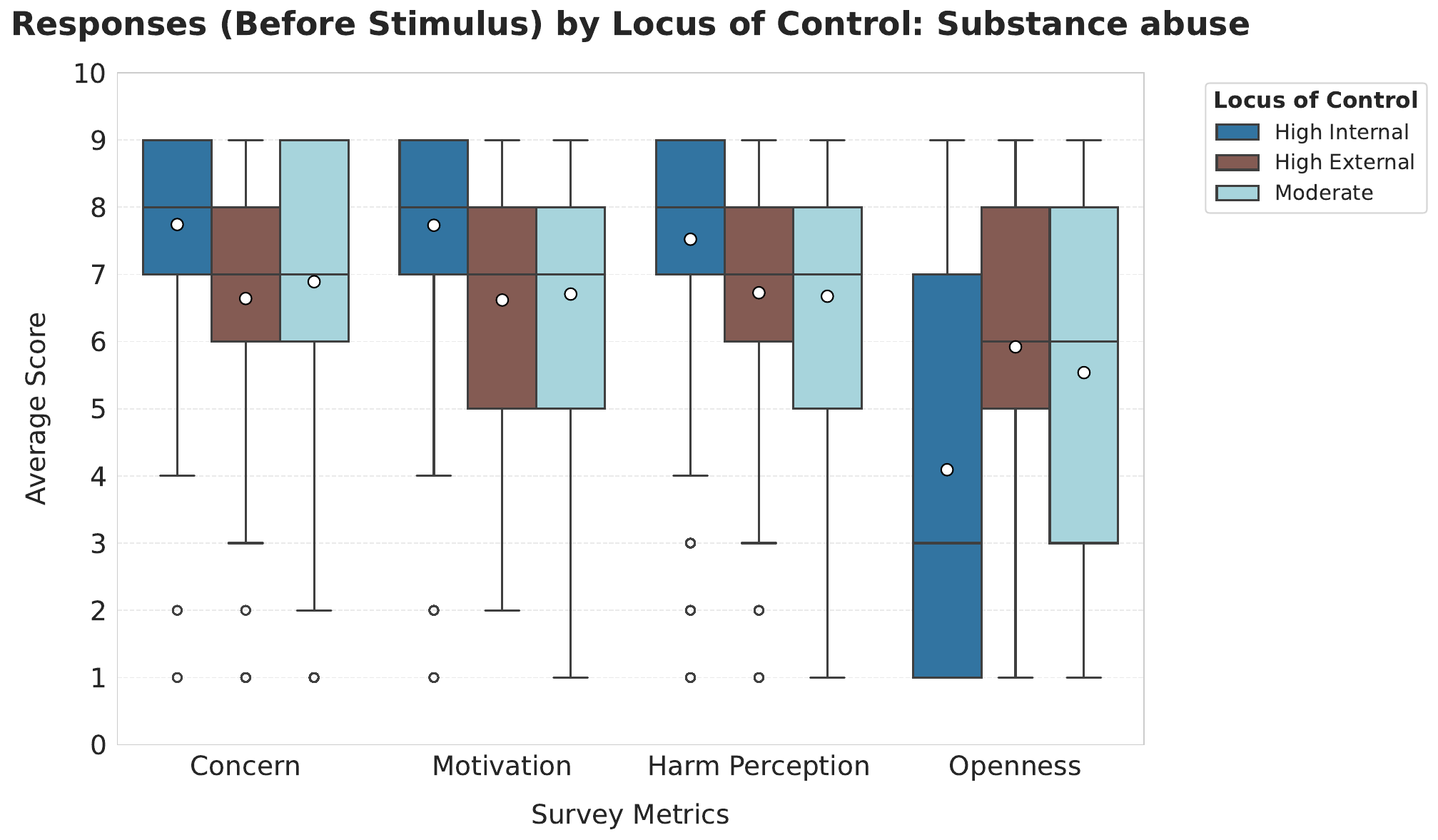}
    \includegraphics[trim={0.3cm 0.85cm 0.2cm 0.0cm},clip,width=0.40\linewidth]{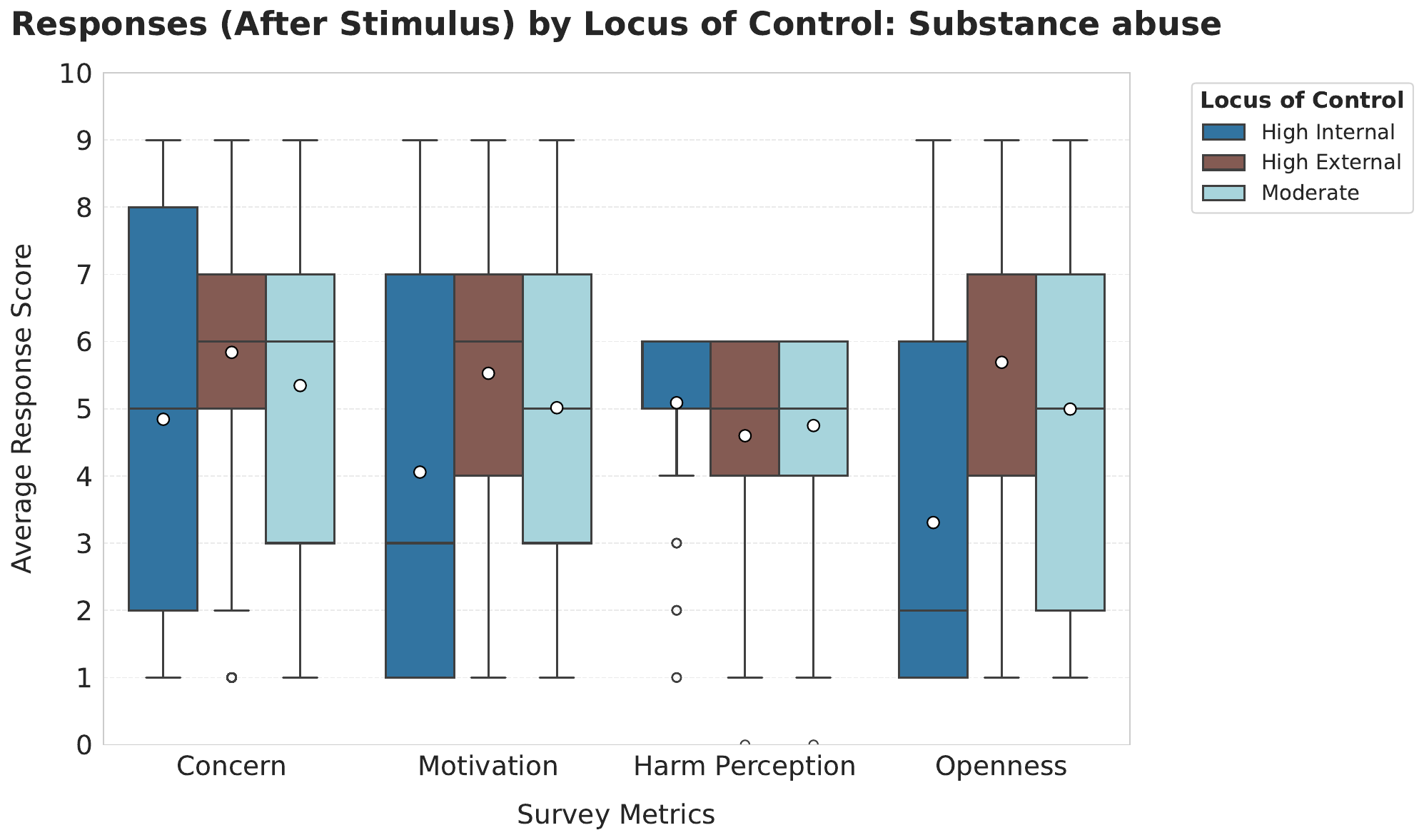}
    \includegraphics[trim={0.3cm 0.85cm 0.2cm 0.0cm},clip,width=0.40\linewidth]{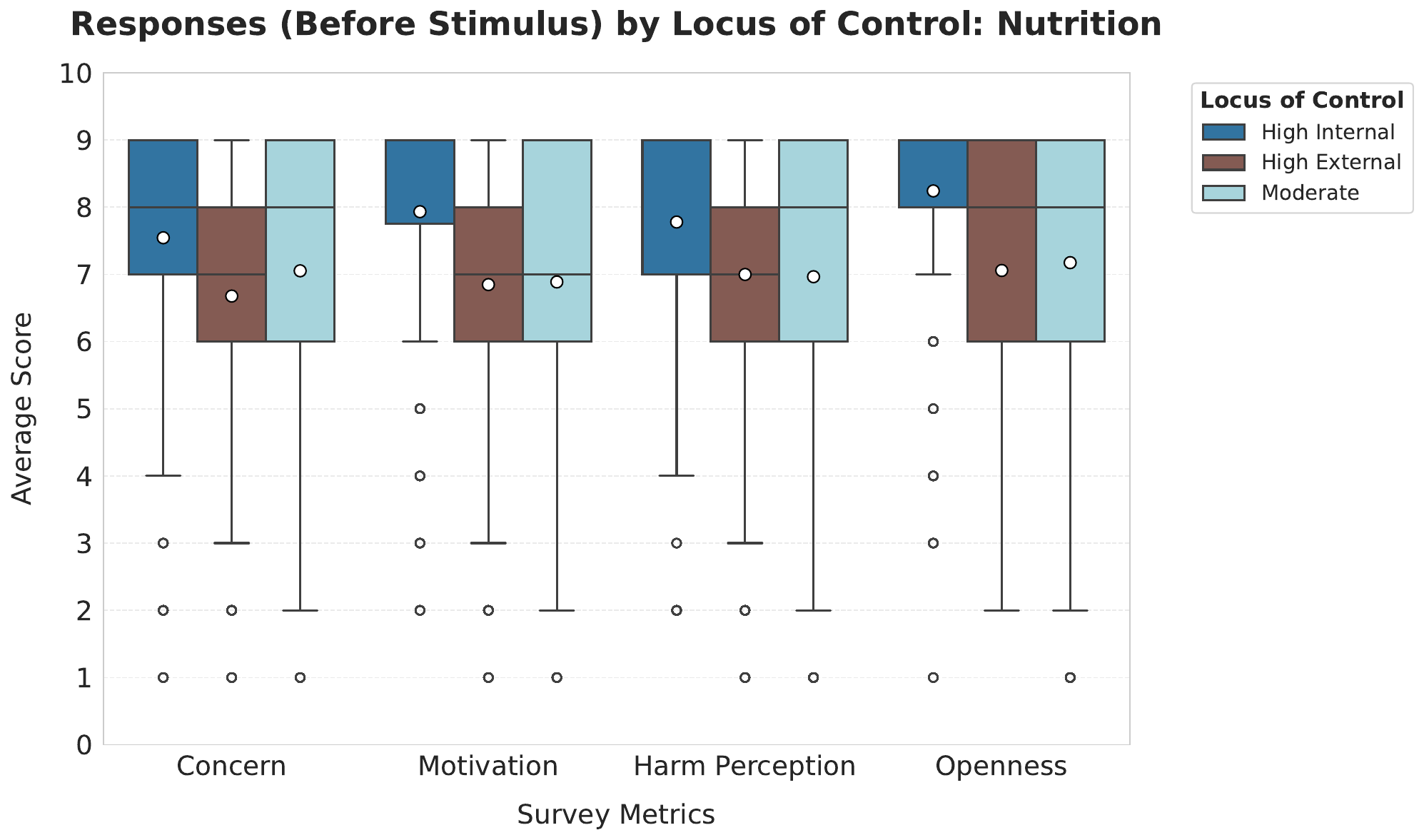}
    \includegraphics[trim={0.3cm 0.85cm 0.2cm 0.0cm},clip,width=0.40\linewidth]{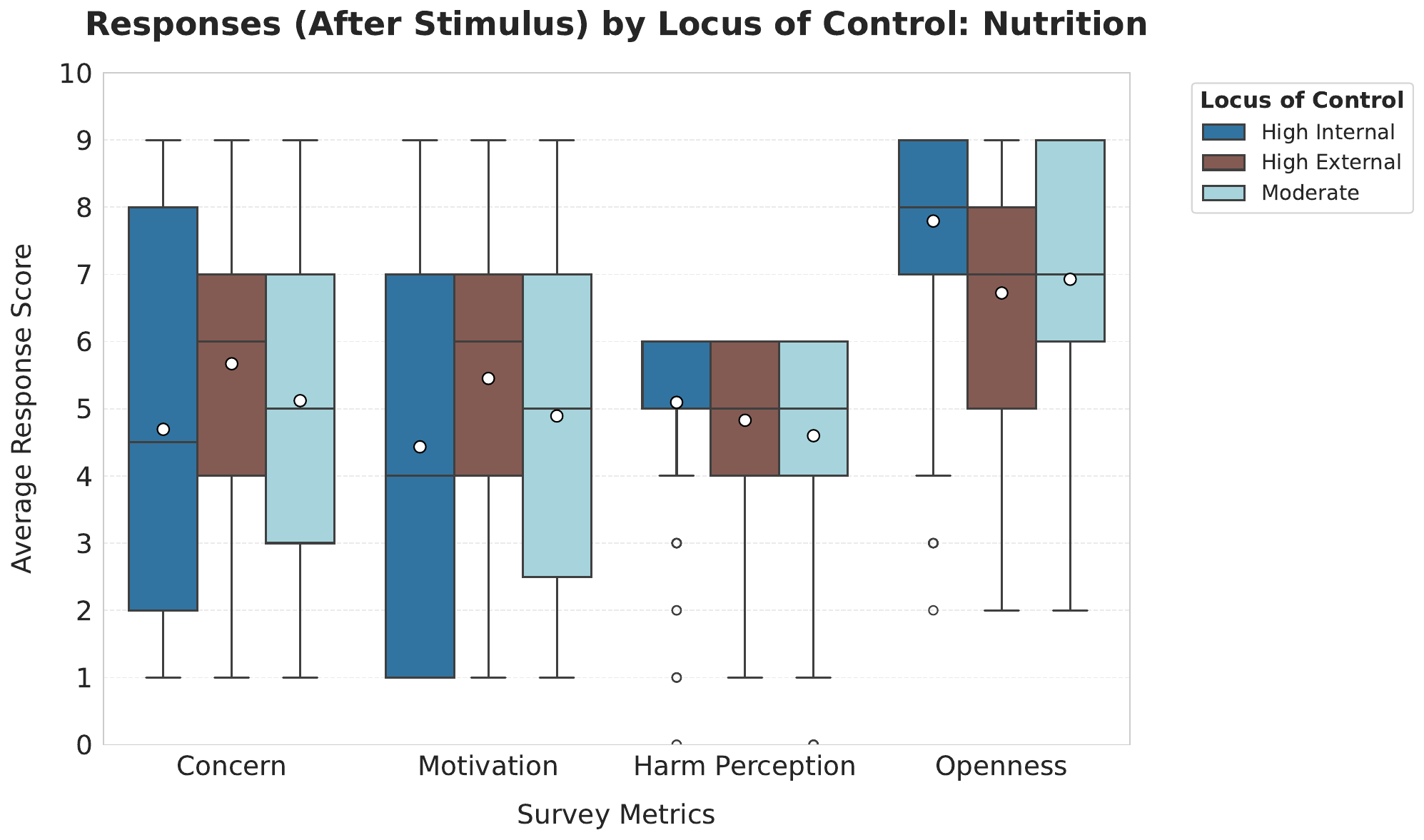}
    \includegraphics[trim={0.3cm 0.85cm 0.2cm 0.0cm},clip,width=0.40\linewidth]{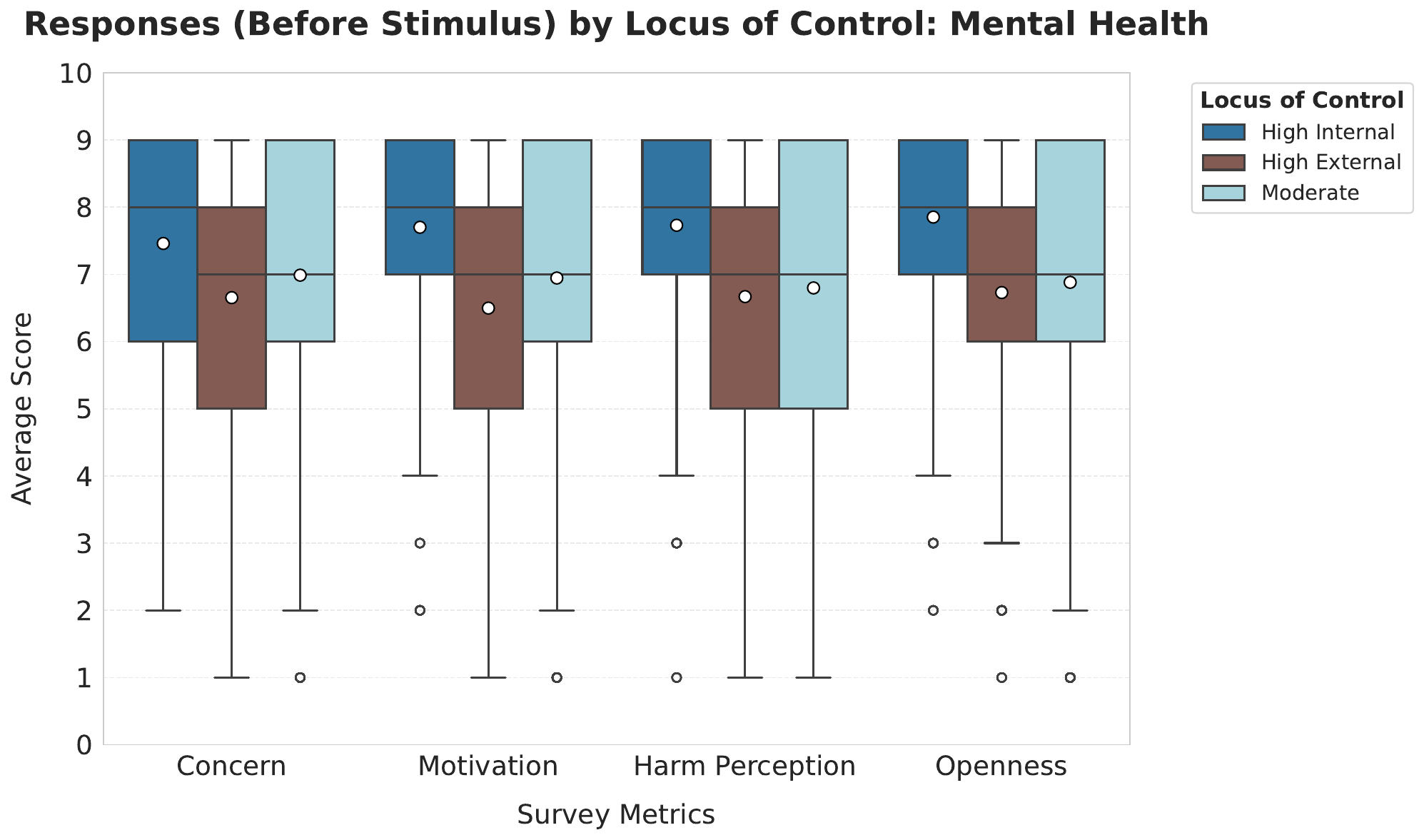}
    \includegraphics[trim={0.3cm 0.85cm 0.2cm 0.0cm},clip,width=0.40\linewidth]{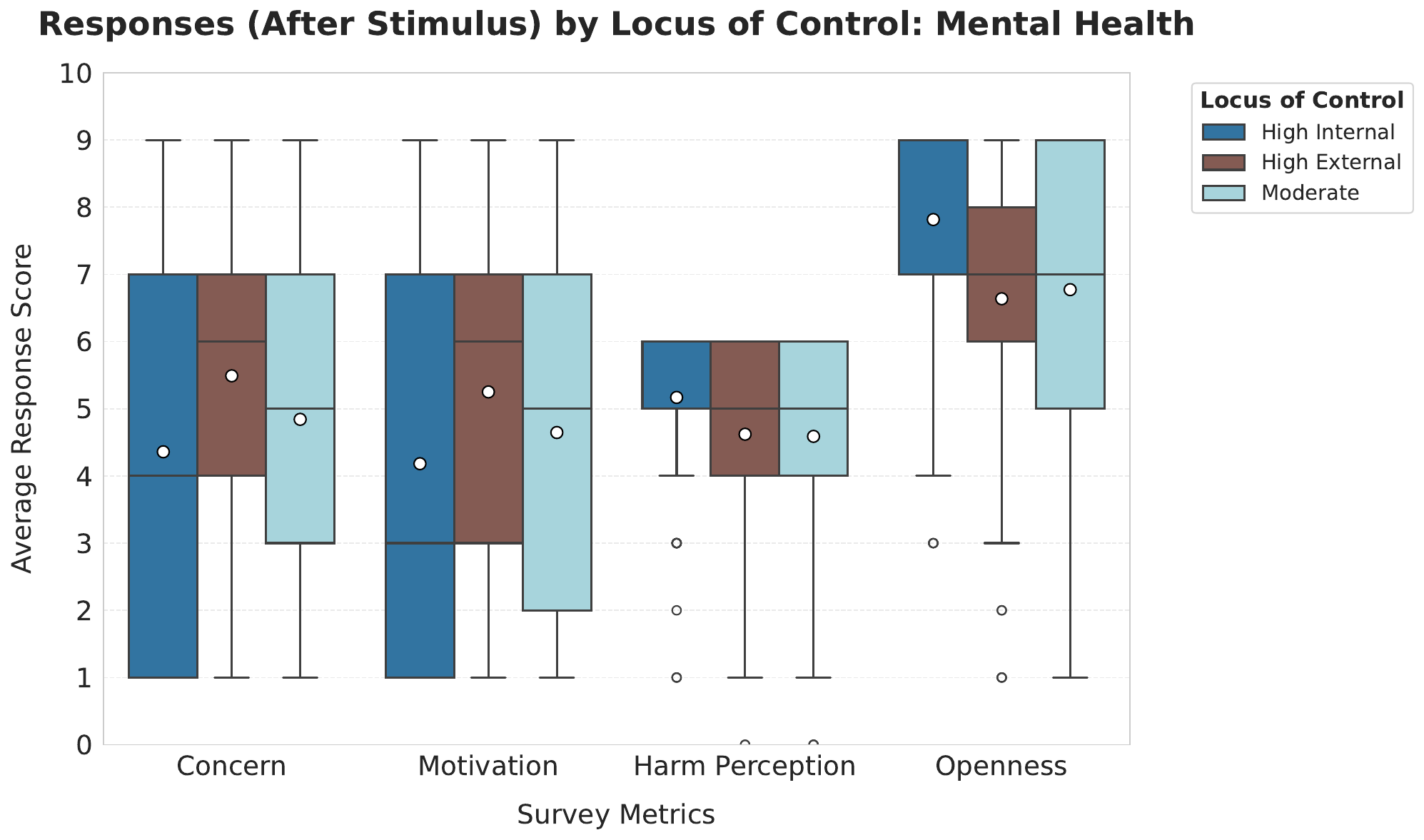}
    
    \caption{Concern, Motivation, Harm Perception and Openness of individuals with different Locus of Control prior to viewing any marketing content. People with a \textit{high internal locus} are more concerned about the risks of health concerns and motivated to practice behaviors to promote their health. This group is less open than others to abuse substances or smoke. People with a \textit{high external locus} are more open to smoking and abuse substances than people with a \textit{moderate locus}.}
    \label{fig:locus-resp}
\end{figure}

As illustrated in Figure \ref{fig:locus-resp}, participants with a high internal locus of control (LOC) consistently exhibit elevated levels of concern, motivation, harm perception, and openness, both pre- and post-stimulus, across all health topics. This finding aligns with the theoretical framework of internal LOC, where individuals' belief in their personal control over outcomes naturally correlates with greater concern, motivation, and risk perception. Conversely, individuals with a high external LOC, who attribute outcomes largely to external factors, tended to display comparatively lower personal concern and motivation.
% We also show the aggregated baseline opinions across all health topics in \ref{}

% \begin{figure}
%     \centering
%     \includegraphics[trim={2cm 0.4cm 2cm 0.2cm},clip,width=0.49\linewidth]{media/supplementary/baseline_distributions/nutrition_distribution.png}
%     \includegraphics[trim={2cm 0.4cm 2cm 0.2cm},clip,width=0.49\linewidth]{media/supplementary/baseline_distributions/MH_distribution.png}
%     \includegraphics[trim={2cm 0.4cm 2cm 0.2cm},clip,width=0.49\linewidth]{media/supplementary/baseline_distributions/SA_distribution.png}
%     \includegraphics[trim={2cm 0.4cm 2cm 0.2cm},clip,width=0.49\linewidth]{media/supplementary/baseline_distributions/smoking_distribution.png}
%     \includegraphics[trim={2cm 0.4cm 2cm 0.2cm},clip,width=0.49\linewidth]{media/supplementary/baseline_distributions/vaccination_distribution.png}
%     \includegraphics[trim={2cm 0.4cm 2cm 0.2cm},clip,width=0.49\linewidth]{media/supplementary/baseline_distributions/chronic_distribution.png}
%     \includegraphics[trim={2cm 0.4cm 2cm 0.2cm},clip,width=0.49\linewidth]{media/supplementary/baseline_distributions/nutrition_distribution.png}
%     \includegraphics[trim={2cm 0.4cm 2cm 0.2cm},clip,width=0.49\linewidth]{media/supplementary/baseline_distributions/hiv_distribution.png}
%     \caption{Caption}
%     \label{fig:enter-label}
% \end{figure}

\subsection{Media Analysis}
\label{supp:media-an}
\begin{figure}
    \centering
    % \includegraphics[trim={1cm 0.4cm 0.2cm 0.2cm},clip,width=0.5\linewidth]{media/possible_teasers/media_effectiveness_by_big5.pdf}
    % \hspace{0.2cm}
    % \includegraphics[trim={5.2cm 3cm 8cm 7cm},clip,width=0.45\linewidth]{media/qal1.pdf}
    \includegraphics[trim={0.72cm 0.8cm 0.4cm 0.2cm},clip,width=0.46\linewidth]{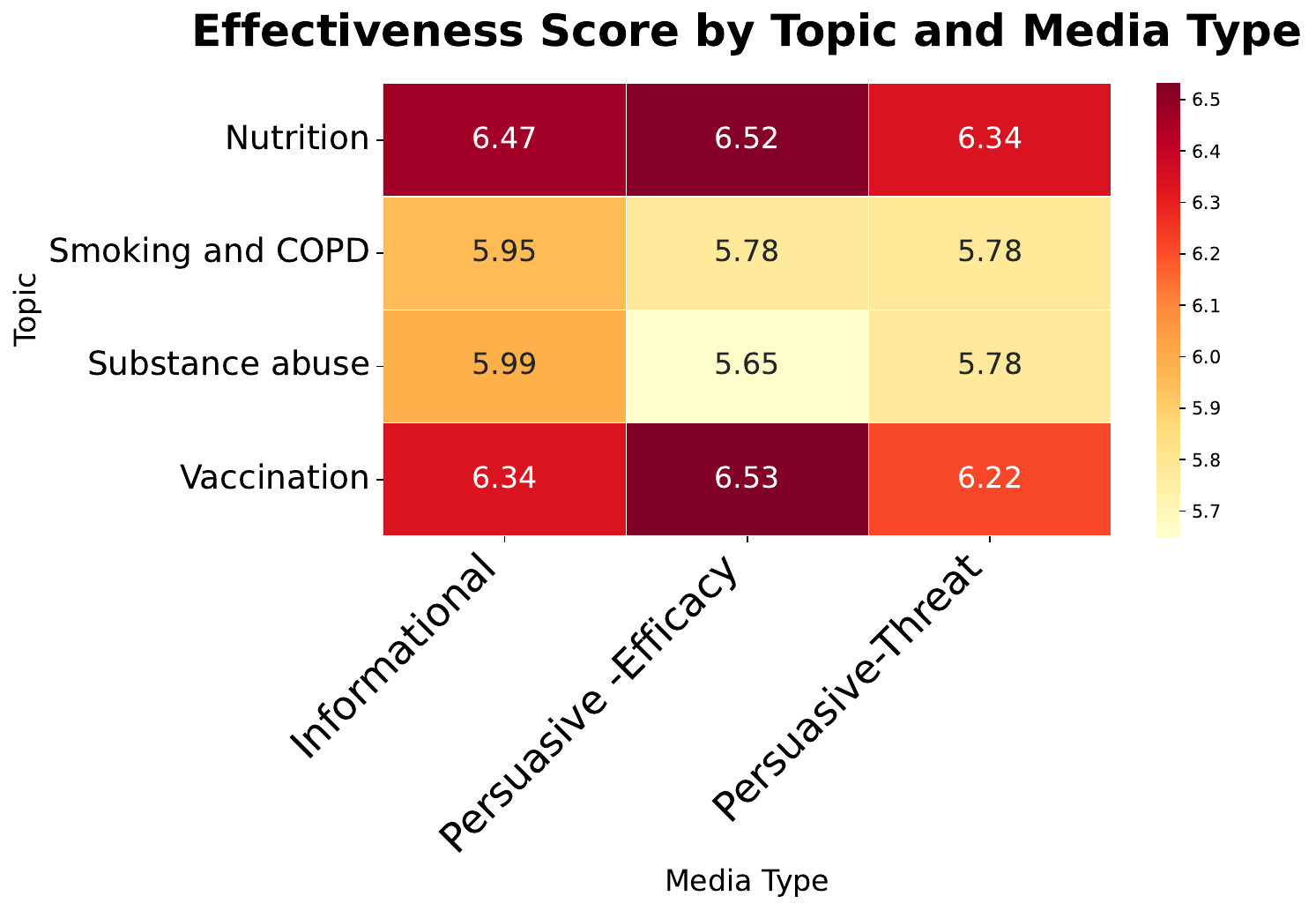} 
    \hspace{0.5cm}
    \includegraphics[trim={0.72cm 0.8cm 0.4cm 0cm},clip,width=0.43\linewidth]{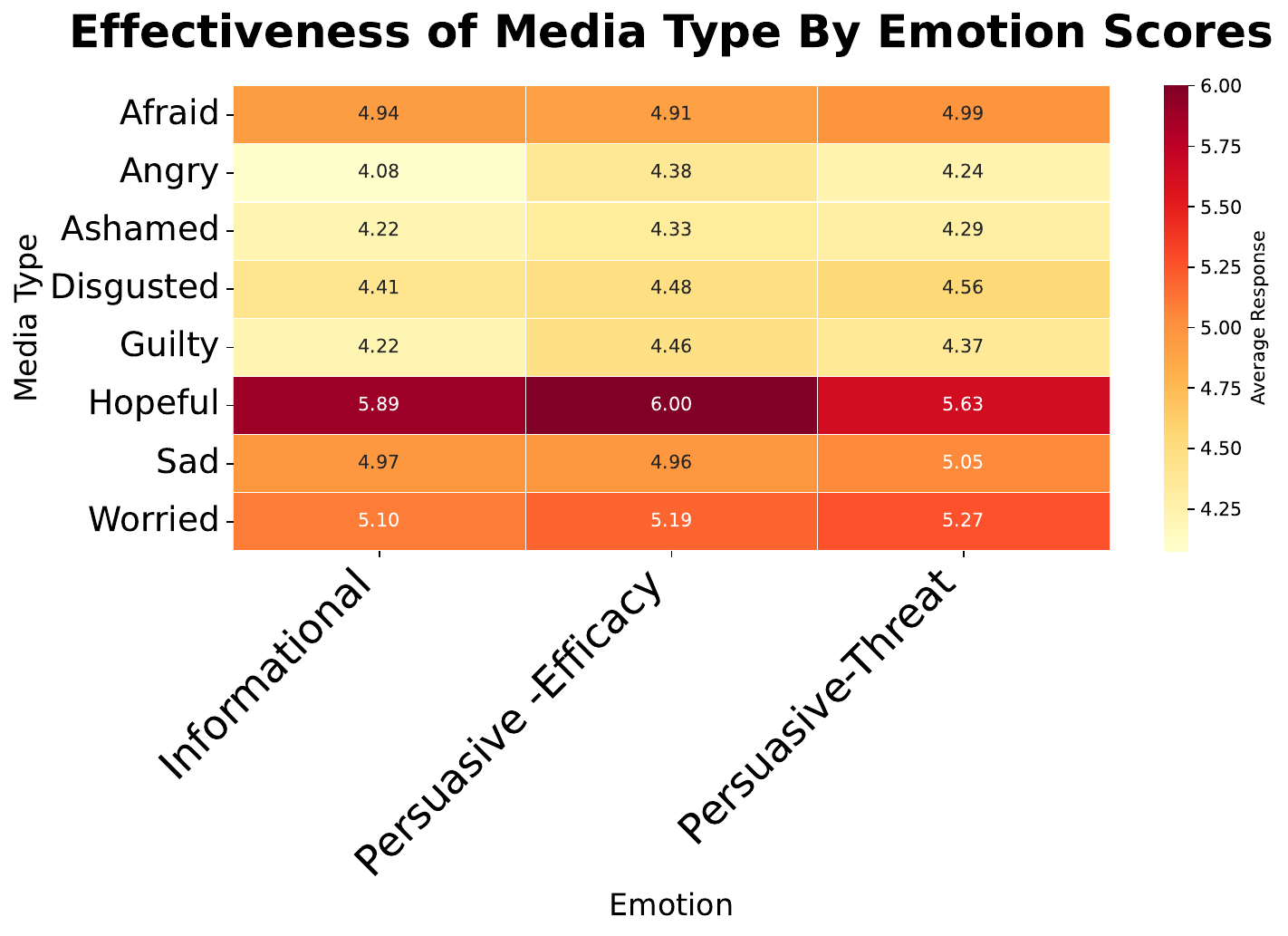}
    \includegraphics[trim={0.2cm 1.1cm 0.2cm 0.2cm},clip,width=1\linewidth]{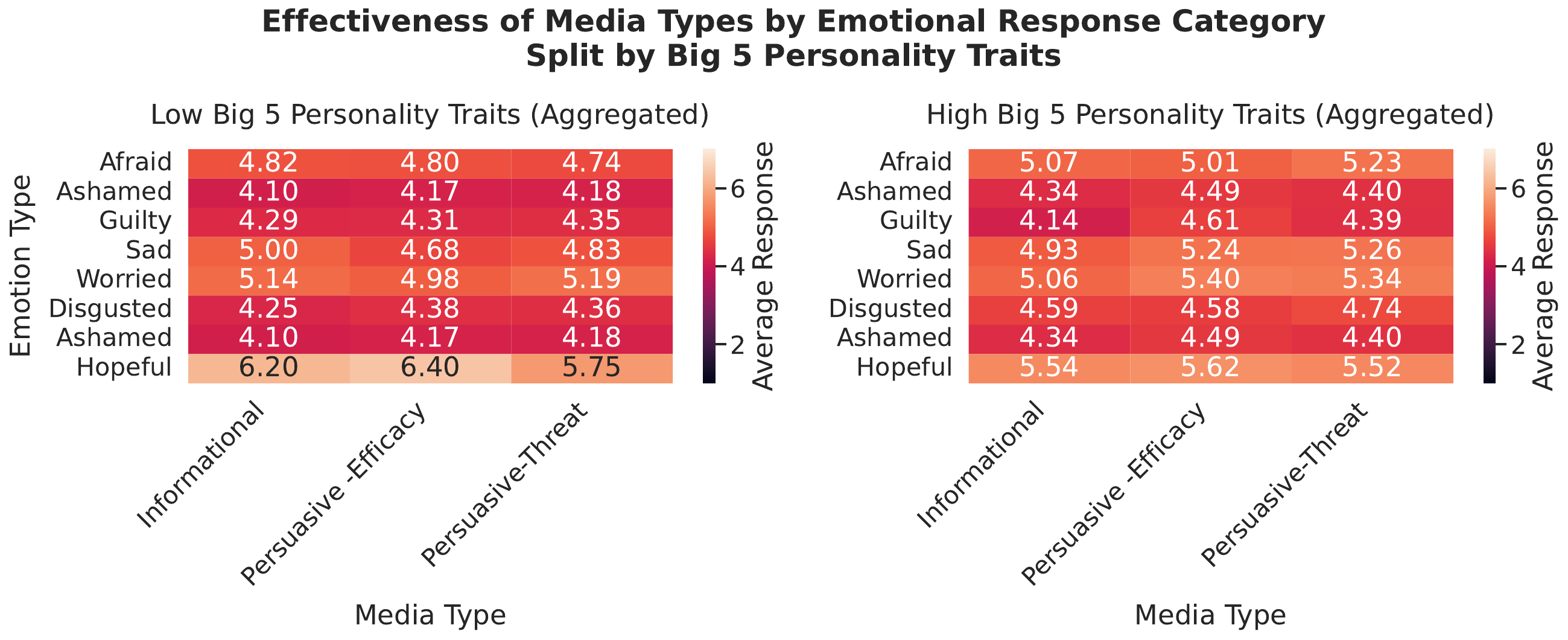}
    \caption{{\bf Motivation and Needs for PHORECAST:  }
    We analyze the concern, motivation, openness and harm perception scores induced by each campaign type to analyze their effect across public health topics.  We observe that for some, like Smoking/COPD and Substance Abuse, Informational (\textit{I}) campaigns are more effective, whereas for well-known public health concerns like nutrition, and vaccination, Persuasive-Efficacy (\textit{PE}) campaigns tend to be more effective. Studies like \cite{MCCLAUGHLIN2023100037} discuss similar results using COVID campaigns.
    Next, we illustrate the effectiveness of different campaign types (Persuasive-Efficacy (\textit{PE}), Persuasive-Threat (\textit{PT}), and Informational (\textit{I}) for eliciting different emotional responses. Overall, \textit{PT} messaging induces the highest levels of fear, anger, disgust, sadness, and worry, while \textit{PE} is particularly useful for evoking hope. Further, our dataset reveals how individuals with varying personality traits respond differently to specific media approaches. We find that PE messaging tends to be more effective for individuals with lower Big 5 personality trait scores (e.g., agreeableness, extraversion), whereas \textit{I} and \textit{PT} based messaging are more effective for those with higher scores.
    }
    \label{fig:big5-responses}
\end{figure}

\begin{figure}
    \centering
    \includegraphics[trim={0.1cm 1cm 0.2cm 0.2cm},clip,width=1\linewidth]{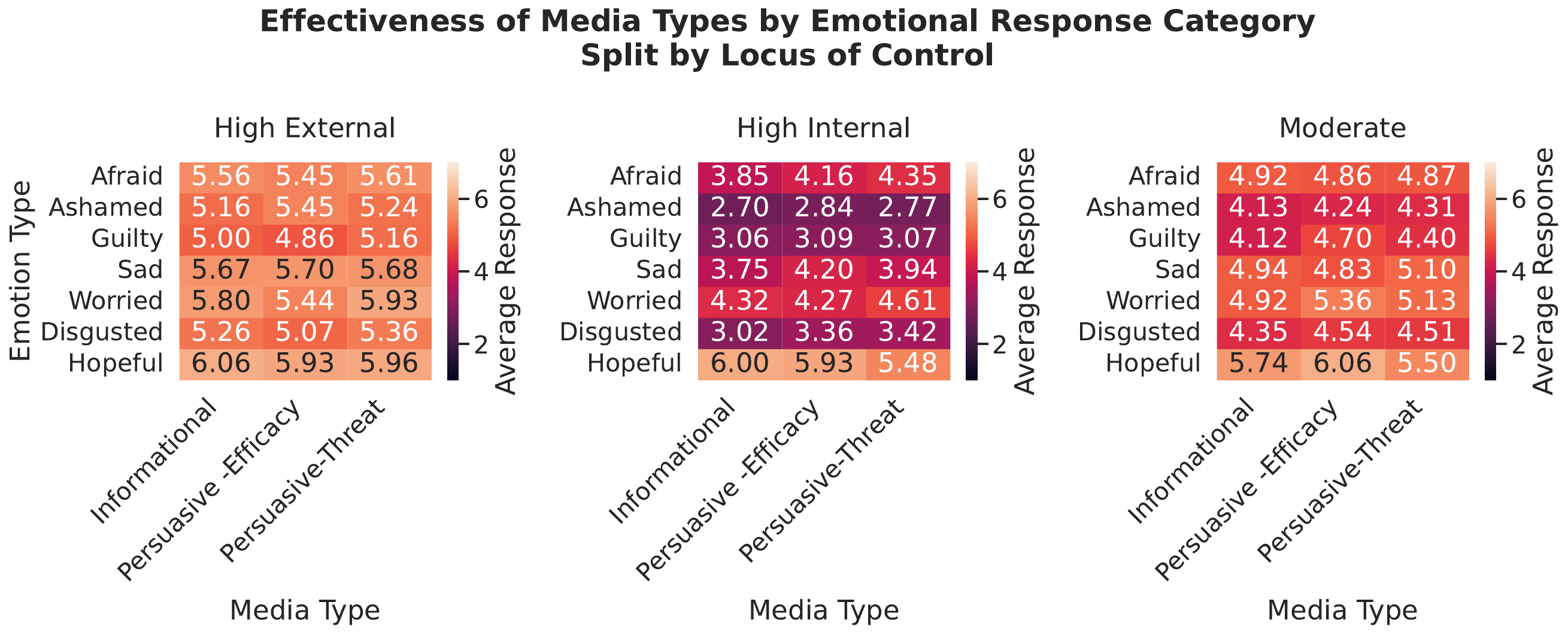}
    \caption{\textbf{Media Effect by Locus of Control:} We illustrate the effect the media type (Persuasive-Efficacy, Persuasive-Threat, and Informational) has on eliciting different emotional responses across different categories of locus of control. We observe that people with a high internal (middle) tend to feel less ashamed and guilty, while those with a high external feel the highest levels of fear, shame, guilt, sadness, worry, disgust and hope. Further, it is not clear which media type is most effective for individuals with different locus of control categories, implying that the impact of media on individuals might be moderated by their pre-existing beliefs about control over their lives. }
    \label{fig:effect-locus}
\end{figure}

\begin{figure}
    \centering
    \includegraphics[trim={1cm 1cm 0.2cm 0.2cm},clip,width=0.32\linewidth]{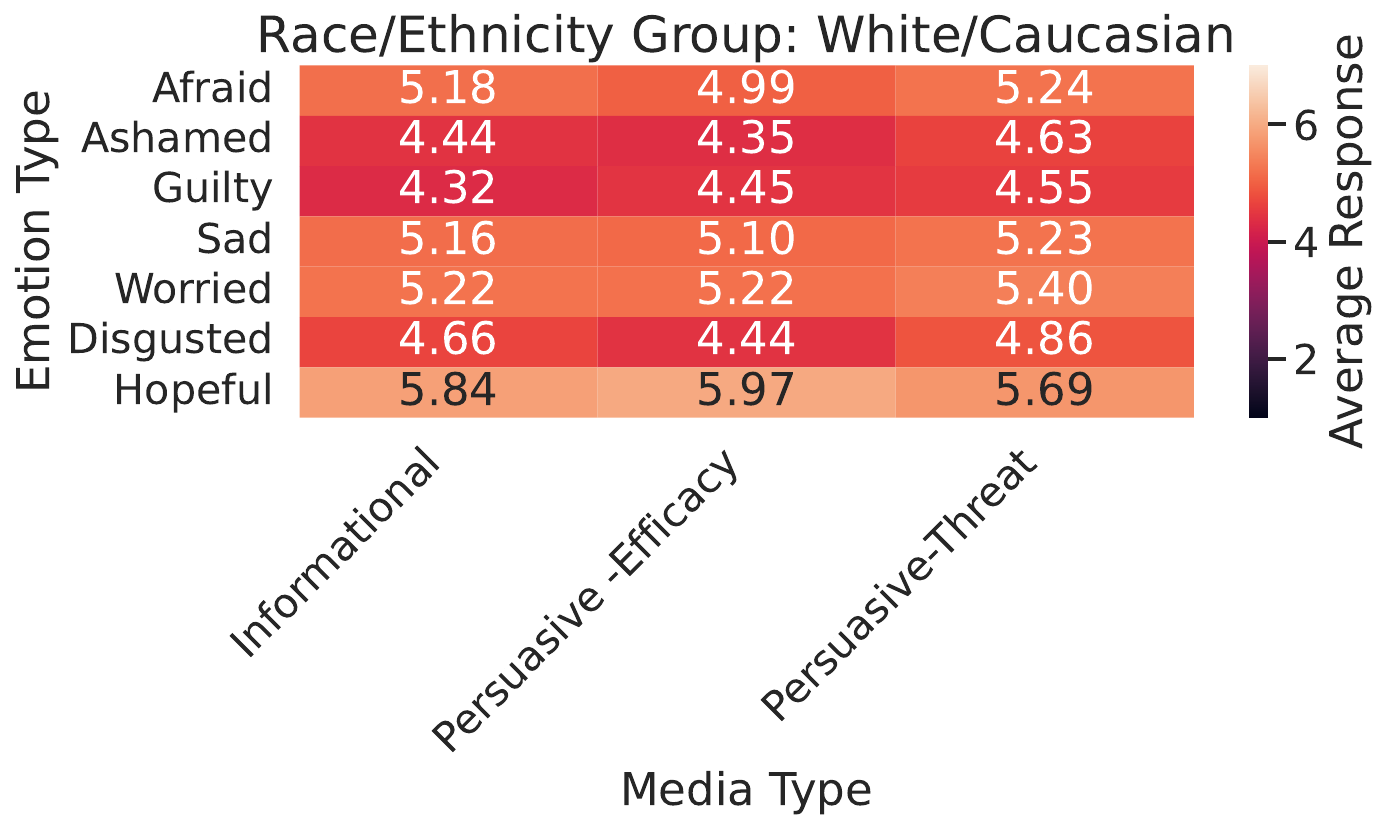}
    \includegraphics[trim={1cm 1cm 0.2cm 0.2cm},clip,width=0.32\linewidth]{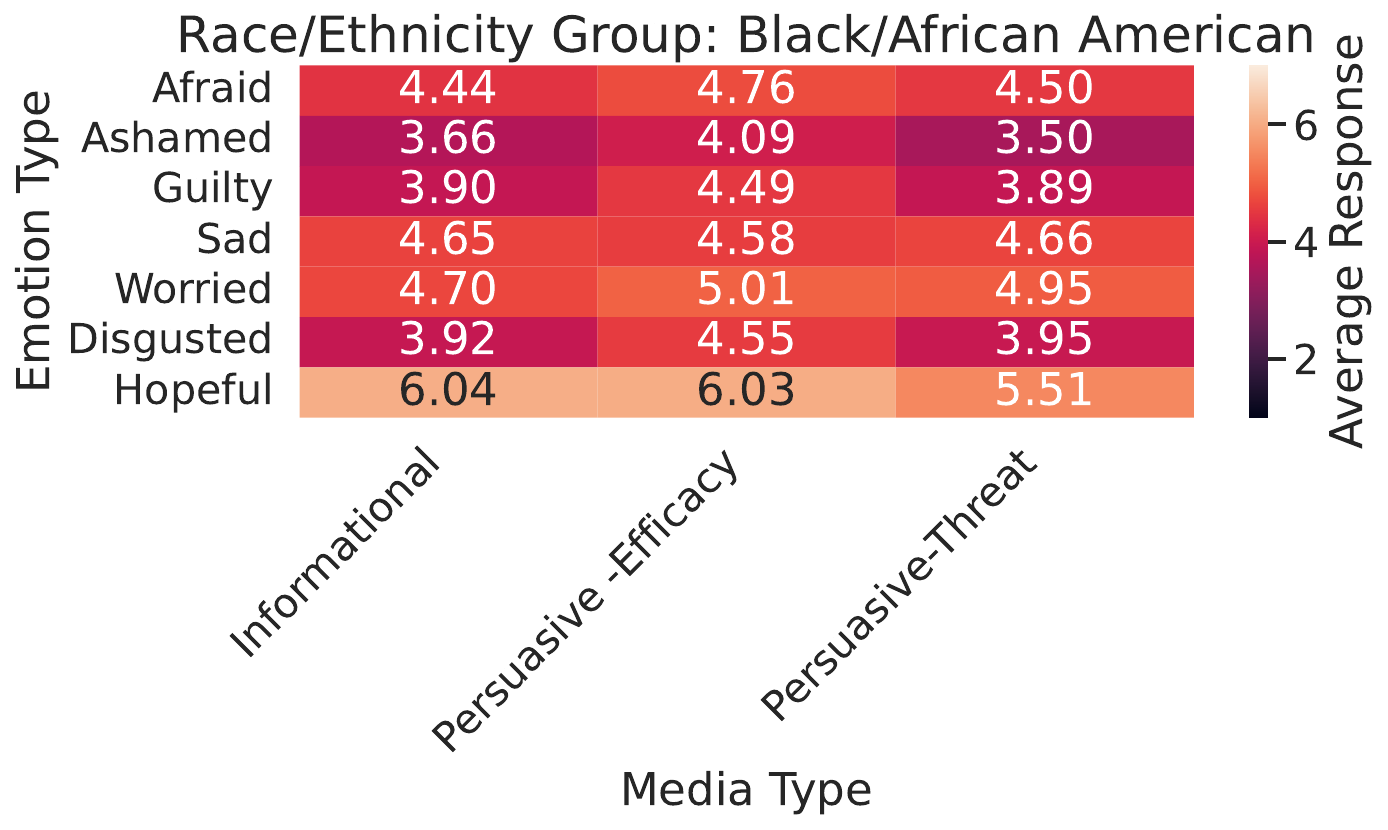}
    \includegraphics[trim={1cm 1cm 0.2cm 0.2cm},clip,width=0.32\linewidth]{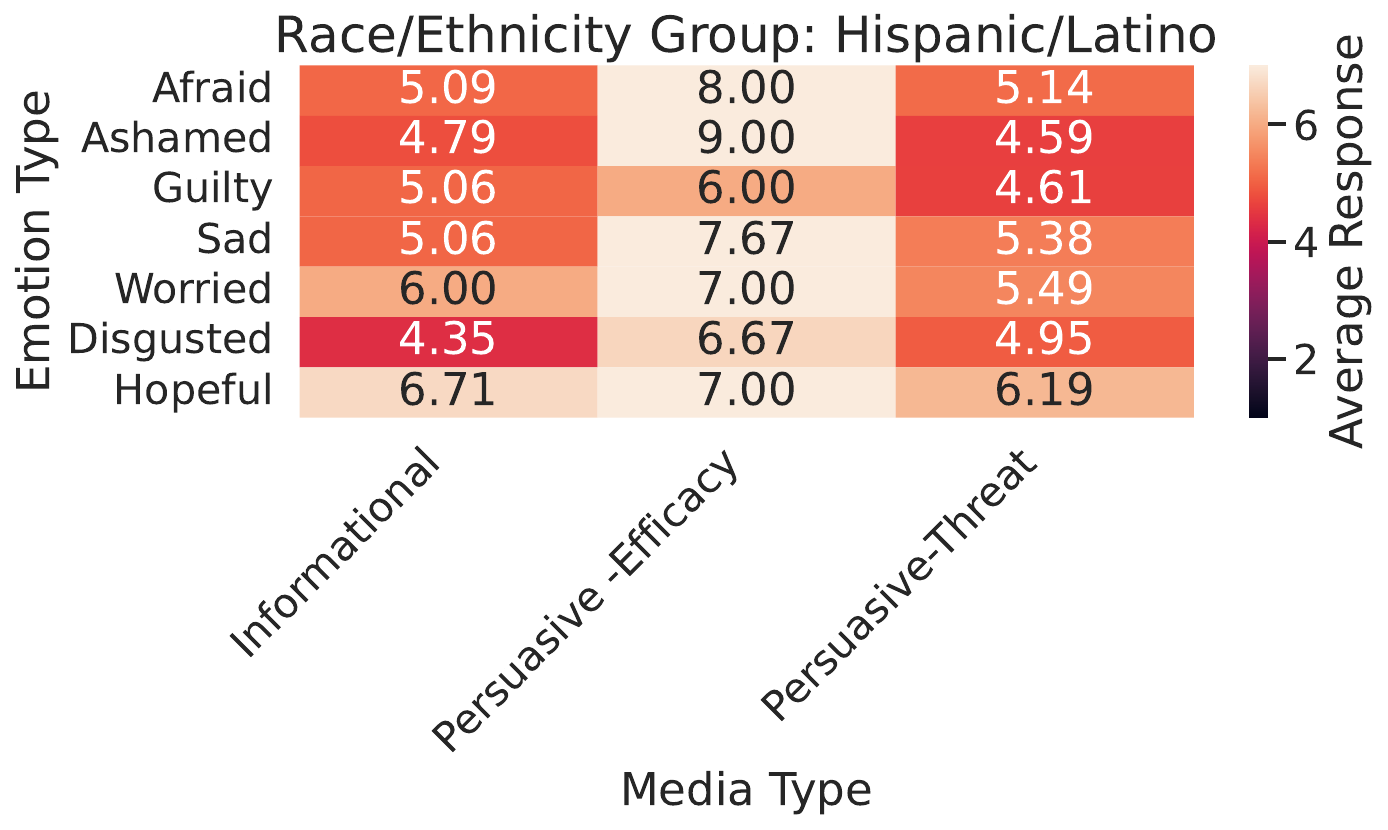}
    \caption{\textbf{Media Effect by Race/Ethnicity: }The effect the media type (Persuasive-Efficacy, Persuasive-Threat, and Informational) has on eliciting different emotional responses across different race/ethnicity groups. We observe that individuals who fall under the Hispanic/Latino group tend to react more strongly to persuasive efficacy messaging. Participants who identify as white/caucasian are likely more effected by persuasive-threat campaigns, while participants that identify as Black/African American react more strongly with persuasive campaigns.}
    \label{fig:effect-race}
\end{figure}

We analyze the effectiveness of different campaign types by (1) analyzing what campaigns are most effective for different health topics, (2) which elicited the highest levels of emotion overall and (3) we split participants with low and high big 5 personality traits and analyze the differences in emotion scores. As seen in Fig.\ref{fig:big5-responses}, different personality traits react and perceive different types of campaigns differently. For example, people with high levels of anxiety tend to feel more afraid, angry, ashamed, guilty, hopeful and worried when viewing persuasive images, as opposed to threatening or informational. On the contrary, people with low anxiety tend to react more strongly to threatening campaigns. We also analyze the effect of the media types across different categories of locus of control in Fig.\ref{fig:effect-locus} and by race/ethnicity in Fig.\ref{fig:effect-race}. Finally, we hope this work inspires researchers to study the aesthetic or psychological features in the pixel space that impact how different individuals react and respond to marketing content.

\newpage
% \clearpage
% \newpage
\section{Training Details and Evaluation Metrics}
\label{supp:part2_extended_results}
\subsection{Training Details}
\label{supp:training}
We use the unsloth \cite{unsloth} framework for training the models with the following hyperparameters: $bs=1$, $lr = 2e-4$, and rank $r=8$ for Lora \cite{hu2021loralowrankadaptationlarge}. We use base models \textit{unsloth/Llama-3.2-11B-Vision-Instruct} and \textit{unsloth/gemma-3-12b-it}. We utilize \textit{paged\_adamw\_8bit} optimizer and a max sequence length of 2048. We always keep the vision layers frozen, and only tune the attention, language and MLP layers. We train for a total of 12k steps. 

\subsection{Dataset Preparation}
\label{supp:part1_data_prep}
The training data comprises each participant's responses to five health advertisements. This is randomly chosen for each participant at the start of the survey. Input to the model is randomized by probabilistically including features: demographics/personality (90\% of the time, for which we then choose a random number of features from 1 to len(avail\_demograhics)), locus of control (75\%), free-form text (30\%), and baseline/in-context opinions (50\%, for which we choose a random number of features to include). This randomization method is used to create the training and validation split for model evaluation, as shown in Table~\ref{tab:data-splits}. After training, we evaluate the model on new advertisements and new individuals. As described in \ref{training-section}, we construct our validation set by ensuring our sampling includes unique representatives for each gender, religion and race/ethnicity. 

The model receives a task description, along with participant demographics, personality traits, and in-context question/answer pairs. Fig. \ref{fig:training_template} illustrates our chat template when all available features are utilized, a scenario that can occur within our feature randomization process.

\begin{table}[h]
\centering
\small
\begin{tabular}{lcc|cc}
\toprule
\textbf{Attribute} & \textbf{Train Count} & \textbf{Train \%} & \textbf{Val Count} & \textbf{Val \%} \\
\midrule
\textit{Number of Individuals} & 1010 & -- & 521 & -- \\
\midrule
\textit{Total Samples} & 27,572 & -- & 8,537 & -- \\
\midrule
\multicolumn{5}{l}{\textit{Gender}} \\
Male        & 15,596 & 56.6\% & 4,422 & 51.8\% \\
Female      & 11,748 & 42.6\% & 4,050 & 47.4\% \\
Other       &    228 &  0.8\% &    65 &  0.8\% \\
\midrule
\multicolumn{5}{l}{\textit{Religion}} \\
Christian   & 20,079 & 72.8\% & 6,432 & 75.3\% \\
Unknown     &  4,087 & 14.8\% & 1,014 & 11.9\% \\
Muslim      &  1,833 &  6.6\% &   663 &  7.8\% \\
Other       &  1,573 &  5.7\% &   428 &  5.0\% \\
\midrule
\multicolumn{5}{l}{\textit{Race}} \\
White       & 17,409 & 63.1\% & 5,568 & 65.2\% \\
Black       &  8,019 & 29.1\% & 2,215 & 25.9\% \\
Hispanic    &  1,039 &  3.8\% &   273 &  3.2\% \\
Asian       &    611 &  2.2\% &   182 &  2.1\% \\
Indigenous  &    481 &  1.7\% &   273 &  3.2\% \\
\midrule
\multicolumn{5}{l}{\textit{Media Count}} \\
Unique Media & 29 & -- & 8 & -- \\
\bottomrule \\
\end{tabular}
\caption{Demographic distribution and media counts across training and validation splits. Percentages are computed within each split. We keep one unique image per topic to validate the model's ability to generalize to unseen images. We ensure we have representative individuals based on three demographical features: gender, religion, and race. This experimental design creates a training split of $1010$ unique individuals with $27,572$ unique samples, and a validation split of $521$ individuals (42 unique individuals) with their corresponding $8537$ samples. Our dataset has comparatively high representation from White, Christian males. }
\label{tab:data-splits}
\end{table}

\begin{figure}[t]
    \centering
    \fbox{\begin{minipage}{0.92\linewidth}
        \small
        \setlength{\parskip}{0.5em}
        \fontfamily{cmtt}\selectfont
        
        \textit{User:}
        You are a helpful assistant trained to interpret user thoughts and feelings and predict how they would react and answer different questions about various health topics. 
        
        Five health topics are randomly selected for you from the following list: Nutrition, Vaccination, Mental Health, Substance abuse, COPD, Chronic Diseases, HIV/aids, Sexual Health.
        
        You are of the following demographics: get\_demographics(row). \\ You have the following personality traits: get\_personality(row).\\ You have a row['locus'].\\ You first answer baseline questions about each health topic.\\ For the topic of row['topic'], you answer as follows: get\_baseline(row). \\ You are then shown the following image and you answer the following: [Q/As].\\ 
        
        Given the question: 'type in every thought that came to mind viewing this material.' What would your response be? 
        
        \textit{Assistant:} "It makes me think about..."
    \end{minipage}} 
    \caption{\textbf{Training Template Structure}: Example of chat format used for training, showing dynamic insertion of (1) user profiles, (2) randomly sampled health topics, and (3) task-specific response targets. Please refer to our explanation and code for a detailed breakdown of how we randomize the number of features seen during training for each sample.}
    \label{fig:training_template}
    \vspace{-0.2cm}
\end{figure}

\newpage
\subsection{Additional Metrics for Free Form Evaluation}
\label{supp:free-form}
In addition to semantic similarity, we utilize the following metrics to analyze LLM generated free form responses:

\paragraph{SDE Score}
\label{free-form-fid}
In keeping with evaluation standards for generative images, 
we compute the statistical distribution of embedding (SDE) features stratified by personality traits to assess distributional alignment across different
subgroups, i.e. between distributions of real and machine-generated responses, stratified by personality trait bins. SDE quantifies the similarity (lower is better) between machine and human responses by embedding them into a feature space, fitting a multivariate Gaussian distribution to each, and measuring the distributional overlap.
Let the set of embeddings from human (ground truth) responses be characterized by mean vector $\mu_r$ and covariance matrix $\Sigma_r$, and let the embeddings of machine-generated responses be described by mean vector $\mu_g$ and covariance matrix $\Sigma_g$. The SDE is defined as:

\[
\text{SDE} = \|\mu_r - \mu_g\|^2 + \text{Tr}\left( \Sigma_r + \Sigma_g - 2\left( \Sigma_r \Sigma_g \right)^{1/2} \right)
\]

% where:
% \begin{itemize}
%   \item $\|\mu_r - \mu_g\|^2$ denotes the squared Euclidean distance between the mean embeddings of real and generated responses,
%   \item $\text{Tr}(\cdot)$ denotes the trace of a matrix (the sum of its diagonal elements),
%   \item $\left( \Sigma_r \Sigma_g \right)^{1/2}$ is the matrix square root of the product of the covariance matrices.
% \end{itemize}

\begin{figure}
    \centering
    \includegraphics[trim={0.2cm 0.2cm 0.2cm 0.2cm},clip,width=1\linewidth]{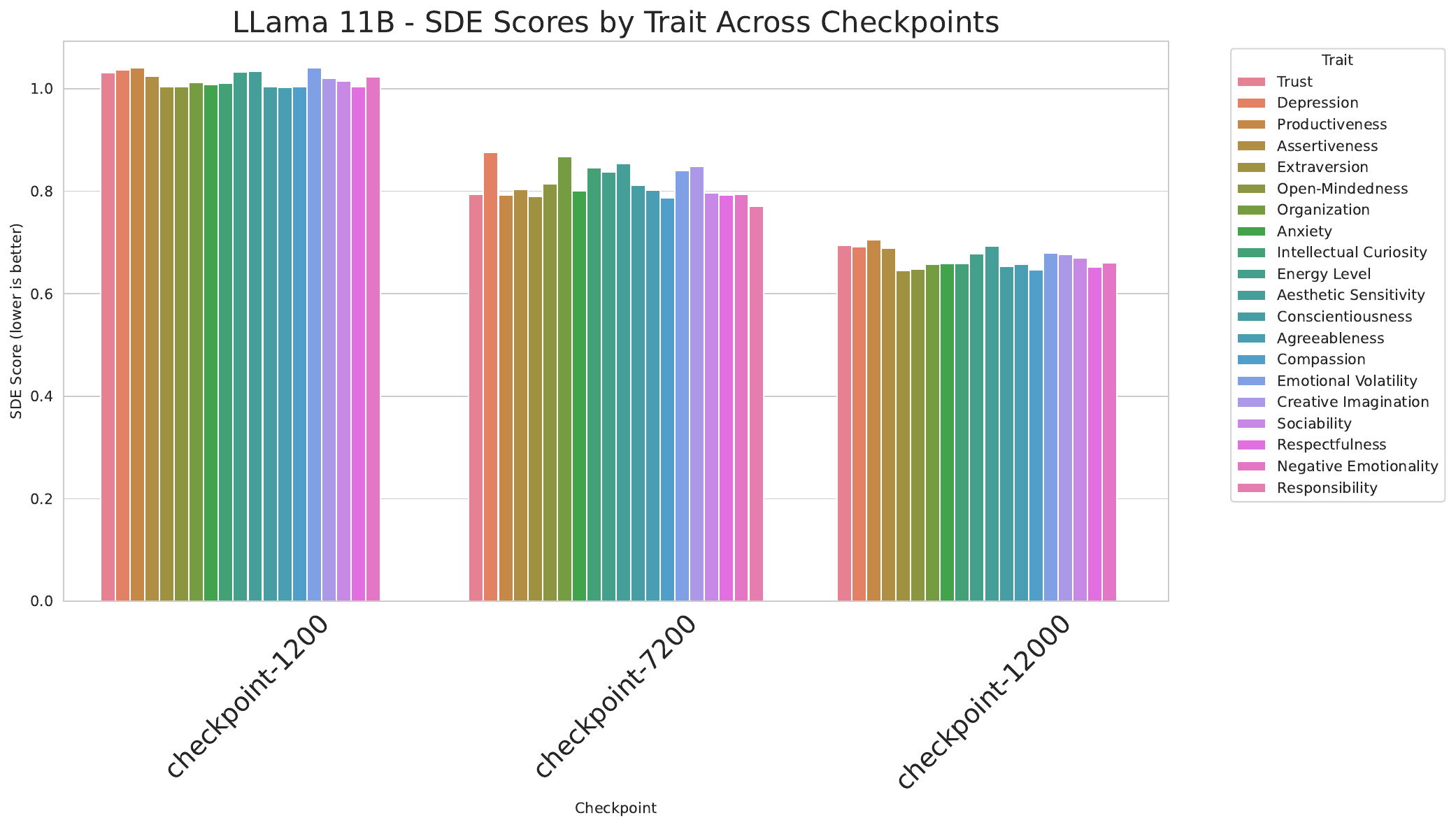}
    \includegraphics[trim={0.2cm 0.2cm 0.2cm 0.2cm},clip,width=1\linewidth]{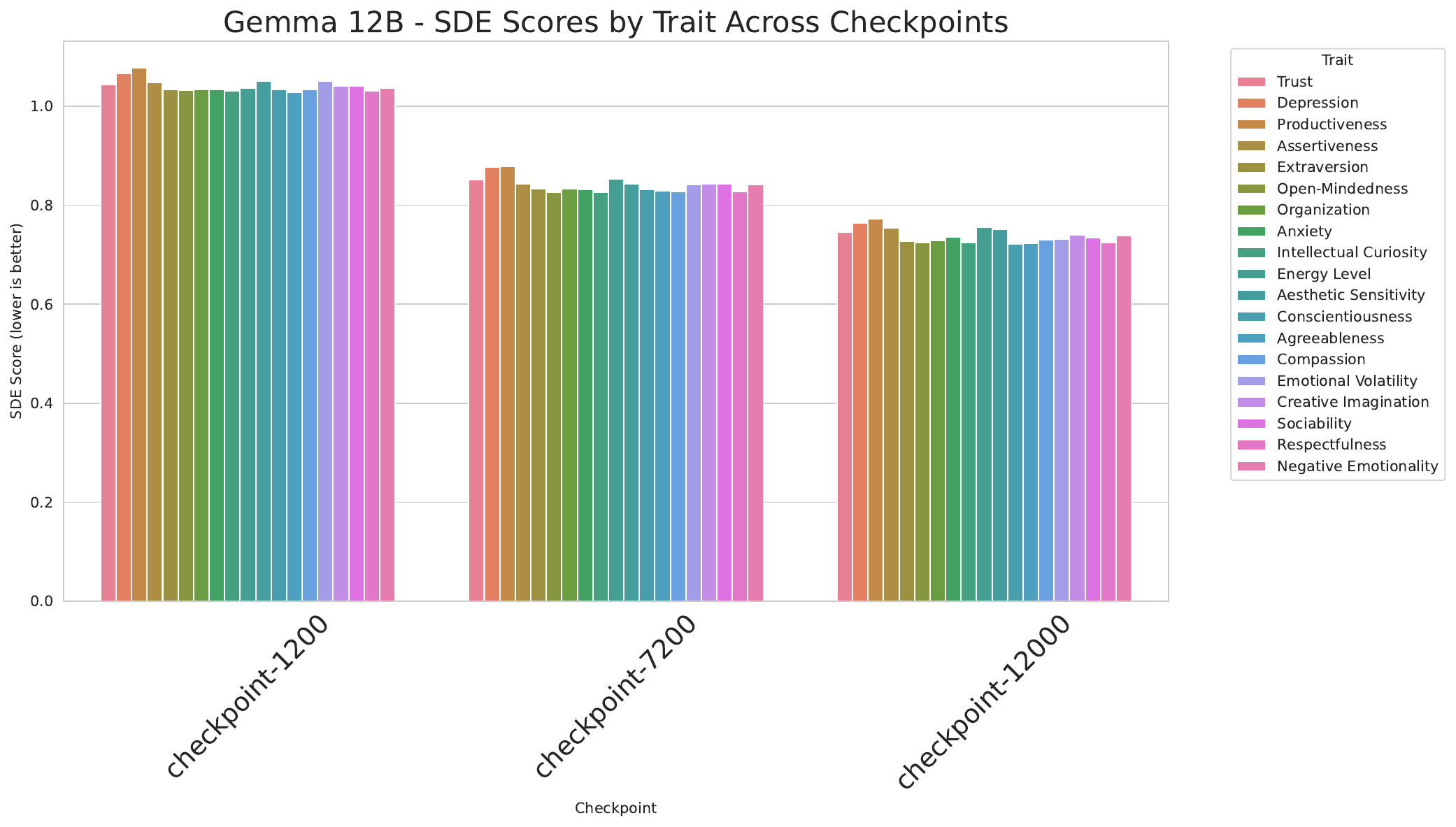}
    \caption{\textbf{SDE Scores Across Checkpoints} of Llama (Top) and Gemma (Bottom). We observe that both models begin with a high SDE score. As the models iterate over our dataset, SDE scores continue to decrease, showing better alignment to individuals. This results in a model that not only aligns better on average but also captures the unique manifestations of different personalities with greater fidelity. }
    \label{fig:sde_across_checkpts}
\end{figure}

A lower SDE indicates that the generated responses more closely resemble the statistical distribution of real human responses in the embedding space. We compute SDE scores separately for subsets of data grouped by trait levels (e.g., low vs.\ high agreeableness). This allows us to measure the model’s fidelity in reproducing nuanced, trait-dependent response patterns. A monotonic decrease in SDE over training steps signals improved generation quality and diversity, and a low trait-specific SDE suggests the model is capturing the distinctive structure of human responses conditioned on that trait—serving as a principled metric for evaluating personality-conditioned generative alignment. We show SDE scores for Llama and Gemma in \ref{tab:model_accuracy} and across different checkpoints in Fig.\ref{fig:sde_across_checkpts}.

% \subsubsection{Accuracy}
\paragraph{Perplexity}
\label{free-form-acc}
We employ `perplexity',  a measure of uncertainty in predicting the next word in a sequence, as a discriminative metric to evaluate the model $\mathcal{M}$'s ability to capture individual-specific language patterns and semantic styles. For any distinct pair of individuals, A and B, with profiles $P_A, P_B$ and observed responses $R_A, R_B$, we compute the full sequence perplexity $\text{PPL}(P_X \circ \texttt{"Response: "} \circ R_Y)$ for $X, Y \in \{A, B\}$, where $\circ$ denotes string concatenation. For each response $R_Y$, a correct attribution is defined if $\text{PPL}(P_Y \circ \texttt{"Response: "} \circ R_Y) < \text{PPL}(P_{X \neq Y} \circ \texttt{"Response: "} \circ R_Y)$. This evaluation measures $\mathcal{M}$'s capacity to recognize and associate responses with their generator's style. We show the results for Llama and Gemma before and after training in \ref{tab:model_accuracy}. 

\begin{table}[ht]
\centering
\begin{tabular}{lcc}
\toprule
\textbf{Model} & \textbf{Perplexity ($\uparrow$) } & \textbf{SDE ($\downarrow$) } \\
\midrule
\multicolumn{3}{l}{\textbf{Llama 11B}} \\
Initial checkpoint (1k steps) & 46.20 & 1.02 \\
After training &\textbf{ 51.67} & \textbf{0.67} \\
\addlinespace
\multicolumn{3}{l}{\textbf{Gemma 12B}} \\
Initial checkpoint (1k steps) & 50.76 & 1.04 \\
After training & 51.52 & 0.74 \\
\bottomrule
\end{tabular}%
\vspace{0.2cm}
\caption{
Comparison of human language emulation performance before (1k-step checkpoint) and after training. 
\textbf{Metrics:} Higher accuracy (\% correct) and lower Statistical Distribution of Embeddings (SDE) indicate better alignment with human responses. 
\textbf{Key findings:} 
(1) Llama shows improvement in SDE (1.02 $\rightarrow$ 0.67) and an increase in accuracy (of over $5\%$). 
(2) Gemma also achieves some gains: accuracy rises from 50.76 to 51.52, and SDE improves from 1.04 $\rightarrow$ 0.74.
\textbf{Qualitative:} Base models often default to uncertain or repetitive language (e.g., Llama's "I'm not sure..." in 80\% of cases).
}
\label{tab:model_accuracy}
\end{table}

\section{Generalization}
\label{supp:generalization}
We use our trained LLama model to investigate how robust we are to distributional shifts. We demonstrate strong generalization to unseen subpopulations in \ref{tab:llama_performance}. 
This evaluation setting tests a difficult case, where each sample in the validation set contains a masked number of features. For instance, a model might need to predict the participant’s response based solely on that person’s locus of control. We use the same practice (methods and hyperparameters) described in the paper but we train the models for approximately 8k steps and test on a random sample of 1k. We report accuracy with ±2 tolerance.

\begin{table}[h!]
\centering
\begin{tabular}{lcc}
\toprule
\textbf{Test Set} & \textbf{Llama (Before Training)} & \textbf{Llama (After Training)} \\
\midrule
Females aged 25--34 & 56.2 & 72.9 \\
Individuals in technology making \$100k+ & 51.0 & 72.7 \\
Non-Christian men & 50.7 & 76.6 \\
\bottomrule
\end{tabular}
\vspace{0.7mm}
\caption{Performance of Llama before and after training on different demographic test sets. When training with data from PHORECAST, models are able to generalize to unseen groups, improving the accuracy by approximately $20\%$. }
\label{tab:llama_performance}
\end{table}

We also investigate model performance on unrelated benchmarks after training on our dataset, to test whether general language modeling capabilities degrade as a result of finetuning. We use the \texttt{lm-evaluation-harness} and evaluate Llama~3.2~11B Instruct and Gemma~3~12B models (batch size $=8$) before and after training with PHORECAST (Table~\ref{tab:performancebench}).

\begin{table}[h!]
\centering
\begin{tabular}{lcc}
\toprule
\textbf{Task} & \textbf{Gemma (Before $\rightarrow$ After)} & \textbf{Llama (Before $\rightarrow$ After)} \\
\midrule
TriviaQA   & 27.6 $\rightarrow$ 46.5 & 51.4 $\rightarrow$ 39.7 \\
ToxiGen    & 56.8 $\rightarrow$ 58.8 & 53.8 $\rightarrow$ 56.8 \\
HellaSwag  & 62.7 $\rightarrow$ 61.1 & 59.2 $\rightarrow$ 57.9 \\
MMLU       & 71.5 $\rightarrow$ 68.6 & 68.0 $\rightarrow$ 61.6 \\
\bottomrule
\end{tabular}
\vspace{1.5mm}
\caption{Generalization performance of Gemma and Llama before and after PHORECAST finetuning. While performance remains stable overall for toxigen, hellaswag, and mmlu, we observe that training with PHORECAST did lead to major changes in triviaqa scores, with Gemma improving substantially and Llama degrading.}
\label{tab:performancebench}
\end{table}

\section{Qualitative Examples}
\label{supp:qual}

We compare model responses before and after training, as illustrated in Figure \ref{fig:qualitative-ex}. Prior to training, the Llama model frequently struggles with task comprehension, often prioritizing the interpretation of visual tokens over emulating the described individual. Post-training, model responses more closely resemble the true human responses.
Furthermore, trained models demonstrate a notable ability to generalize to individuals who did not respond, accurately predicting an 'none' output. This intriguing capability motivates further investigation into theories of fear appeal effectiveness. Our future work will focus on identifying fear-based and danger-based responses, as well as refining the classification of non-responses, informed by these theoretical frameworks. 

\begin{figure}
    \centering
    \includegraphics[trim={1.2cm 1cm 2cm 1cm},clip,width=1\linewidth]{media/supplementary/qual_suppl.pdf}
    \includegraphics[trim={1.2cm 4cm 2cm 1cm},clip,width=1\linewidth]{media/supplementary/qual_suppl2.pdf}
    \caption{\textbf{Qualitative Examples:} We compare responses of different individuals using one of the public health campaigns related to nutrition (top), and mental health (bottom), before and after training with our dataset. Before training, Gemma is able to attempt the task of emulating the different individuals, while Llama just tries to understand and explain the visual. After training using PHORECAST, both model responses are more aligned and human-like.}
    \label{fig:qualitative-ex}
\end{figure}

% \begin{figure}
%     \centering
%     \includegraphics[width=0.5\linewidth]{media/fid_temp.png}
%     \caption{SDE score as a function of iteration steps for Gemma4B.}
%     \label{fig:temp1}
% \end{figure}

\newpage

%%%%%%%%%%%%%%%%%%%%%%%%%%%%%%%%%%%%%%%%
%%% Appendix B: Anh's experiments    %%%
%%%%%%%%%%%%%%%%%%%%%%%%%%%%%%%%%%%%%%%%
%%%%%%%%%%%%%%%%%%%%%%%%%%%%%%%%%%%%%%
%%%     MODEL GENERALIZATION       %%%   
%%%%%%%%%%%%%%%%%%%%%%%%%%%%%%%%%%%%%%
% \newpage
\clearpage
\section{Generalizing Response Predictions to Unseen Communication Strategies}
\label{unseen-comm}
\subsection{Problem Description}
In practice, public health campaigns may have communication strategies that were not covered in the model's training data. We further extend our experiments to test whether a VLM trained on a set of communication strategies can effectively generalize to health messages with novel communication strategies. The core difference from the prior use case is the \textit{train/test partition approach}: instead of splitting randomly based on specific posters, we split based on \textit{communication strategies}. The motivation is to evaluate whether the VLMs trained with our dataset can generalize to a new campaign message with a communication strategy not covered in the training set.

Specifically, for each VLM architecture, we trained 3 different set of weights:
\begin{enumerate}
\item \textbf{Set 1:} Test set includes only messages using \textit{Self-Efficacy} strategies.
\item \textbf{Set 2:} Test set includes only messages using \textit{Informational/Educational/Neutral} strategies.
\item \textbf{Set 3:} Test set includes only messages using \textit{Threatening/Fear-driven} strategies.
\end{enumerate}

This training and evaluation setup enables us to investigate the model’s capacity to generalize to campaign messages with unseen communication strategies in practice.

\begin{figure}[htbp]
    \centering
    \subfigure[Self-Efficacy]{
        \includegraphics[width=0.31\textwidth]{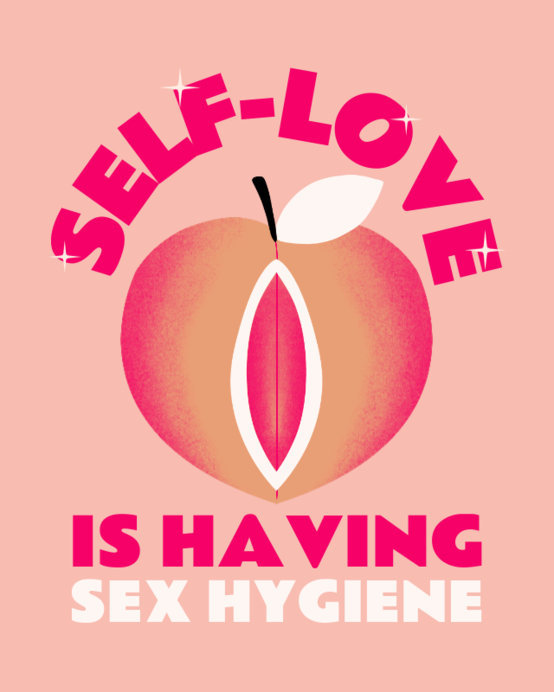}
    }
    \subfigure[Threatening/Fear-Driven]{
        \includegraphics[width=0.29\textwidth]{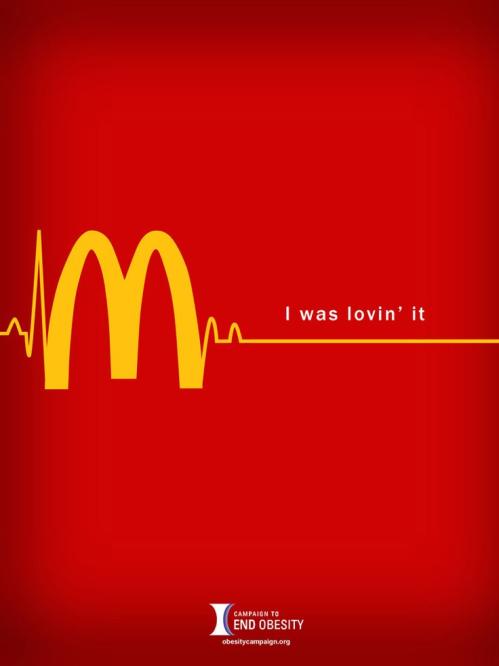}
    }
    \subfigure[Informational/Educational/Neutral]{
        \includegraphics[width=0.465\textwidth]{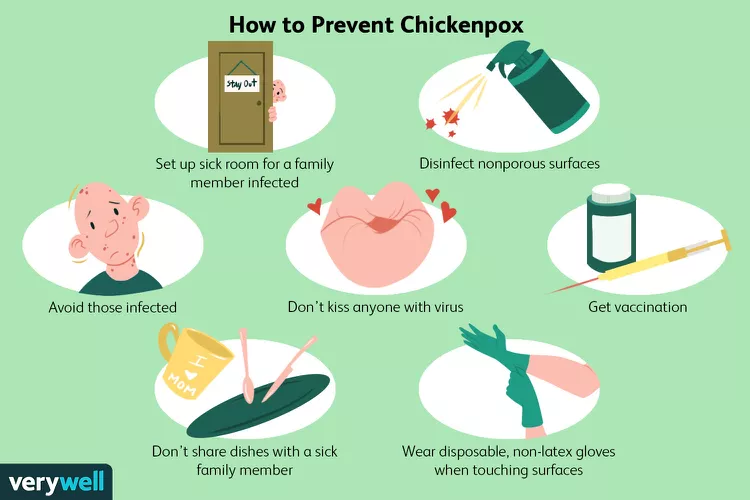}
    }
    \caption{{\bf Examples of Health Campaign Message Communication Strategies.} (a) \textit{Self-Efficacy} (Health Topic: Sexual Practice); (b) \textit{Threatening/Fear-Driven} (Health Topic: Nutrition); (c) \textit{Informational/Educational/Neutral} (Health Topic: Vaccination).}
    \label{fig:message-styles}
\end{figure}

%%%%%%%%%%%%%%%%%%%%%%%%%%%%%%%%%%%%%%%%%%%%%%%%%%%%
%%%             EXAMPLE POSTER STYLES            %%%  
%%%%%%%%%%%%%%%%%%%%%%%%%%%%%%%%%%%%%%%%%%%%%%%%%%%%
Some representative health messages for each of the communication strategies are shown in Fig.~\ref{fig:message-styles}.

\subsection{Experiments \& Analysis}
% \Anh{Change figure captions, color theme, and style}
To demonstrate the value of our dataset in improving personality- and demographic-conditioned response prediction, we compare the performance of zero-shot vision-language models (VLMs) with VLMs fine-tuned on our dataset. Both models are evaluated using the same set of system and instruction prompts to ensure a fair comparison. 

\subsubsection{Personality-specific Evaluations: Generalization to Unseen \textcolor{blue}{``Informational/Neutral''} Strategy.}
% \Anh{Intellectual Curiousity}
In this section, we evaluate whether fine-tuned VLMs can generalize to predict responses to \textbf{Informational/Neutral} health campaign messages held out during training. We compare the predicted response distributions of fine-tuned models against both zero-shot, untrained baselines and ground-truth responses. 

To assess how well the models capture group-specific patterns, we investigate the true and predicted distributions across personality traits partitioned into \emph{low, moderate, and high} levels. To ensure the statistical significance of each group-specific distribution, we excluded any personality group having fewer than 20 individuals. 

\begin{figure}[h!]
    \centering
    \includegraphics[width=0.75\linewidth]{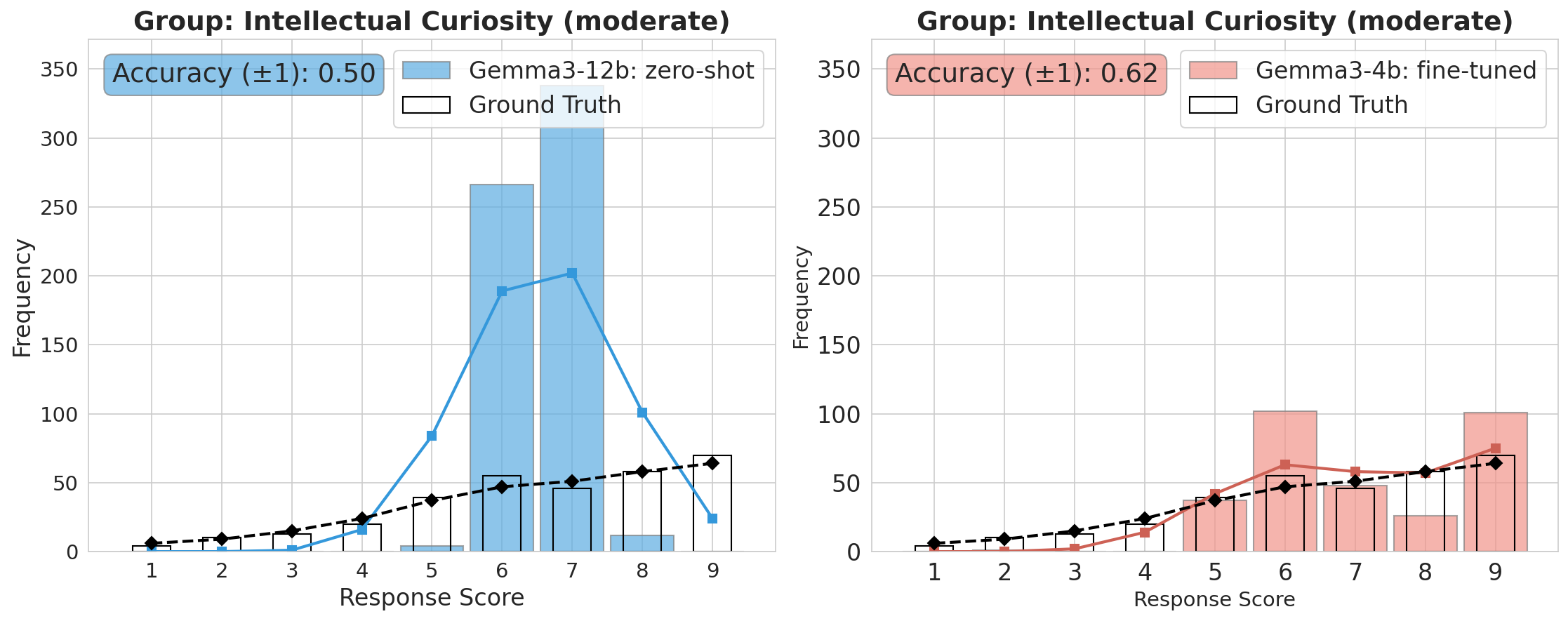}
    \includegraphics[width=0.75\linewidth]{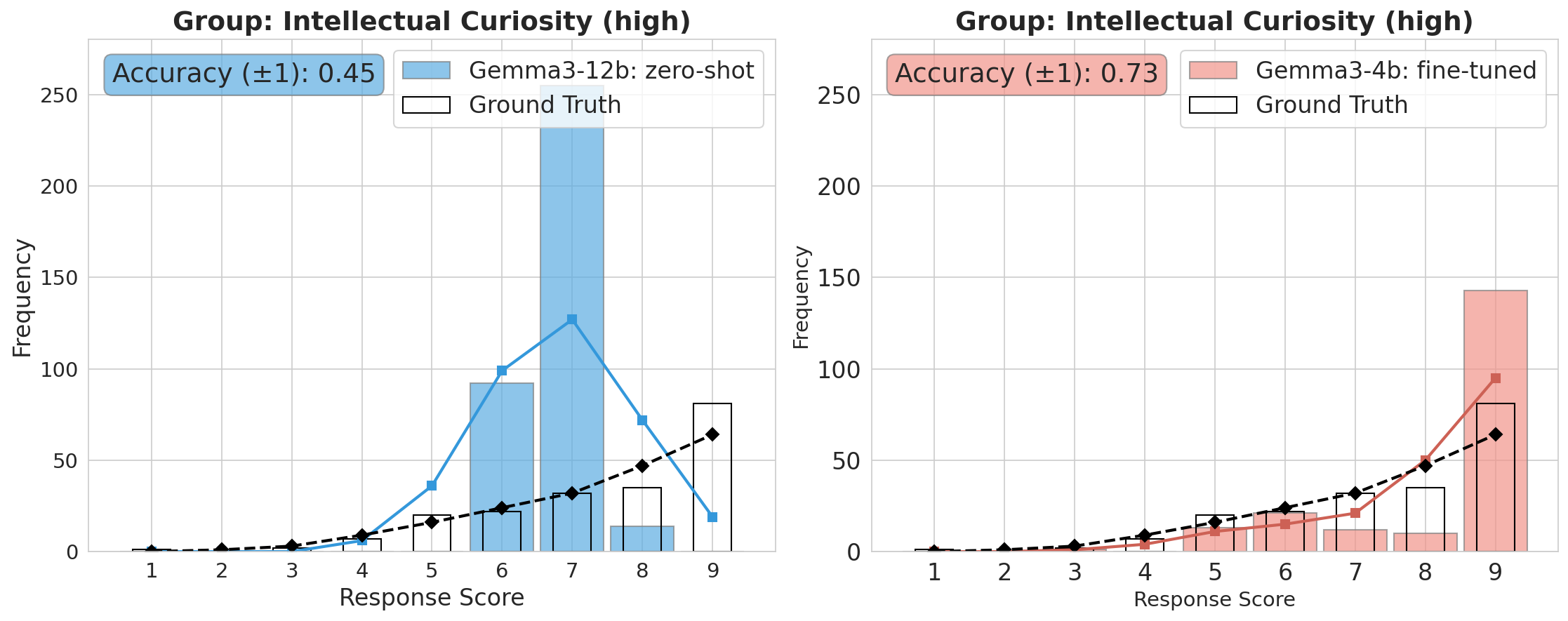}
    \caption{Comparison of sentimental response distributions from \textbf{Gemma} models on unseen \emph{Informational/Neutral} messages, evaluated across \emph{Intellectual Curiosity} personality groups (\emph{moderate, high}). The personality group ``Intellectual Curiosity: \emph{low}'' is not included since it has fewer than 20 samples in the test set. While the zero-shot model (left) shows limited sensitivity to group differences and fails to capture the true response distribution, the trained model (right) using PHORECAST closely aligns with the true personality-conditioned response patterns. Specifically, the $\pm$1 accuracy improved from \textbf{0.50} to \textbf{0.62} for the \emph{moderate} ``Intellectual Curiosity'' group (\textbf{improved by 24\%}), and from \textbf{0.45} to \textbf{0.73} for the \emph{high} group (\textbf{improved by 62.2\%}).}
    \label{fig:eval-style:neutral-persona:intel}
\end{figure}

Comparisons between the predicted response distribution of the zero-shot and trained Gemma model are shown in Fig.~\ref{fig:eval-style:neutral-persona:intel} for varying ``Intellectual Curiosity'' personality groups. After trait-conditioned training, the VLM shows substantially improved alignment with ground-truth distributions, more accurately capturing personality- and demographic-specific sentimental response patterns. In particular, $\pm$1 accuracy increased from 0.50 to 0.62 for the \emph{moderate} ``Intellectual Curiosity'' group (a 24\% gain), and from 0.45 to 0.73 for the \emph{high} group (a 62.2\% gain). Zero-shot baselines tend to give more moderate responses (6--7 out of 9), failing to capture trait-conditioned variations. In contrast, PHORECAST-trained models can capture such response distribution, such as the high-scoring (8–9) responses to Informational/Educational/Neutral messages by those \emph{high} in Intellectual Curiosity.

% INTELLECTUAL-CURIOUSITY groups

\clearpage
\subsubsection{Personality-specific Evaluations: Generalization to Unseen \textcolor{blue}{``Self-Efficacy''} Strategy.}
% \Anh{Trust}
In this section, we evaluate whether fine-tuned VLMs can generalize to predict responses to \textbf{Self-Efficacy} health campaign messages held out during training. We compare the predicted response distributions of fine-tuned models against both zero-shot, untrained baselines and ground-truth responses. 

To assess how well the models capture group-specific patterns, we investigate the true and predicted distributions across personality traits partitioned into \emph{low, moderate, and high} levels. To ensure the statistical significance of each group-specific distribution, we excluded any personality group having fewer than 20 individuals. 

\begin{figure}[h!]
    \centering
    \includegraphics[width=0.75\linewidth]{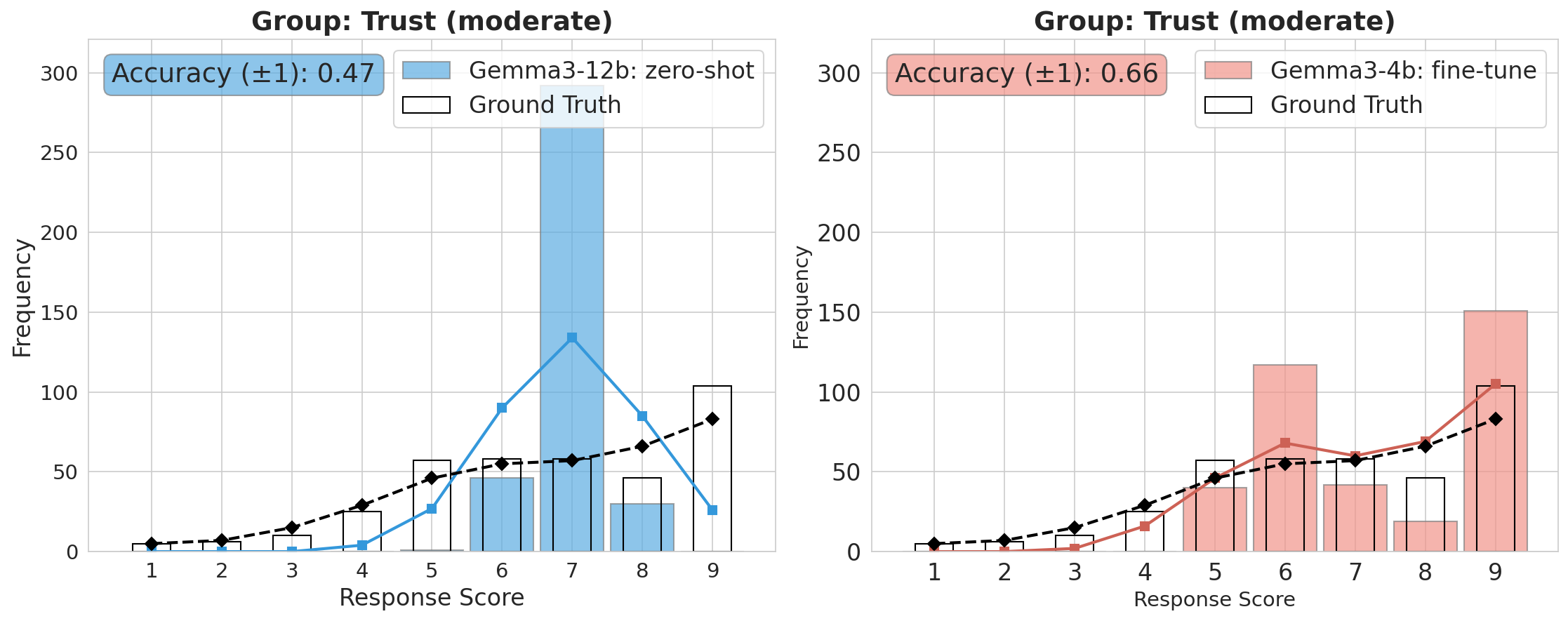}
    \includegraphics[width=0.75\linewidth]{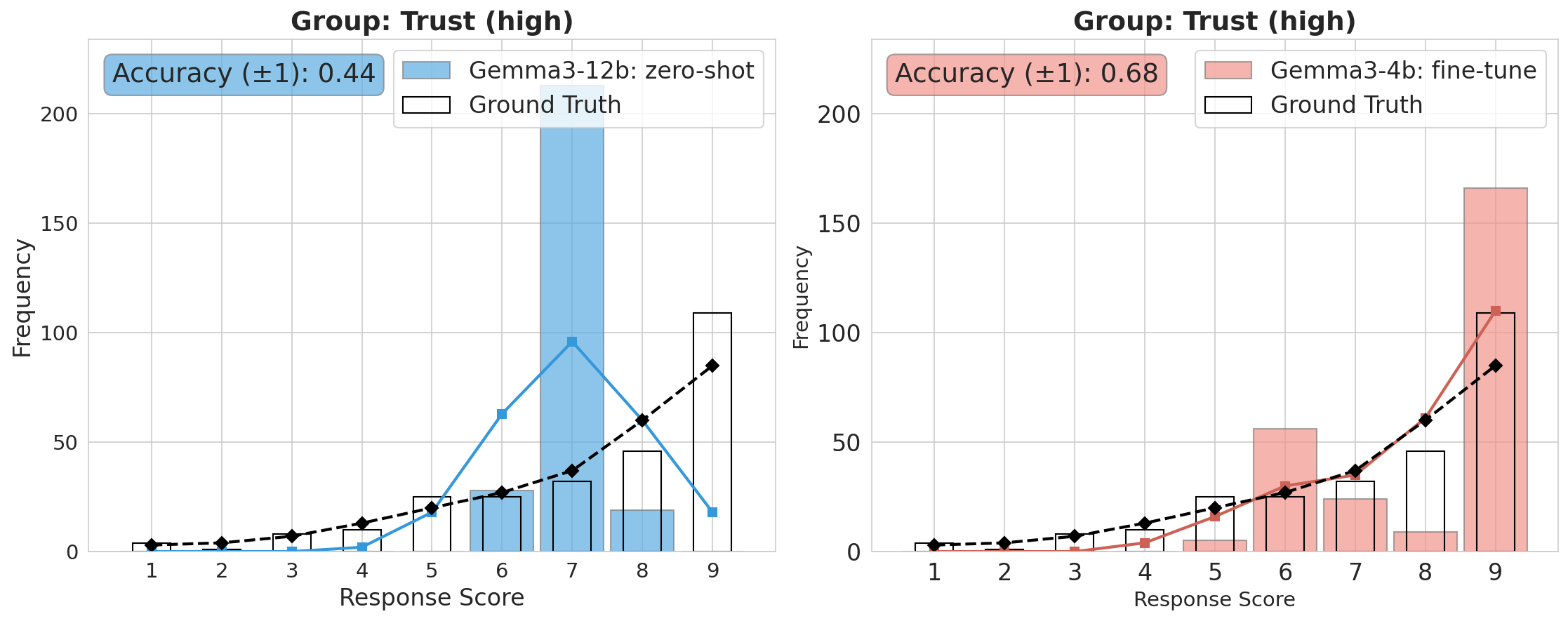}
    \caption{Comparison of sentimental response distributions from \textbf{Gemma3} models on unseen \emph{Self-Efficacy} messages, evaluated across \emph{Trust} personality groups (\emph{moderate, high}). The personality group ``Trust: \emph{low}'' is not included since it has fewer than 20 samples in the test set. While the zero-shot model (left) shows limited sensitivity to the group differences and fails to capture the true response distribution, the trained model using PHORECAST (right) closely aligns with the true personality-conditioned response patterns. Specifically, the $\pm$1 accuracy improved from \textbf{0.47} to \textbf{0.66} for the \emph{moderate} ``Trust'' group (\textbf{improved by 40.4\%}), and from \textbf{0.44} to \textbf{0.68} for the \emph{high} group (\textbf{improved by 54.5\%}).}
    \label{fig:eval-style:efficacy-persona:trust}
\end{figure}

Comparisons between the predicted response distribution of the zero-shot and trained Gemma3 model are shown in Fig.~\ref{fig:eval-style:efficacy-persona:trust}  for varying ``Trust'' personality groups. After trait-conditioned training, the VLM shows substantially improved alignment with ground-truth distributions, more accurately capturing personality- and demographic-specific sentimental response patterns. In particular, $\pm$1 accuracy increased from 0.47 to 0.66 for the \emph{moderate} ``Trust'' group (a 40.4\% gain), and from 0.44 to 0.68 for the \emph{high} group (a 54.5\% gain).

Qualitatively, the zero-shot, pretrained baselines tend to give more moderate responses (6--7 out of 9) with some highly positive responses (8 out of 9). It fails to correctly capture the tendency to have very positive sentimental responses (9 out of 9) among the \emph{high} ``Trust'' group. In contrast, PHORECAST-trained models can capture such response distribution and patterns for both \emph{moderate} and \emph{high} ``Trust'' groups. Since the sentimental response score are often imprecise by nature, there is practically little difference between prediction sentimental score of 8 versus 9 (or 5 versus 6), showing that the trained VLM is able to capture the overall trend in different ``Trust'' groups, as reflected in the similar distribution shapes and high ±1 accuracy.

% TRUST groups

\clearpage
\subsubsection{Personality-specific Evaluations: Generalization to Unseen \textcolor{blue}{``Threatening/Fear-driven''} Strategy.}
% \Anh{Neurocitism}
In this section, we evaluate whether fine-tuned VLMs can generalize to predict responses to \textbf{Threatening/Fear-driven} health campaign messages held out during training. We compare the predicted response distributions of fine-tuned models against both zero-shot, untrained baselines and ground-truth responses. 

To assess how well the models capture group-specific patterns, we investigate the true and predicted distributions across personality traits partitioned into \emph{low, moderate, and high} levels. To ensure the statistical significance of each group-specific distribution, we excluded any personality group having fewer than 20 individuals. 

\begin{figure}[h!]
    \centering
    \includegraphics[width=0.75\linewidth]{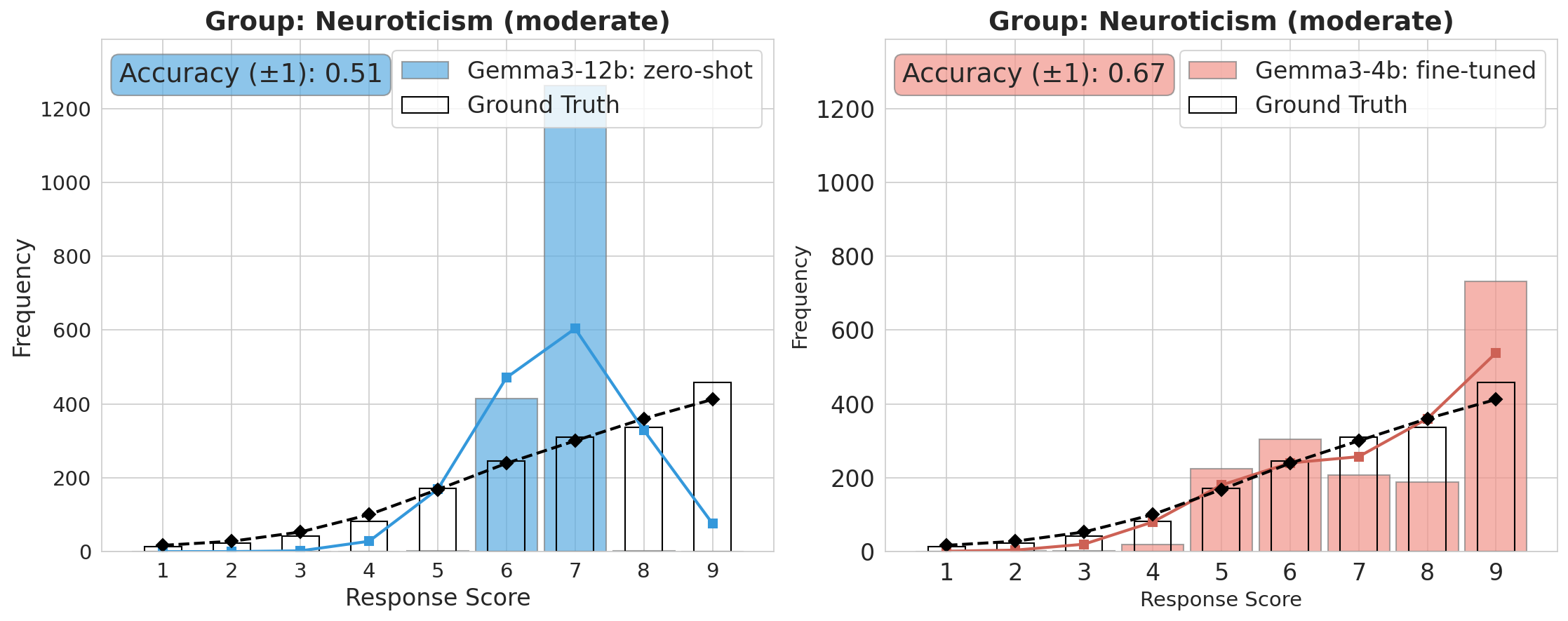}
    \includegraphics[width=0.75\linewidth]{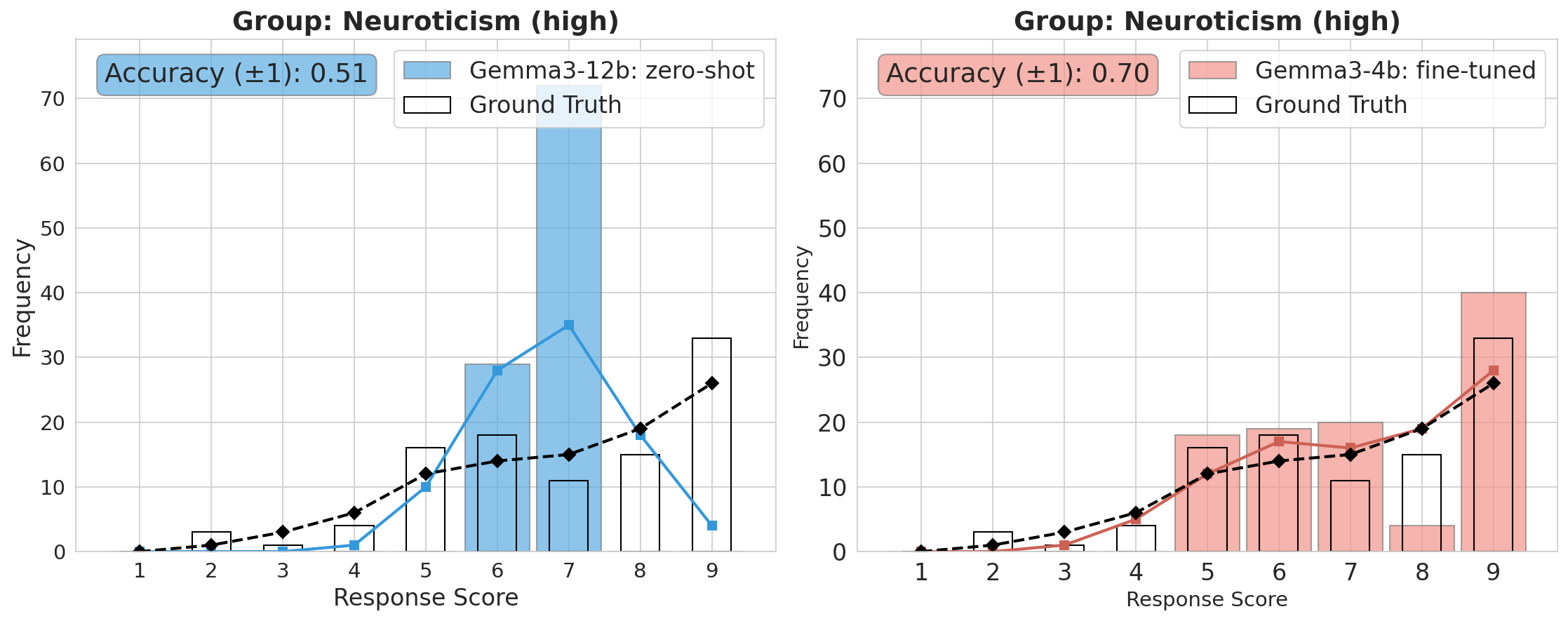}
    \caption{Comparison of sentimental response distributions from \textbf{Gemma3} models on unseen \emph{Threatening/Fear-driven} messages, evaluated across \emph{Neurocitism} personality groups (\emph{moderate, high}). While the zero-shot model (left) shows limited sensitivity to the group differences and fails to capture the true response distribution, the trained model using PHORECAST (right) closely aligns with the true personality-conditioned response patterns. Specifically, the $\pm$1 accuracy improved from \textbf{0.51} to \textbf{0.67} for the \emph{moderate} ``Neurocitism'' group (\textbf{improved by 31.4\%}), and from \textbf{0.51} to \textbf{0.70} for the \emph{high} group (\textbf{improved by 37.3\%}).}
    \label{fig:eval-style:threat-persona:neuro}
\end{figure}

Comparisons between the predicted response distribution of the zero-shot and trained Gemma3 model are shown in Fig.~\ref{fig:eval-style:threat-persona:neuro}  for varying ``Neurocitism'' personality groups. After trait-conditioned training, the VLM shows substantially improved alignment with ground-truth distributions, more accurately capturing personality- and demographic-specific sentimental response patterns. In particular, $\pm$1 accuracy increased from 0.51 to 0.67 for the \emph{moderate} ``Neurocitism'' group (a 31.4\% gain), and from 0.51 to 0.70 for the \emph{high} group (a 37.3\% gain).

The zero-shot pretrained baseline again shows poor alignment with true sentimental responses, predominantly predicting moderate scores of 6–7 out of 9. Additionally, its predicted response distribution shows little variation between the \emph{moderate} and \emph{high} neuroticism groups. In contrast, the PHORECAST-trained model captures sentiment variations both within and between different ``Neuroticism'' personality groups.

% EMOTIONAL VOLATILITY groups

\clearpage
%%%%%%%%%%%%%%%%%%%%%%%%%%%%%%%%%%%%%%
%%%     MODEL GENERALIZATION       %%%   
%%%%%%%%%%%%%%%%%%%%%%%%%%%%%%%%%%%%%%
% \newpage
\clearpage
\section{Future Practical Use Case: VLM-enabled Health Communication Strategy Recommendation}
\label{comm-rec}
\subsection{Problem Description and Prediction Pipeline}
% \Anh{Motivation for strategy recommendation} \\
% \Anh{Describe high-level model pipeline: inputs, predictions} \\

\begin{figure}[h!]
    \centering
    \includegraphics[width=1.0\linewidth]{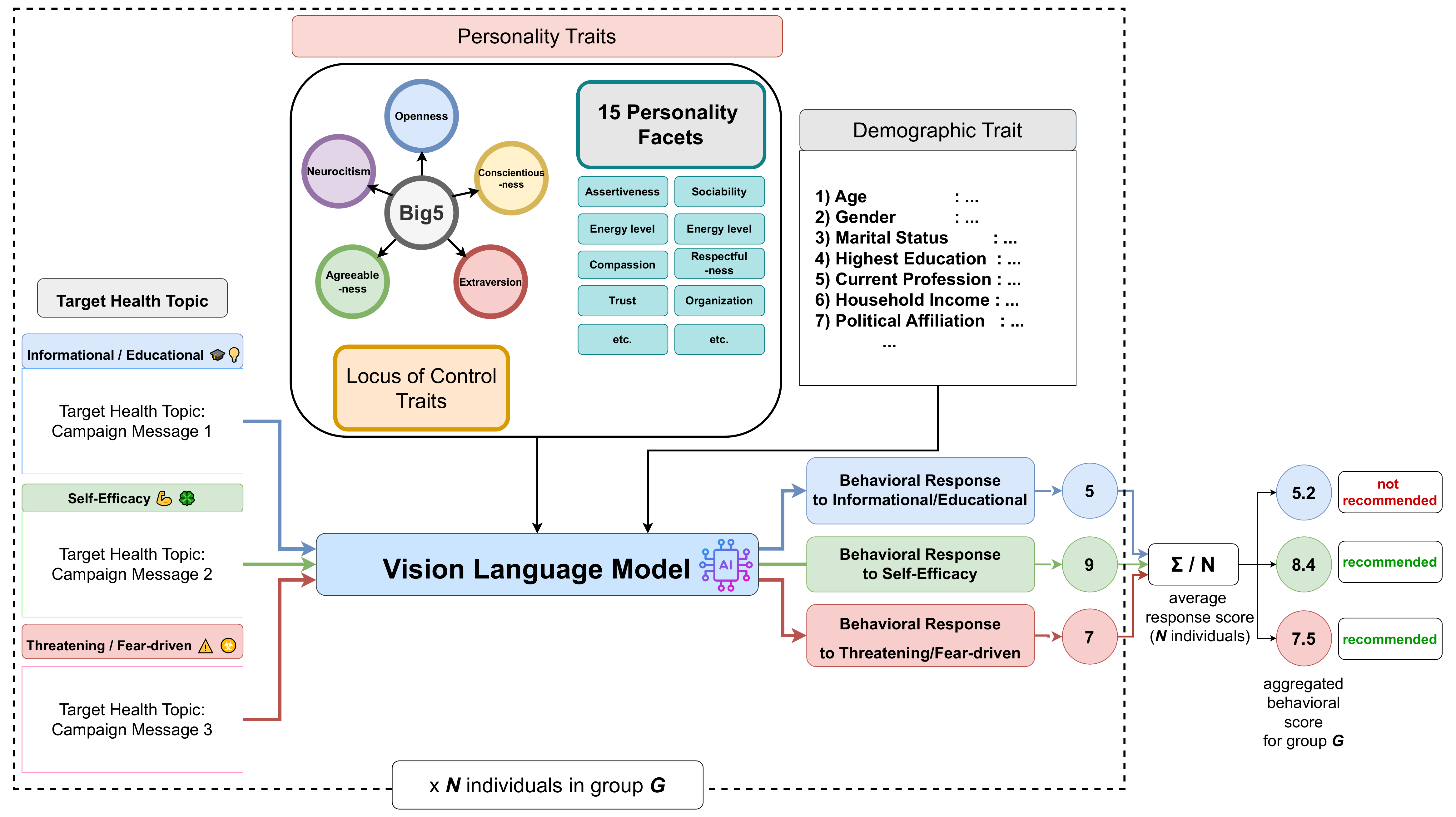}
    \caption{VLM-enabled Personality- and Demographic-conditioned   Communication Strategy Recommendation to different Individuals/Communities. The model aims to recommend the potentially most effetive communication strategy that likely have positive impacts on a given individual/community with a particular personality and demographic traits in a health topic.}
    \label{fig:model-pipeline-recommendation}
\end{figure}

In this section, we describe a potential practical use case of VLM-enabled trait-conditioned response prediction as an interesting line of future work:  VLM-enabled communication strategies recommendation tailored to specific personality or demographic groups. Based on predicted responses to different health campaign posters, we can aggregate VLM-predicted reactions across individuals within a group to identify the most effective messaging strategy. An overview of the VLM-enabled recommendation pipeline is shown in Fig.~\ref{fig:model-pipeline-recommendation}.
% At a high level, the formulation is as follows:
% $$s = VLM(x_{persona}, x_{demo}, x_{topic})$$
% where $s$ is the outputted recommendation strategy ($s \in $ \{ \textit{Threatening/Fear-driven}, \textit{Self-Efficacy}, \textit{Informational/Neutral} \} ); $x_{persona}, x_{demo}, x_{topic}$ are the personality traits, demographic information, and target health topic, respectively. 

The high-level idea is to aggregate the responses of each person in a group $G$ to different communication strategies in the targeted health topic. 
Given a health topic for which behavior change is targeted, we consider a set of health campaign messages $V^s$, each associated with a communication strategy $s \in { \textit{Threatening/Fear-driven}, \textit{Self-Efficacy}, \textit{Informational/Neutral} }$. Using a trained VLM, we predict how individuals with given traits respond to each strategy:
$$\hat{y}^{(i),s} = VLM(V^{s}, x_{persona}^{(i)}, x_{demo}^{(i)}) \quad \forall i \in [1,N] $$
where $\hat{y}^{(i)}$ is the individual $i$'s behavioral response to visual health campaign message $V^{s}$, conditioned on their personality $x_{persona}^{(i)}$ and demographic information $x_{demo}^{(i)}$, and $N$ is the total individuals in group or community $G$. The behavioral responses of the group $G$ to different health messages $V^s$ with different strategies $s$ are aggregated as:
$$y^{group,s} = \sum \hat{y}^{(i),s} / N$$
As discussed in the previous sections, the behavioral responses in PHORECAST are measured on a 9-point Likert Scale, in which scores $\geq 7$ correspond to ``positive'' responses. Therefore, to recommend the best communication strategies for a group/community $G$, the VLM outputs all communication strategies $s$ such that $y^{group,s} \geq 7$.

By leveraging a VLM-enabled communication strategy recommendation system, we can tailor public health messages to specific personality and demographic profiles, thereby maximizing message effectiveness for diverse target groups. This line of future work has the potential to enhance awareness of health issues, spread important health information, and promote healthier behaviors at scale. Beyond the immediate applications in public health, this framework is also applicable to future research in other disciplines such as political science, education, and social marketing. These applications not only highlight interesting lines of future technical works but also the direct societal impacts of our PHORECAST dataset and the models presented in this paper.

To account for multiple effective strategies, we suggest using an evaluation pipeline that considers any recommended strategies to be "correct" if it is among the true effective strategies. For example, if all strategies for the health topic "Nutrition" are effective for personality "Open-Mindedness: \emph{high}", any recommended strategy in "Nutrition" is considered to be valid for this personality group.

\clearpage

%%%%%%%%%%%%%%%%%%%%%%%%%%%%%%%%%%%%%%%%%%%%%%%%%%%%%%%%%%%%
% \input{tex/neurips_checklist}

\end{document}